\pdfoutput=1
\documentclass[preprint]{imsart}
\usepackage[top=1in,bottom=1in,right=0.9in,left=0.9in]{geometry}
\usepackage{placeins}
\usepackage{array}
\usepackage{amsthm,amsmath}
\makeatletter
\let\c@author\relax
\makeatother
\usepackage[backend=biber, style=numeric]{biblatex}
\bibliography{bibliography}
\usepackage{hyperref}
\usepackage{bbm}
\hypersetup{
  colorlinks,
  citecolor= blue,
  linkcolor=blue}
\usepackage{subfigure,tikz}
\usetikzlibrary{backgrounds,automata}

\usepackage{multirow}
\usepackage{color,soul}
\usepackage{arydshln}

\usepackage{bbold}
\usepackage{rotating}
\usepackage{multicol}
\newcommand\independent{\protect\mathpalette{\protect\independenT}{\perp}}
\def\independenT#1#2{\mathrel{\rlap{$#1#2$}\mkern2mu{#1#2}}}
\usepackage{mathtools}
\usepackage{blindtext}
\usepackage{graphicx}
\usepackage{caption}
\usepackage{subfigure}
\usepackage[]{graphicx}\usepackage[]{color}
\usepackage{makecell}
\usepackage{booktabs}
\usepackage{enumerate}

\newtheorem{theoremp}{Proposition}

\usepackage{lmodern}
\begin{document}

\begin{frontmatter}
\title{Bayesian Record Linkage with Variables in One File}
\author{\fnms{Gauri} \snm{Kamat}\thanksref{t1}\thanksref{t3},}
\author{\fnms{Mingyang} \snm{Shan}\thanksref{t2}}
\and
\author{\fnms{Roee} \snm{Gutman}\thanksref{t1}}

\thankstext{t1}{Department of Biostatistics, Brown University}
\thankstext{t2}{Eli Lilly and Company}
\thankstext{t3}{Correspondence to: gaurik1.edu@gmail.com}


\bigskip
\begin{abstract}
In many healthcare and social science applications, information about units is dispersed across multiple data files. Linking records across files is necessary to estimate the associations of interest. Common record linkage algorithms only rely on similarities between linking variables that appear in all the files. Moreover, analysis of linked files often ignores errors that may arise from incorrect or missed links. Bayesian record linking methods allow for natural propagation of linkage error, by jointly sampling the linkage structure and the model parameters. We extend an existing Bayesian record linkage method to integrate associations between variables exclusive to each file being linked. We show analytically, and using simulations, that this method can improve the linking process, and can yield accurate inferences. We apply the method to link Meals on Wheels recipients to Medicare Enrollment records.
\end{abstract}


\end{frontmatter}

\section{Introduction}
Access to large data files with rich information is becoming increasingly prevalent. Such data files enable researchers to investigate complex scientific questions. However, confidentiality restrictions, or the nature of data collection, may partition individuals' information across two or more files. When unique identifiers are available, linking individuals' records across files is a relatively simple task. This task is less straightforward when such identifiers
are unavailable, or only weakly informative  \cite{Winkler1993b}. Record linkage is a statistical technique that identifies overlapping records across files, in the absence of unique identifiers. Record linkage methods have been used to combine and aggregate data in medical studies  \cite{Newcombe1988,Gill1977}, sociological studies  \cite{Sadinle2018}, and US Census Bureau applications  \cite{Winkler1994,Winkler1995}. 

Record linkage methods can be broadly categorized into two classes, namely deterministic and probabilistic. Deterministic methods link records via exact matching on the linking variables that exist in all the files  \cite{Gomatam2002}. These methods identify a high proportion of true links when the linking variables are informative. However, when the variables are subject to errors, such as variations in spelling, data entry inaccuracies, or incompleteness, deterministic methods may miss a large number of true links  \cite{Campbell2008}. Probabilistic record linkage is based on the likelihood that each pair of records represents a true link. This likelihood can be computed using a mixture model proposed by Fellegi and Sunter \cite{Fellegi1969}, microclustering models  \cite{zanellabetancourt}, or using machine learning algorithms like random forests  \cite{cochinwala2001,chen2009}.




In practice, record linkage is often not the end goal; instead, researchers seek to estimate associations between variables exclusive to the files being linked. Many current inference methods for linked files assume that linking probabilities for every record-pair are known, or can be estimated from the available data. Moreover, they are restricted to specific regression models  \cite{Scheuren1993,Scheuren1997,Lahiri2005,wang2022}. Other approaches attempt to adjust for linkage errors in generalized estimating equations \cite{kimchambers2012,chambersdasilva2020}. However, these approaches also rely on possibly unknown linkage error rates for estimation.

Bayesian record linkage methods address some of these limitations. One set of Bayesian methods build on the Fellegi-Sunter model, and posit that similarity measures derived from the linking variables form a mixture of links and non-links \cite{Fortini2001,Sadinle2017}. An alternative Bayesian method models the measurement error in the linking variables, to identify records that belong to the same latent entity  \cite{Tancredi2011,Steorts2015}. All of these Bayesian methods introduce a latent linkage structure, that is sampled along with the parameters of the linkage model. This can effectively propagate linkage errors into the downstream statistical analysis \cite{binette22}. \par


The Fellegi-Sunter-based models and the measurement error models only consider comparisons of variables common to all the files. Gutman et al. \cite{Gutman2013} propose a Bayesian method that models the relationship between variables that exist in either file. This method assumes that all records in the smaller file exist in the larger file. The model in \cite{Tang2020} builds on the model of Sadinle \cite{Sadinle2017}, and incorporates associations between variables in each file among the true links. However, \cite{Tang2020} does not discuss adjustments with blocking. Moreover, it ignores possible associations among non-links, which may arise when the blocking variables are associated with variables in the downstream analysis. \par

In this article, we propose a Bayesian record linkage method that extends the model of Sadinle \cite{Sadinle2017}, to incorporate associations between variables exclusive to the files being linked. We provide theoretical justification for the improvements in linkage, and discuss possible adjustments for blocking. Using simulations, we show that this method can result in increased linkage accuracy when the linking variables are not highly informative. 

The remainder of this article is organized as follows. Section \ref{secbg} provides background on Bayesian Fellegi-Sunter modeling. Section \ref{brlvof} describes the proposed method. Section \ref{sim} describes simulations evaluating the method. Section \ref{data} applies the method to link a data file from Meals on Wheels America to a Medicare claims file. Section \ref{disc} concludes with a discussion.

\section{Background} \label{secbg}
\subsection{Data structure and notation}
Let $\mathbf{A}$ and $\mathbf{B}$ represent two files containing $n_A$ and $n_B$ records, respectively. Let $\mathbf{Z}_{Ai}=(Z_{Ai1}, \dots, Z_{AiK})$ represent $K$ linking variables for record $i \in \mathbf{A}$, where $i=1, \dots, n_A$. Similarly, let $\mathbf{Z}_{Bj}=(Z_{Bj1}, \dots, Z_{BjK})$ represent the same $K$ linking variables for record $j \in \mathbf{B}$, where $j = 1,\dots,n_B$. Let $\mathbf{X}_{Ai}=(X_{Ai1}, \dots, X_{AiP})$ denote $P$ variables that are exclusive to file $\mathbf{A}$. Let $\mathbf{X}_{Bj}=(X_{Bj1}, \dots, X_{BjQ})$ denote $Q$ variables exclusive to file $\mathbf{B}$. Further, let $\mathbf{C}=\{ C_{ij}\}$ denote a latent $n_A \times n_B$ binary matrix that represents the linkage structure between files $\mathbf{A}$ and $\mathbf{B}$. The  $(i,j)^{th}$ entry  of $\mathbf{C}$ is given by
\begin{equation} \label{eq:bg1}
    C_{ij}=
     \begin{cases} 
      1 & \text{if records $i \in \mathbf{A}$ and $j \in \mathbf{B}$ represent the same entity,} \\
      0 & \text{otherwise. }
   \end{cases}
\end{equation}
To obtain a linkage estimate that is one-to-one, where a record in $\mathbf{A}$ can be linked with at most one record in $\mathbf{B}$ and vice versa, the following constraints are placed on $\mathbf{C}$: $\sum_i C_{ij}\leq 1$,$\forall j \in \mathbf{B}$, and  $\sum_j C_{ij} \leq 1$, $\forall i \in \mathbf{A}$.

To assess the similarity between records $i \in \mathbf{A}$ and $j \in \mathbf{B}$ on the $K$ linking variables, comparison vectors $\mathbf{\Gamma}(\mathbf{Z}_{Ai},\mathbf{Z}_{Bj})=(\gamma_{ij1}, \dots, \gamma_{ijK})$ are constructed. The similarity on field $k$ is defined using $L_k$ levels of ordinal agreement, where 1 represents complete disagreement and $L_k$ indicates complete agreement \cite{Winkler1990}. Formally, $\gamma_{ijk}$ is represented by a set of indicator variables $\{\gamma_{ijkl_k} : l_k=1, \dots, L_k\}$, where
\begin{equation} \label{eq:bg2}
    \gamma_{ijkl_k}=
     \begin{cases} 
      1 & \text{if $Z_{Aik}$ and $Z_{Bjk}$ have the $l_k^{th}$ level of agreement}, \\
      0 & \text{otherwise. }
   \end{cases} 
\end{equation}

\subsection{Bayesian Fellegi-Sunter model}
The Fellegi and Sunter \cite{Fellegi1969} framework considers the set of all record pairs in $\mathbf{A} \times \mathbf{B}$ as the union of two disjoint sets: true links, $\mathbf{M}=\{(i,j): i \in \mathbf{A}, j \in \mathbf{B}, C_{ij}=1 \}$, and non-links, $\mathbf{U}=\{(i,j): i \in \mathbf{A}, j \in \mathbf{B}, C_{ij}=0 \}$. Let $\theta_{M}=\{\theta_{Mk}:k=1, \dots, K \}$ and $\theta_{U}=\{\theta_{Uk}: k=1, \dots, K \}$ be parameters governing the distributions of $\mathbf{\Gamma}(\mathbf{Z}_{Ai},\mathbf{Z}_{Bj})$ for true links and non-links, respectively. The distribution of $\mathbf{\Gamma}(\mathbf{Z}_{Ai},\mathbf{Z}_{Bj})$ given the latent linkage status $C_{ij}$ is 

\begin{align} \label{eq:bg3}
\begin{split}
   \mathbf{\Gamma}(\mathbf{Z}_{Ai},\mathbf{Z}_{Bj})| C_{ij}=1 & \sim f_M(\mathbf{\Gamma}(\mathbf{Z}_{Ai},\mathbf{Z}_{Bj})|\theta_{M}), \\
    \mathbf{\Gamma}(\mathbf{Z}_{Ai},\mathbf{Z}_{Bj})| C_{ij}=0 & \sim f_U(\mathbf{\Gamma}(\mathbf{Z}_{Ai},\mathbf{Z}_{Bj})|\theta_{U}).  
\end{split}
\end{align}

Let $\theta_{Mk} = \{\theta_{Mkl_{k}}\}$ and $\theta_{Uk} = \{\theta_{Ukl_{k}}\}$, where   $\theta_{Mkl_k}=Pr(\gamma_{ijk}=l_k|C_{ij}=1)$ and $\theta_{Ukl_k}=Pr(\gamma_{ijk}=l_k|C_{ij}=0)$, for $k=1, \dots, K$. A common simplifying assumption is that each $\theta_{Mk}$ and each $\theta_{Uk}$ are conditionally independent given $\mathbf{C}$ \cite{Winkler1989,Jaro1989}. Under this assumption, the Bayesian record linkage likelihood is
\begin{align} \label{eq:bg4}
    \mathcal{L}^{BRL}(\mathbf{C},\theta_M, \theta_U,| \mathbf{Z}_A, \mathbf{Z}_B)= \prod_{i=1}^{n_A} \prod_{j=1}^{n_B} &\bigg[\prod_{k=1}^{K} \prod_{l_k=1}^{L_k} \theta_{Mkl_{k}}^{\mathbbm{1}(\gamma_{ijk}=l_k)} \bigg]^{C_{ij}}
     \bigg[\prod_{k=1}^{K} \prod_{l_k=1}^{L_k} \theta_{Ukl_{k}}^{\mathbbm{1}(\gamma_{ijk}=l_k)} \bigg]^{1-C_{ij}}.
\end{align}

To complete the Bayesian model, we assume independent Dirichlet prior distributions, $\theta_{Mk} \sim \mathrm{Dirichlet}(\alpha_{Mk1}, \dots, \alpha_{MkL_k})$ and $\theta_{Uk} \sim \mathrm{Dirichlet}(\alpha_{Uk1}, \dots, \alpha_{UkL_k})$, for $k=1, \dots, K$. In addition, we specify a prior distribution for $\mathbf{C}$ that is independent of $\theta_M$ and  $\theta_U$ \cite{Sadinle2017}. Let $n_m=\sum_{i=1}^{n_A} \sum_{j=1}^{n_B} C_{ij}$ be the number of true links, such that $n_m \sim \mathrm{Binomial}(\min(n_A,n_B),\pi)$ and $\pi \sim \mathrm{Beta}(\alpha_{\pi}, \beta_{\pi})$, where $\alpha_{\pi}$ and $\beta_{\pi}$ are known \textit{a priori}. Conditional on $n_m$, the prior distribution for $\mathbf{C}$ is uniform over all $\binom{n_A}{n_m} \binom{n_B}{n_m} n_m!$ linking configurations that satisfy the one-to-one linkage constraint with $n_m$ links. The probability mass function for $\mathbf{C}$, marginalized over $\pi$, is \cite{Sadinle2017}
\begin{align}
    &p(\mathbf{C},n_{m}|\alpha_{\pi},\beta_{\pi})=\dfrac{(\max(n_{A},n_{B})-n_{m})!}{\max(n_{A},n_{B})!}  \text{ } \frac{ \textit{B}(n_{m}+\alpha_{\pi},\min(n_{A},n_{B}) - n_{m}+\beta_{\pi})}{\textit{B}(\alpha_{\pi},\beta_{\pi})},
\end{align}
where $\textit{B}(.)$ represents the Beta function. This prior distribution was introduced by Larsen\cite{Larsen2005}, with later uses by Tancredi and Liseo \cite{Tancredi2011} and Sadinle \cite{Sadinle2017}. We refer to this Bayesian record linkage model as BRL throughout. \par

Suppose that the goal of the study is to estimate the association between $\mathbf{X}_{A}$ and $\mathbf{X}_{B}$ among the identified links. Under BRL, estimation proceeds under the assumption that $P(\mathbf{C}|\mathbf{Z}_{A},\mathbf{Z}_{B},\mathbf{X}_{A},\mathbf{X}_{B}) = P(\mathbf{C}|\mathbf{Z}_{A},\mathbf{Z}_{B})$.  In other words, the linkage is independent of $\mathbf{X_A}$ and $\mathbf{X_B}$ given the information contained in $\mathbf{Z}_{A}$ and $\mathbf{Z}_{B}$.

Let $\beta$ be a set of parameters governing the joint distribution of $(\mathbf{X_A}, \mathbf{X_B},\mathbf{Z}_{A},\mathbf{Z}_{B})$. The posterior distribution of $\beta$ is
\begin{align} \label{xx}
P(\beta|\mathbf{Z}_{A},\mathbf{Z}_{B},\mathbf{X}_{A},\mathbf{X}_{B}) = \sum_{\mathbf{C}} P(\beta|\mathbf{Z}_{A},\mathbf{Z}_{B},\mathbf{X}_{A},\mathbf{X}_{B},\mathbf{C}) \text{ } P(\mathbf{C}|\mathbf{Z}_{A},\mathbf{Z}_{B}).
\end{align}
Based on Equation \eqref{xx}, Bayesian inference on $\beta$ can be accomplished using the following steps  \cite{Sadinle2018}: (i) Sample $\tilde{\mathbf{C}}$ from  $P(\mathbf{C}|\mathbf{Z}_{A},\mathbf{Z}_{B})$; (ii) Sample $\beta$ from  $P(\beta|\mathbf{Z}_{A},\mathbf{Z}_{B},\mathbf{X}_{A},\mathbf{X}_{B},\tilde{\mathbf{C}})$; and (iii) Repeat steps (i) and (ii) sufficiently large number of times. The resultant samples of $\beta$ can be used for point and interval estimation.

\section{Bayesian Record Linkage with Variables in One File} \label{brlvof}
We augment the BRL model by introducing the relationship between $\mathbf{X}_{A}$ and $\mathbf{X}_{B}$, in addition to the comparisons between $\mathbf{Z}_{A}$ and $\mathbf{Z}_{B}$. Formally, this relationship can be modeled  as $f_M(\mathbf{X}_{Ai}, \mathbf{X}_{Bj}|\mathbf{Z}_{Ai},\mathbf{Z}_{Bj}, \beta_{M})$ and $f_{U}(\mathbf{X}_{Ai},\mathbf{X}_{Bj}|\mathbf{Z}_{Ai},\mathbf{Z}_{Bj},\beta_{U})$, for record pair $(i,j)$ in $\mathbf{M}$ and $\mathbf{U}$, respectively. The exact specification of these distributions would vary depending on the application. Using these distributions, the mixture model in Equation \eqref{eq:bg3} can be re-expressed as:
\begin{align} \label{eq:brlf1}
\begin{split}
     \mathbf{X}_{Ai},\mathbf{X}_{Bj},\mathbf{Z}_{Ai},\mathbf{Z}_{Bj}| C_{ij}=1 & \sim f_M(\mathbf{X}_{Ai},\mathbf{X}_{Bj}|\mathbf{Z}_{Ai},\mathbf{Z}_{Bj},\beta_{M})\text{ }f_M(\mathbf{\Gamma}(\mathbf{Z}_{Ai},\mathbf{Z}_{Bj})|\theta_{M}),  \\
    \mathbf{X}_{Ai},\mathbf{X}_{Bj},\mathbf{Z}_{Ai},\mathbf{Z}_{Bj}| C_{ij}=0 & \sim f_U(\mathbf{X}_{Ai},\mathbf{X}_{Bj}|\mathbf{Z}_{Ai},\mathbf{Z}_{Bj},\beta_{U})\text{ }f_U(\mathbf{\Gamma}(\mathbf{Z}_{Ai},\mathbf{Z}_{Bj})|\theta_{U}). 
\end{split}
\end{align}
Assuming $\theta_{Mk}$ and $\theta_{Uk}$ are conditionally independent given $\mathbf{C}$, the likelihood will be 
\begin{align} \label{eq:brlf2}
    \mathcal{L}^{BRLVOF}(\mathbf{C},\theta_M, \theta_U, \beta_M, \beta_U|\mathbf{X}_A, \mathbf{X}_B, \mathbf{Z}_A, \mathbf{Z}_B)= \prod_{i=1}^{n_A} \prod_{j=1}^{n_B} &\bigg[f_M(\mathbf{X}_{Ai},\mathbf{X}_{Bj}|\mathbf{Z}_{Ai},\mathbf{Z}_{Bj},\beta_M) \prod_{k=1}^{K} \prod_{l_k=1}^{L_k} \theta_{Mkl_{k}}^{\mathbbm{1}(\gamma_{ijk}=l_k)} \bigg]^{C_{ij}} \nonumber \\
     \times &\bigg[f_U(\mathbf{X}_{Ai},\mathbf{X}_{Bj}|\mathbf{Z}_{Ai},\mathbf{Z}_{Bj},\beta_U) \prod_{k=1}^{K} \prod_{l_k=1}^{L_k} \theta_{Ukl_{k}}^{\mathbbm{1}(\gamma_{ijk}=l_k)} \bigg]^{1-C_{ij}}. 
\end{align}
The prior distributions for $\theta_{Mk}$, $\theta_{Uk}$, and $\mathbf{C}$ remain the same as in BRL, while the prior distributions of $\beta_M$ and $\beta_U$ conditional on $\theta_{Mk}$, $\theta_{Uk}$, and $\mathbf{C}$ will be specific to the application, and the distributional form of $f_M(\mathbf{X}_A,\mathbf{X}_B|\mathbf{Z}_A,\mathbf{Z}_B,\beta_M)$ and $f_U(\mathbf{X}_A,\mathbf{X}_B|\mathbf{Z}_A,\mathbf{Z}_B,\beta_U)$. We refer to this method as Bayesian record linkage with variables in one file (BRLVOF) throughout.

\subsection{Theoretical insights}
We present results that demonstrate the added benefit of BRLVOF over BRL in detecting true links and non-links. We base the results on the evidentiary strength of the likelihood ratio under the two methods \cite{Guha2022}.

\begin{theoremp} \label{prop1} Let $LR_{BRLVOF} = \frac{\mathcal{L}^{BRLVOF}_{(i,j) \in \mathbf{M}}}{\mathcal{L}^{BRLVOF}_{(i,j) \in \mathbf{U}}}$ and $LR_{BRL} = \frac{\mathcal{L}^{BRL}_{(i,j) \in \mathbf{M}}}{\mathcal{L}^{BRL}_{(i,j) \in \mathbf{U}}}$.

(a) If record pair $(i,j) \in \mathbf{M}$, then the following holds:
\begin{align}
    E_{(\mathbf{\Gamma},\mathbf{X}_A,\mathbf{X}_B)} \text{log } LR_{BRLVOF} \geq \text{  } E_{(\mathbf{\Gamma},\mathbf{X}_A,\mathbf{X}_B)} \text{log }LR_{BRL}. 
\end{align}

(b) If record pair $(i,j) \in \mathbf{U}$, then the following holds:
\begin{align}
    E_{(\mathbf{\Gamma},\mathbf{X}_A,\mathbf{X}_B)} \text{log } LR_{BRLVOF} \leq \text{  } E_{(\mathbf{\Gamma},\mathbf{X}_A,\mathbf{X}_B)} \text{log }LR_{BRL}.
\end{align}
\end{theoremp}

Proposition 1(a) asserts that for true links, the likelihood ratio is larger under BRLVOF than under BRL, which facilitates better identification of correct links. Conversely, Proposition 1(b) states that for non-links, the likelihood ratio is more attenuated under BRLVOF, which helps classify non-linking record pairs. We provide proofs for the above proposition in the supplementary materials. 

To exemplify Proposition \ref{prop1}, let $f_{M}(\mathbf{X}_{Ai},\mathbf{X}_{Bj}|\mathbf{Z}_{Ai},\mathbf{Z}_{Bj},\beta_M)$ and $f_{U}(\mathbf{X}_{Ai},\mathbf{X}_{Bj}|\mathbf{Z}_{Ai},\mathbf{Z}_{Bj},\beta_U)$ be bivariate normal distributions $N_2(\boldsymbol{\mu_{M}},\boldsymbol{\Sigma_{M}})$ and $N_2(\boldsymbol{\mu_{U}},\boldsymbol{\Sigma_{U}})$, respectively. For simplicity, assume that $\boldsymbol{\mu_M}=\boldsymbol{\mu_U}$, $\Sigma_{M}= \bigl( \begin{smallmatrix}1 & \rho_M \\\ \rho_M  & 1\end{smallmatrix}\bigr)
$, and  $\Sigma_{U}= \bigl( \begin{smallmatrix}1 & \rho_U\\\ \rho_U  & 1\end{smallmatrix}\bigr)
$. As shown in the supplementary materials, for record pair $(i,j)\in \mathbf{M}$, 
\begin{align}
    E_{(\mathbf{\Gamma},\mathbf{X}_A,\mathbf{X}_B)} \text{log }LR_{BRLVOF}  - E_{(\mathbf{\Gamma},\mathbf{X}_A,\mathbf{X}_B)} \text{log }LR_{BRL}  = \mathbb{K}, 
\end{align}
where $\mathbb{K}$ denotes the Kullback-Leibler (KL) divergence between the densities $f_{M}(\mathbf{X}_{Ai},\mathbf{X}_{Bj}|\mathbf{Z}_{Ai},\mathbf{Z}_{Bj},\beta_M)$ and $f_{U}(\mathbf{X}_{Ai},\mathbf{X}_{Bj}|\mathbf{Z}_{Ai},\mathbf{Z}_{Bj},\beta_U)$. Table S1 in the supplementary materials displays the values of $\mathbb{K}$ for different values of $\rho_{M}$ and $\rho_{U}$. For example, when  $\rho_{M}=0.65$ and $\rho_{U}=0.05$, $\mathbb{K} = 0.24$. Using Jensen's inequality, $\mathbb{K} = 0.24$ implies that  $LR_{BRLVOF}$ is at least $e^{0.24}=1.27$ times larger than $LR_{BRL}$, on average. Similar calculations can be performed when record pair $(i,j)\in \mathbf{U}$. This example shows that BRLVOF can better discriminate between the true links and false links compared to BRL.

\subsection{Modifications with blocking}
With large files, comparing all record pairs across the two files becomes computationally intensive, and may result in many false links. To improve the scalability and accuracy of the linkage process, blocking is a commonly implemented pre-processing technique. Blocking requires records to agree on a set of blocking variables, that are accurately recorded in both files. Records that do not agree on these variables are considered non-links. When files are partitioned into blocks, typical practice is to implement linkage algorithms independently within each block \cite{Tancredi2011,Murray2015,Sadinle2017}.

Let $\mathbf{Q} = \{Q_{ij}\}$ denote a vector of length $n_A \times n_B$, where 
\begin{equation} \label{eq:brlf3}
    Q_{ij}=
     \begin{cases} 
      1 & \text{if records $i \in \mathbf{A}$ and $j \in \mathbf{B}$ belong to the same block,} \\
      0 & \text{otherwise. }
   \end{cases}
\end{equation}
We adjust for blocking in BRLVOF by modifying Equation (\ref{eq:brlf2}) to
\begin{align} \label{eq:brlf4}
    \mathcal{L}^{BRLVOF_{Block}}(\mathbf{C},\theta_M, \theta_U, \beta_M, \beta_U|\mathbf{X}_A, \mathbf{X}_B, \mathbf{Z}_A, \mathbf{Z}_B,\mathbf{Q})= \prod_{i=1}^{n_A} \prod_{j=1}^{n_B} &\bigg[f_M(\mathbf{X}_{Ai},\mathbf{X}_{Bj}|\mathbf{Z}_{Ai},\mathbf{Z}_{Bj},\beta_M) \prod_{k=1}^{K} \prod_{l_k=1}^{L_k} \theta_{Mkl_{k}}^{\mathbbm{1}(\gamma_{ijk}=l_k)} \bigg]^{C_{ij}Q_{ij}} \nonumber \\
     \times &\bigg[f_U(\mathbf{X}_{Ai},\mathbf{X}_{Bj}|\mathbf{Z}_{Ai},\mathbf{Z}_{Bj},\beta_U) \prod_{k=1}^{K} \prod_{l_k=1}^{L_k} \theta_{Ukl_{k}}^{\mathbbm{1}(\gamma_{ijk}=l_k)} \bigg]^{(1-C_{ij})Q_{ij}}.
\end{align}
In Equation \eqref{eq:brlf4}, we assume that the parameters $\theta_{Mk}$ and $\theta_{Uk}$ do not vary between blocks. This allows the linking algorithm to aggregate information across blocks, while enforcing that true links can only be identified within a block. The prior distributions in BRLVOF without blocking can be used for all the linkage model parameters and $\mathbf{C}$. The prior on $\mathbf{C}$ can also be a uniform distribution that adheres to the blocking restrictions. A possible extension of the prior on $\mathbf{C}$ is to allow the proportion of links to vary across blocks. 

In some applications, the blocking variables may be inconsistently recorded, leading to erroneous assignments of records to blocks. In Section 2 of the supplementary materials, we briefly illustrate possible models to account for these errors in BRLVOF. 

\subsection{Posterior sampling algorithms}
Sampling from the joint posterior of the model parameters and $\mathbf{C}$ is not analytically tractable, even with a relatively small number of records. Thus, we use a data augmentation procedure \cite{tannerwong1987}, that iterates between sampling $\mathbf{C}$ and the model parameters. Starting with a random sample of $\mathbf{C}$, at iteration $[t+1]$, we perform the following steps to obtain samples of $\theta_M$, $\theta_U$, $\beta_M$, $\beta_U$, and $\mathbf{C}$:
\begin{enumerate} \label{eq:post1}
    \item Sample new values $\theta_{Mk}^{[t+1]}$ and $\theta_{Uk}^{[t+1]}$ for $k=1, \dots, K$ from 
    \begin{flalign*}
    &\theta_{Mk}^{[t+1]}|\mathbf{\Gamma}(\mathbf{Z}_A,\mathbf{Z}_B),\mathbf{C}^{[t]} \sim \mathrm{Dirichlet}(\alpha_{Mk1}+ \sum_{i=1}^{n_A} \sum_{j=1}^{n_B}C_{ij}^{[t]}\mathbbm{1}(\gamma_{ijk}=1), \dots, \alpha_{MkL_k}+ \sum_{i=1}^{n_A} \sum_{j=1}^{n_B}C_{ij}^{[t]}\mathbbm{1}(\gamma_{ijk}=L_k))&&
    \end{flalign*}
    and
    \begin{flalign*} \label{eq:post2}
    &\theta_{Uk}^{[t+1]}|\mathbf{\Gamma}(\mathbf{Z}_A,\mathbf{Z}_B),\mathbf{C}^{[t]} \sim
    \mathrm{Dirichlet}(\alpha_{Uk1}+ \sum_{i=1}^{n_A} \sum_{j=1}^{n_B}(1-C_{ij}^{[t]})\mathbbm{1}(\gamma_{ijk}=1), \dots, \alpha_{UkL_k}+ \sum_{i=1}^{n_A} \sum_{j=1}^{n_B}(1-C_{ij}^{[t]})\mathbbm{1}(\gamma_{ijk}=L_k)).&&
    \end{flalign*}
    \item Sample new values $\beta_M^{[t+1]}$ and $\beta_U^{[t+1]}$ from the conditional posterior distributions $f(\beta_M|\mathbf{X}_A,\mathbf{X}_B,\mathbf{Z}_A,\mathbf{Z}_B,\mathbf{C}^{[t]})$ and $f(\beta_U|\mathbf{X}_A,\mathbf{X}_B,\mathbf{Z}_A,\mathbf{Z}_B,\mathbf{C}^{[t]})$, respectively.
    \item Sample a new linking configuration $\mathbf{C}^{[t+1]}$ from $f(\mathbf{C}|\mathbf{X}_A,\mathbf{X}_B,\mathbf{Z}_A,\mathbf{Z}_B,\theta_M^{[t+1]},\theta_U^{[t+1]},\beta_M^{[t+1]},\beta_U^{[t+1]})$.
\end{enumerate}

Directly sampling from the posterior distribution of $\mathbf{C}$ in Step 3 would require computing the linkage likelihood for all $\binom{n_A}{n_m} \binom{n_B}{n_m} n_m!$ linking configurations with $n_m$ links. This is computationally intensive, even with a small number of records in each file. To overcome this limitation, several sampling procedures have been proposed to iterate through the rows of $\mathbf{C}$ and update its configuration, while preserving the one-to-one linking constraint.  One method adopts a version of the Metropolis Hastings algorithm \cite{hastings,Wu1995} to propose three possible updates for every record $i \in \mathbf{A}$ depending on its link status at iteration $[t]$. These updates are: \cite{MardiaGreen2006,Larsen2005}
\begin{enumerate} \label{eq:post3}
    \item If record $i \in \mathbf{A}$ does not form a link with any record in $\mathbf{B}$, randomly select a record $j \in \mathbf{B}$ to form a true link, so that $C_{ij}^{[t+1]}=1$.
    \item If record $i \in \mathbf{A}$ is linked with $j \in \mathbf{B}$ at iteration $[t]$, propose dropping the linked record pair so that $C_{ij}^{[t+1]}=0$.
    \item If record $i \in \mathbf{A}$ is linked with $j \in \mathbf{B}$ at iteration $[t]$, propose swapping link designations with another linked record pair $((r,s): r \in \mathbf{A}, s \in \mathbf{B},C_{rs}^{[t]}=1)$ to form new linked pairs $(i,s)$ and $(r,j)$, so that $C_{is}^{[t+1]}=1$ and $C_{rj}^{[t+1]}=1$.
\end{enumerate}
An alternative sampling procedure updates the link designation of $i \in \mathbf{A}$ through an adaptive multinomial distribution \cite{Sadinle2017}. This method requires fewer iterations to reach convergence; however, more computations are required at each iteration. We describe both procedures in the supplementary materials.

\subsection{Inference with linked files} \label{sec:MI}
A fully Bayesian approach for inference would incorporate the scientific model of interest in the sampling scheme, and iterate between estimating the linkage structure and the model parameters. However, the scientific model of interest and the models used in the linkage, namely $f_{M}(\mathbf{X}_{Ai},\mathbf{X}_{Bj}|\mathbf{Z}_{Ai},\mathbf{Z}_{Bj},\beta_{M})$ and $f_{U}(\mathbf{X}_{Ai},\mathbf{X}_{Bj}|\mathbf{Z}_{Ai},\mathbf{Z}_{Bj},\beta_{U})$, may not necessarily be the same. In some cases, an analysis, not necessarily foreseen at the time of linkage, may be of interest. Estimating the linkage is computationally demanding, and incorporating additional models within the Bayesian framework can be computationally prohibitive. A possible approximation is to generate multiple imputations of the linkage structure, analyze each linked file separately, and derive point and interval estimates using common combination rules \cite{Rubin1987}. This procedure properly propagates errors in the linkage, while simplifying the computation \cite{Gutman2013,Sadinle2018}. 

Let $\beta$ represent the estimand of interest. Let $\hat{\beta}^{(m)}$ be the point estimate of $\beta$, and let $U^{(m)}$ be its sampling variance in the $m^{th}$ posterior sample of $\mathbf{C}$, where $m = 1,\dots,M$. A point estimate for $\beta$ across the $M$ samples is \cite{Rubin1987}
\begin{equation}
    \hat{\beta}=\dfrac{1}{M}\sum_{m=1}^{M}\hat{\beta}^{(m)},
\end{equation}
and an estimate of its variance is
\begin{equation}
    T=\bar{U}+(1+\frac{1}{M})B,
\end{equation}
where
\begin{align}
    \bar{U}=&\dfrac{1}{M}\sum_{m=1}^M U^{(m)},\\
    B=& \dfrac{1}{M-1}\sum_{m=1}^M (\hat{\beta}^{(m)}-\hat{\beta})^{2}.
\end{align}
Interval estimates for $\beta$ can be obtained using a Student's-t approximation, $(\hat{\beta}-\beta)/ \sqrt{T} \sim t_{\nu}$, where $\nu=(M-1)(1+1/r)^2$, and $r=(1+M^{-1})B/\bar{U}$ \cite{Rubin1987}. \par

The above procedure may suffer from potential uncongeniality \cite{Meng1994} between the linkage model and the analysis model. Uncongeniality arises when the imputation model for the linkage structure is more restrictive than the scientific model of interest \cite{Meng1994}. Uncongenial imputation models can lead to biased and possibly inefficient estimates \cite{xiemeng}. One solution is to define $f_{M}(\mathbf{X}_{Ai},\mathbf{X}_{Bj}|\mathbf{Z}_{Ai},\mathbf{Z}_{Bj},\beta_{M})$
 and $f_{U}(\mathbf{X}_{Ai},\mathbf{X}_{Bj}|\mathbf{Z}_{Ai},\mathbf{Z}_{Bj},\beta_{U})$ to include the largest possible number of relationships between variables that may be of interest in the analysis \cite{Rubin1996}.

\section{Simulation Study} \label{sim}
\subsection{Data generation}
We compare the performance of BRLVOF to BRL using simulations. We consider files of sizes $n_A=500$ and $n_B=1000$, with $n_m=300$ true links. We generate three linking variables to represent an individual's gender, three-digit ZIP code, and date of birth (DOB). Table \ref{tab:datagen} depicts the distributions used to generate the linking variables. For true links, we generate the linking variables $\mathbf{Z}_{A}$ and replicate their values in file $\mathbf{B}$. For the non-linking record pairs, we generate $\mathbf{Z}_{A}$ and $\mathbf{Z}_{B}$ independently. 


We generate variables exclusive to file $\mathbf{B}$ from a multivariate normal distribution, $\mathbf{X}_{Bj} \sim N_P(1, 4\mathbf{I}_P)$, where $P \in \{1,2,4\}$. We generate a univariate $\mathbf{X}_A$ under three settings. In the first setting, $X_{Aj} \sim N(10 + \mathcal{B}_M^{T}\mathbf{X}_{Bj},\sigma^2)$  if $j \in \mathbf{M}$, and $X_{Aj} \sim N(5 + \mathcal{B}_U^{T}\mathbf{X}_{Bj},\sigma^2)$  if $j \in \mathbf{U}$, where $\mathcal{B}_M$ and $\mathcal{B}_U$ are P-dimensional vectors of $\beta_M\in \{0.05,0.1,0.2,0.5,1\}$ and $\beta_U=0.05$, respectively. We let $\sigma \in \{0.1,0.5\}$, to investigate how the strength of the association between $\mathbf{X}_A$ and $\mathbf{X}_B$ influences the inferences. In the second setting, we let $X_{Aj} \sim N(10 + \mathcal{B}_M^{T}\mathbf{X}_{Bj}+0.1 W_{j},\sigma^2)$ if $j \in \mathbf{M}$, and $X_{Aj} \sim N(5 + \mathcal{B}_U^{T}\mathbf{X}_{Bj}+ 0.1 W_j,\sigma^2)$ if $j \in \mathbf{U}$, where $W_j \sim N(1,4)$ is an unobserved variable. In the third setting, we let $X_{Aj} \sim N(10 + \mathcal{B}_M^{T} \mathbf{X}_{Bj} + 0.1\mathbf{X}_{Bj}^2 ,\sigma^2)$ if $j \in \mathbf{M}$, and $X_{Aj} \sim N(5 + \mathcal{B}_U^{T} \mathbf{X}_{Bj}+ 0.03\mathbf{X}_{Bj}^2,\sigma^2)$ if $j \in \mathbf{U}$, so that the true relationship between $\mathbf{X}_A$ and $\mathbf{X}_{B}$ is non-linear. 

We consider error rates of $\epsilon \in \{0.0,0.2,0.4\}$ for the DOB and the ZIP code. We generate errors in the ZIP code by re-sampling the value of each digit of the ZIP code with probability $\epsilon/3$. We introduce errors in the DOB by randomly altering each of its components (day, month, year) with probability $\epsilon/3$. \par
Table \ref{tab:datagen2} summarizes the factors that are varied in the simulation design, resulting in $5 \times 3^3 \times 2 =270$ configurations, which are implemented like a full factorial design. 

\subsection{Record linkage methods}
For each level of $\epsilon, P, \beta_M, \sigma$, and the distributional forms of $\mathbf{X}_A$, we generate 100 simulation replications. In each replication, we link the two files using BRL, BRLVOF, and the method of \cite{Tang2020} (abbreviated BRLVOF$_{ind}$). For all procedures, we generate 1000 MCMC samples, and discard the first 100 as burn-in. 

We use the same similarity functions for $\mathbf{Z}_{A}$ and  $\mathbf{Z}_{B}$ under all the three methods. We use exact agreement to compare gender. We use four levels of similarity to compare ZIP codes: disagreement on the first ZIP digit, agreement on the first ZIP digit only, agreement on the first and second ZIP digit only, and agreement on all three ZIP digits. We use four levels of similarity to compare the elements of DOB: no agreement on DOB year, agreement on DOB year only, agreement on DOB year and month only, and agreement on all elements of the DOB. Assuming conditional independence between the linking variables, the likelihood under BRL takes the form of Equation \eqref{eq:bg4}. We use a Dirichlet $(1, \dots, 1)$ prior for each component of $\theta_{Mk}$ and $\theta_{Uk}$.

\begin{table}[t] 
\caption{Description of linking variables used in the simulations.} \label{tab:datagen}
\centering
\begin{tabular}{lll}
\hline
Field & Distribution & Type \\ \hline
Gender  & Bernoulli(.5) & Categorical with 2 levels\\ 
ZIP Code  & $1^{st}$ Digit $\sim$ Discrete uniform & Numeric with 3 values \\
& $2^{nd}$ Digit $\sim$ Discrete Uniform & Numeric with 4 values\\
& $3^{rd}$ Digit $\sim$ Discrete Uniform & Numeric with 5 values\\
Date of Birth & Age $\sim$ N(50, $5^2$) & Converted to year, month, day \\
\hline
\end{tabular}
\end{table}

Under BRLVOF, we model $f_M(X_{Ai}|\mathbf{X}_{Bj},\mathbf{Z}_{Ai},\mathbf{Z}_{Bj},\beta_M)$ and $f_{U}(X_{Ai}|\mathbf{X}_{Bj},\mathbf{Z}_{Ai},\mathbf{Z}_{Bj},\beta_{U})$ using a linear model $X_{Ai}=\mathbf{X}_{Bj} \beta_M + {\delta}_{M(i,j)}$ for $(i,j) \in \mathbf{M}$, and $X_{Ai}=\mathbf{X}_{Bj} \beta_U + {\delta}_{U(i,j)}$ for $(i,j) \in \mathbf{U}$, where  ${\delta}_{M(i,j)} \overset {iid}\sim  N(0, \sigma_M^2)$ and ${\delta}_{U(i,j)} \overset {iid}\sim N(0, \sigma_U^2)$. For computational simplicity, we assume that the marginal distribution of $\mathbf{X}_B$ is the same among the linked and unlinked records \cite{Tang2020}. This assumption obviates the need to specify   $f_M(\mathbf{X}_{Bj}|\mathbf{Z}_{Ai},\mathbf{Z}_{Bj})$ and $f_U(\mathbf{X}_{Bj}|\mathbf{Z}_{Ai},\mathbf{Z}_{Bj})$, because they do not contribute to the partitioning of record pairs into true links and non-links. To complete the Bayesian model, we assume improper prior distributions $p(\beta_M,\sigma_M) \propto \sigma_M^{-2}$ and $p(\beta_U, \sigma_U) \propto \sigma_U^{-2}$.

Let $\mathbf{X}_{AM} = \{X_{Ai}\}$ and $\mathbf{X}_{BM} = \{(1,\mathbf{X}_{Bj})\}$ be the variables exclusive to file $\mathbf{A}$ and $\mathbf{B}$, respectively, for $(i,j) \in \mathbf{M}$. Similarly, let $\mathbf{X}_{AU} = \{X_{Ai}\}$ and $\mathbf{X}_{BU} = \{(1,\mathbf{X}_{Bj})\}$ be  the variables exclusive to file $\mathbf{A}$ and $\mathbf{B}$, respectively, for $(i,j) \in \mathbf{U}$. The conditional posterior distributions of $\beta_M$, $\beta_U$, $\sigma_M^2$, and $\sigma_U^2$ are
\begin{align} \label{rlmod}
\begin{split}
    \beta_M|\mathbf{X}_A,\mathbf{X}_B,\mathbf{C},\sigma_M &\sim N\bigg(\big(\mathbf{X}_{BM}^{T}\mathbf{X}_{BM} \big)^{-1} \mathbf{X}_{BM}^{T}\mathbf{X}_{AM}, \sigma_M^2 \big(\mathbf{X}_{BM}^{T} \mathbf{X}_{BM} \big)^{-1} \bigg) \\
    \beta_U|\mathbf{X}_A,\mathbf{X}_B,\mathbf{C},\sigma_U & \sim N \bigg(\big(\mathbf{X}_{BU}^{T}\mathbf{X}_{BU}\big)^{-1}\mathbf{X}_{BU}^{T}\mathbf{X}_{AU}, \sigma_U^2 \big(\mathbf{X}_{BU}^{T}\mathbf{X}_{BU}\big)^{-1} \bigg)\\
    \sigma_M^2|\mathbf{X}_A,\mathbf{X}_B,\mathbf{C},\beta_M & \sim Inv-Gamma \bigg(\frac{n_m}{2}, \frac{1}{2}\big(\mathbf{X}_{AM} - \mathbf{X}_{BM} \beta_M\big)^{T}\big(\mathbf{X}_{AM} - \mathbf{X}_{BM} \beta_M\big) \bigg)\\
    \sigma_U^2|\mathbf{X}_A,\mathbf{X}_B,\mathbf{C},\boldsymbol{\beta_U} &\sim Inv-Gamma \bigg(\dfrac{n_An_B-n_m}{2}, \dfrac{1}{2}\big(\mathbf{X}_{AU}-\mathbf{X}_{BU} \beta_U \big)^{T}\big(\mathbf{X}_{AU}-\mathbf{X}_{BU} \beta_U \big) \bigg).
\end{split}
\end{align}
 Under BRLVOF$_{ind}$, $f_{M}(X_{Ai}|\mathbf{X}_{Bj},Z_{Ai},Z_{Bj},\beta_{M})$ takes the same form as under BRLVOF. However, $f_{U}(X_{Ai}|\mathbf{X}_{Bj},Z_{Ai},Z_{Bj},\beta_{U})$ is specified assuming $X_{Ai} \independent \mathbf{X}_{Bj}$. We describe the exact model specification in Section 5.4 of the supplementary materials.
 
When the data generating model is linear with $\mathbf{X}_{B}$ as the independent variable and $\mathbf{X}_{A}$ as the dependent variable, the model in Equation \eqref{rlmod} is correctly specified. Equation \eqref{rlmod} is misspecified when the data generating model includes $W$, or non-linear terms.

\begin{table}[t] 
\caption{Simulation factors.} \label{tab:datagen2}
\centering
\begin{tabular}{lc}
\hline
Factor & Levels  \\ \hline
$P$ & \{1, 2, 4\} \\
$\sigma$  & \{0.1, 0.5\} \\
$\beta_M$  & \{0.05, 0.1, 0.2, 0.5, 1\}\\ 
$\epsilon$ & \{0.0, 0.2, 0.4\} \\
Model for $\mathbf{X}_A$ & \{linear, linear with $W$, non-linear\}\\
\hline
\end{tabular}
\end{table}

\subsection{Evaluation metrics}
We assess the linking performance of BRLVOF, BRL, and BRLVOF$_{ind}$ within each MCMC iteration by calculating the number of linked records ($n_m$), the true positive rate ($TPR$), the positive predictive value ($PPV$), and the average $F1$ score. The $TPR$ is calculated as the proportion of true links that are correctly identified, $TPR=\frac{\sum_{i=1}^{n_A} \sum_{j=1}^{n_B} \mathbbm{1}(C_{ij}=1|(i,j) \in \mathbf{M})}{\sum_{i=1}^{n_A} \sum_{j=1}^{n_B} \mathbbm{1}((i,j)\in \mathbf{M})}$, the positive predictive value is the proportion of linked records that are true links, $PPV=\frac{\sum_{i=1}^{n_A} \sum_{j=1}^{n_B} \mathbbm{1}(C_{ij}=1|(i,j) \in \mathbf{M})}{n}$, and the F1 score is equal to $2 \times \dfrac{TPR \times PPV}{TPR + PPV}$. 

In each linked sample (implied by the posterior draw of $\mathbf{C}$), we estimate the marginal association, $\hat\rho$, between $\mathbf{X}_{A}$ and $\mathbf{X}_{B1}$. We compute its sampling variance using Fisher's z-transformation \cite{fisher1915}. We obtain a point estimate and a 95\% interval estimate across the posterior draws of $\mathbf{C}$ using the multiple imputation combining rules 
from Section \ref{sec:MI}.
In each simulation replication, we record the absolute bias, $|\hat{\rho}-\rho|$, the squared error, $(\hat{\rho}-\rho)^{2}$, and whether the 95\% confidence interval covers the true value of $\rho$. We summarize results across all replications in terms of the mean absolute bias, $\overline{Bias} = \frac{1}{100}\sum|\hat{\rho}-\rho|$, the root mean squared error, $RMSE = \sqrt{\frac{1}{100}\sum(\hat{\rho}-\rho)^2}$, and the coverage of the 95\% confidence intervals.

We present results for additional quantities in the supplementary materials. For selected simulation scenarios, we report the $\overline{Bias}$ and $RMSE$ for $\beta_M$ and $\beta_U$, and the slope of the regression of  $\mathbf{X}_{A}$ on $\mathbf{X}_{B1}$ among the links, $\beta$.

\subsection{Results}
The linkage and estimation accuracy under BRLVOF and BRLVOF$_{ind}$ (Supplementary Tables S6-S11) are similar in these simulations. Thus, we focus on comparing BRLVOF to BRL, and provide additional information in the Discussion.

Table \ref{tab:res1} displays the results under BRLVOF when the model for $\mathbf{X}_A$ given $\mathbf{X}_B$ is correctly specified, and $\sigma=0.1$. The $\overline{TPR}$, $\overline{PPV}$, and the $\overline{F1}$ score are the average $TPR$, $PPV$, $F1$ score over 100 replications across different levels of $\epsilon$, $P$, and $\beta_M$. In configurations with $\epsilon=0$, BRL and BRLVOF perform similarly in terms of $n_m$ and $\overline{TPR}$. The number of links identified under both methods is close to 300. However, BRL identifies more false links compared to BRLVOF because of the limited information in the linking variables. This results in lower $\overline{PPV}$ and $\overline{F1}$ scores. The false links that are included in the linked file result in higher $\overline{Bias}$ and $RMSE$ for estimates of $\rho$ when using BRL. 

The $\overline{TPR}$, $\overline{PPV}$, and $\overline{F1}$ score of BRL decrease when $\epsilon$ increases, resulting in estimates of $\rho$ with high $\overline{Bias}$ and $RMSE$. Due to the increased error level, the total number of links identified under both BRL and BRLVOF are fewer than 300. However, BRLVOF achieves significant improvements to the $\overline{TPR}$, $\overline{PPV}$, and $\overline{F1}$ score. This improvement is greater as $\beta_M$ increases, i.e., when the associations between $\mathbf{X}_{A}$ and $\mathbf{X}_{B}$ are stronger. 

Table \ref{tab:res2} displays results when $\sigma=0.1$, and the true model for $\mathbf{X}_A$ given $\mathbf{X}_B$ includes $W$. Table \ref{tab:res3} depicts the results when $\sigma=0.1$, and the true model for $\mathbf{X}_A$ includes non-linear terms. In both cases, $f_{M}(X_{Ai},\mathbf{X}_{Bj}|\mathbf{Z}_{Ai},\mathbf{Z}_{Bj},\beta_{M})$ and $f_{U}(X_{Ai},\mathbf{X}_{Bj}|\mathbf{Z}_{Ai}|\mathbf{Z}_{Bj},\beta_{U})$ are misspecified. When $\epsilon=0$, BRLVOF performs better than BRL  in terms of $\overline{TPR}$, $\overline{PPV}$, $\overline{F1}$ score, $\overline{Bias}$ and $RMSE$. Compared to the correctly specified configurations, there is an increase in $\overline{Bias}$ and $RMSE$ under BRLVOF. The $\overline{Bias}$ and $RMSE$ under BRL are practically similar, because BRL does not consider relationships between variables exclusive to each of the files. When $\epsilon$ increases to 0.2 and 0.4, BRLVOF only shows marginal improvements to the $\overline{TPR}$ over BRL, but more significant improvements are observed for the $\overline{PPV}$ and $\overline{F1}$ score. Similar to the correctly specified model configurations, the improvements are greater for larger values of $\beta_{M}$ within each error rate. 

The results when $\sigma=0.5$ are qualitatively similar to the configurations with $\sigma = 0.1$ (Tables S2-S4 in the supplementary materials). BRLVOF demonstrates  improvements over BRL in terms of the $\overline{TPR}$, $\overline{PPV}$, $\overline{F1}$, as well as the $\overline{Bias}$ and $RMSE$. However, the gains are less pronounced than when $\sigma = 0.1$, because the strength of association between $\mathbf{X}_A$ and $\mathbf{X}_B$ is attenuated.

\FloatBarrier
\begin{table}[]
\caption {Results for estimation of $\rho$ under BRLVOF and BRL, when models for $\mathbf{X}_A$ given $\mathbf{X}_B$  are linear, and $\sigma=0.1$. Values in parentheses represent standard deviations of the corresponding quantities.} \label{tab:res1}
\centering
\begin{tabular}{ccccccccccc}

$\epsilon$ & Method& P & $\beta_M$ & $n_m$ & $\overline{TPR}$ & 
$\overline{PPV}$ & $\overline{F1}$ & $\overline{Bias}$ & $RMSE$ & Coverage \\ \hline
\multirow{10}{*}{0} & BRL &  &  &301 (0.17) & .9984(.0016) &.8845(.0418) &.9200(.0240)& .075(.028)& 0.080 & 0.90 \\ \cline{2-11} 
 & \multirow{15}{*}{BRLVOF} &\multirow{3}{*}{1} & 0.05 & 300 (0.12) & {\ul .9994(.0384)} & .9995(.0123) & .9994(.0147) & .000(.000) & 0.001 & 1.00 \\
 &&  & 0.1 &302 (0.81) &{\ul .9931(.0358)} & .9923(.0333) & .9926(.0341) & .001(.000) & 0.002 & 1.00 \\
 && & 0.2  &  304 (2.72) & {\ul .9997(.0006)} & .9998(.0005) & .9997(.0005) & .000(.000) & 0.000 &0.99 \\
 &&  & 0.5 &  302 (0.12) & {\ul .9999(.0005)} & .9999(.0003) & .9999(.0003) & .000(.000) & 0.000&0.99 \\
 &&  & 1.0 &  302 (0.88) & .9999(.0004) & .9998(.0008) & .9998(.0005) & .000(.000) & 0.000&0.99 \\ \cline{3-11} 
 && \multirow{3}{*}{2} & 0.05 & 304 (0.11) & {\ul .9797(.0154)} & .9796(.0122) & .9994(.0138) & .017(.017) & 0.018 & 1.00 \\
 &&  & 0.1 &300 (0.11) & {\ul .9995(.0006)} & .9995(.0005) & .9995(.0005) & .000(.000) & 0.000&1.00 \\
 && & 0.2  &  300 (0.11) & .9997(.0006) & .9998(.0004) & .9997(.0005) & .000(.000) & 0.000 &1.00 \\
 &&  & 0.5 &  318 (0.09) & .9999(.0005) & .9999(.0003) & .9999(.0004) & .000(.000) & 0.000& 0.91 \\
 &&  & 1.0 &  312 (0.38)& .9999(.0004) & .9989(.0016) & .9994(.0008) & .001(.002) & 0.002 &0.94\\ \cline{3-11} 
 && \multirow{3}{*}{4} & 0.05 &308 (0.14)& {\ul .9599(.0156)} & .9599(.0111) & .9998(.0132) & .028(.019) & 0.030 &1.00 \\
 &&  & 0.1 &326 (2.28) & {\ul .9704(.0718)} & .9693(.0677) & .9997(.0031) & .024(.010) & 0.023 &0.99 \\
 && & 0.2  &  332 (0.88) & .9997(.0006) & .9951(.0004) & .9968(.0004) & .004(.003) & 0.005&0.91 \\
 &&  & 0.5 &  325 (1.12) & .9999(.0005) & .9998(.0006) & .9999(.0004) & .001(.002) & 0.001&0.91 \\
 &&  & 1.0 &  314 (0.49) & .9999(.0004) & .9985(.0019) & .9992(.0010) & .002(.003) & 0.003&0.94 \\ \hline
\multirow{10}{*}{0.2} & BRL&  &  &288 (2.47) & .7780(.0151) &.7168(.0587)& .7424(.0326) &.297(.045) & 0.301&0.64 \\ \cline{2-11} 
 & \multirow{15}{*}{BRLVOF}& \multirow{3}{*}{1} & 0.05 &287 (3.25)& {\ul .8713(.0315)} & .9118(.0355) & .8911(.0317) & .011(.013) & 0.013 & 1.00 \\
 &&  & 0.1& 282 (3.57)& {\ul .8807(.0092)} & .9358(.0121) & .9074(.0089) & .004(.002) & 0.005 & 1.00 \\
 && & 0.2  &  279 (3.41)& .8874(.0086) & .9626(.0098) & .9234(.0071) & .001(.001) & 0.001&0.99  \\
 &&  & 0.5  & 273 (2.96)& .8932(.0083) & .9833(.0065) & .9360(.0056) & .000(.000) & 0.000&1.00 \\
 &&  & 1.0 &  273 (2.88)& .8986(.0084) & .9892(.0054) & .9416(.0053) & .000(.000) & 0.000&1.00 \\ \cline{3-11} 
 && \multirow{3}{*}{2} & 0.05 &285 (3.42)& {\ul .8751(.0313)} & .9220(.0337) & .8979(.0313) & .013(.030) & 0.016&1.00 \\
 &&  & 0.1 &280 (3.56) & {\ul .8844(.0088)} & .9493(.0112) & .9156(.0081) & .010(.005) & 0.011&1.00 \\
 && & 0.2  &  275 (3.27) &.8895(.0085) & .9716(.0085) & .9286(.0065) & .008(.005) & 0.010&1.00 \\
 &&  & 0.5 &  279 (2.75) & .8959(.0082) & .9867(.0057) & .9391(.0054) & .007(.004) & 0.008&0.97 \\
 &&  & 1.0 &  277 (2.97) & .9049(.0083) & .9874(.0058) & .9443(.0052) & .009(.005) & 0.010&0.99 \\ \cline{3-11} 
 && \multirow{3}{*}{4} & 0.05 &291 (3.34)& {\ul .8508(.0307)} & .9012(.0326) & .9012(.0306) & .044(.032) & 0.044 &1.00 \\
 &&  & 0.1 &292 (3.18) & {\ul .8617(.0009)} & .9104(.0095) & .9176(.0073) & .042(.061) & 0.045 & 1.00 \\
 && & 0.2  &  288 (3.83) & .8908(.0085) & .9779(.0074) & .9322(.0061) & .007(.001) & 0.008&0.95 \\
 &&  & 0.5 &  275 (3.53) & .8979(.0082) & .9881(.0055) & .9407(.0053) & .001(.001) & 0.002&0.99 \\
 &&  & 1.0 &  276 (3.24) & .9096(.0089) & .9854(.0062) & .9459(.0054) & .001(.001) & 0.001&1.00 \\ \hline
\multirow{10}{*}{0.4} & BRL & &  & 275 (2.99)& .5583(.0212)& .5665(.0642)& .5539(.0344)& .482(.046)& 0.484&0.56 \\ \cline{2-11} 
 & \multirow{15}{*}{BRLVOF}& \multirow{3}{*}{1} & 0.05 &281 (3.88)& {\ul .6904(.0034)} & .7304(.0036) & .7243(.0035) & .025(.031) & 0.029&0.99 \\
 &&  & 0.1 &276 (4.65)& {\ul .7291(.0136)} & .7916(.0183) & .7589(.0145) & .007(.005) & 0.009 & 1.00 \\
 && & 0.2  &  263 (5.51)& .7522(.0124) & .8596(.0186) & .8022(.0128) & .002(.001) & 0.002&1.00 \\
 &&  & 0.5 &  255 (3.53)& .7671(.0115) & .9351(.0138) & .8426(.0095) & .000(.000) & 0.000&1.00\\
 &&  & 1.0 &  244 (4.45)& .7789(.0115) & .9598(.0109) & .8597(.0085) & .000(.000) & 0.000&1.00 \\ \cline{3-11} 
 && \multirow{3}{*}{2} & 0.05 &281 (4.19)& {\ul .7112(.0314)} & .7603(.0363) & .7351(.0348) & .024(.034) & 0.028&1.00 \\
 &&  & 0.1 &275 (5.94)& {\ul .7310(.0399)} & .8072(.0381) & .7711(.0213) & .033(.058) & 0.037&1.00 \\
 && & 0.2  &  256 (5.44)&.7568(.0121) & .8892(.0172) & .8175(.0114) & .014(.008) & 0.016&1.00 \\
 &&  & 0.5 &  247 (4.63)& .7707(.0118) & .9486(.0126) & .8503(.0091) & .013(.007) & 0.014&0.99 \\
 &&  & 1.0 &  246 (4.64)& .7887(.0119) & .9593(.0111) & .8655(.0085) & .014(.007) & 0.016&1.00 \\ \cline{3-11} 
 && \multirow{3}{*}{4} & 0.05 &281 (4.19)& {\ul .7214(.0608)} & .7715(.0501) & .7444(.0301) & .029(.019) & 0.031 & 0.99 \\
 &&  & 0.1 &275 (4.94) & {\ul .7418(.0501)} & .8172(.0301) & .8100(.0103) & .023(.011) & 0.024 &0.99 \\
 && & 0.2  &  265 (5.03) & .7595(.0118) & .9151(.0158) & .8299(.0102) & .020(.011) & 0.020&0.94 \\
 &&  & 0.5 &  243 (4.60) & .7761(.0118) & .9576(.0115) & .8571(.0087) & .020(.010) & 0.020&1.00 \\
 &&  & 1.0 &  250 (4.99) & .8002(.0127) & .9578(.0116) & .8717(.0088) & .018(.010) & 0.022&1.00 \\ \hline
\end{tabular}
\end{table}
\FloatBarrier

\FloatBarrier
\begin{table}[]
\caption {Results for estimation of $\rho$ under BRLVOF and BRL, when models for $\mathbf{X}_A$ given $\mathbf{X}_B$ include $W$, and $\sigma=0.1$.Values in parentheses represent standard deviations of the corresponding quantities.} \label{tab:res2}
\centering
\begin{tabular}{ccccccccccc}
$\epsilon$ & Method &P & $\beta_M$ &$n_m$ & $\overline{TPR}$ & $\overline{PPV}$ & $\overline{F1}$ & $\overline{Bias}$ & $RMSE$ &Coverage\\ \hline
\multirow{10}{*}{0.0} & BRL& & & 301 (0.17) & .9984(.0016) &.8845(.0418) &.9200(.0240) & .072(.027) & 0.077 &0.91 \\ \cline{2-11} 
 &\multirow{15}{*}{BRLVOF}& \multirow{3}{*}{1} 
     &  0.05 &300 (0.19)  & .9992(.0009) & .9994(.0004) & .9993(.0006) & .001(.001) & 0.001 & 1.00 \\ 
 &&  &  0.1   &300 (0.19) &   .9992(.0009) & .9994(.0004) & .9993(.0006) & .000(.001) & 0.001 & 1.00 \\ 
 &&  &  0.2   &300 (0.18) & {\ul .9994(.0008)} & .9996(.0005) & .9995(.0006) & .000(.001) & 0.001&   1.00 \\
 &&  &  0.5   &302 (0.17) & {\ul .9996(.0007)} & .9998(.0003) & .9997(.0005) & .000(.000) & 0.000&   0.99 \\
 &&  &  1.0   &300 (0.29) & .9997(.0006) & .9994(.0010) & .9996(.0006) & .000(.000) & 0.000    &     1.00  \\ \cline{3-11} 
 && \multirow{3}{*}{2} 
     &0.05 &300 (0.19)    & .9993(.0009) & .9995(.0004) & .9994(.0006) & .001(.001) & 0.002 & 1.00   \\
 &&  & 0.1   &302 (0.19)  &   .9995(.0009) & .9997(.0004) & .9995(.0006) & .001(.001) & 0.002 & 1.00    \\
 &&& 0.2     &301 (1.11)  & .9997(.0008) & .9998(.0004) & .9997(.0005) & .001(.001) & 0.001         &1.00     \\
 &&  & 0.5   &314 (0.17)  & .9997(.0007) & .9998(.0004) & .9997(.0004) & .000(.001) & 0.001         &0.93      \\
 &&  & 1.0   &310 (0.46)  & .9998(.0005) & .9990(.0016) & .9994(.0008) & .001(.002) & 0.002         &0.95      \\ \cline{3-11} 
 && \multirow{3}{*}{4} 
     & 0.05 & 306 (0.20)   & {\ul .9994(.0006)} & .9998(.0005) & .9997(.0005) & .006(.006) & 0.008 &1.00 \\
 &&  & 0.1    &316 (0.19)  & {\ul .9995(.0005)} & .9999(.0003) & .9999(.0003) & .006(.008) & 0.007   &0.98 \\
 &&& 0.2     & 323 (1.60)  & .9996(.0008) & .9954(.0004) & .9969(.0005) & .004(.003) & 0.005         &0.99  \\
 &&  & 0.5    &330 (0.24)  & .9997(.0006) & .9951(.0005) & .9968(.0005) & .003(.003) & 0.005         &0.97   \\
 &&  & 1.0    &314 (0.63)  & .9998(.0005) & .9938(.0022) & .9963(.0011) & .004(.003) & 0.005         &0.94    \\ \hline
\multirow{10}{*}{0.2} &BRL&&  & 288 (2.47) & .7780(.0151) &.7168(.0587)& .7424(.0326) & .291(.045) & 0.295 &0.66 \\ \cline{2-11} 
 &\multirow{15}{*}{BRLVOF}& \multirow{3}{*}{1} 
     & 0.05 &278 (3.93)  & .8565(.0105) & .9228(.0135) & .8884(.0102) & .020(.011) & 0.024 & 1.00      \\
 &&  & 0.1    &277 (3.94)&   .8604(.0103) & .9335(.0128) & .8954(.0096) & .013(.010) & 0.015 & 1.00      \\
 &&& 0.2     & 274 (1.28)  & .8656(.0100) & .9540(.0108) & .9076(.0082) & .005(.003) & 0.006          &0.99      \\
 &&  & 0.5    &268 (3.49)  & .8719(.0099) & .9773(.0076) & .9215(.0068) & .001(.001) & 0.001          &1.00      \\
 &&  & 1.0    &268 (3.18)  & .8801(.0094) & .9851(.0060) & .9295(.0061) & .000(.000) & 0.000          &1.00       \\ \cline{3-11} 
 && \multirow{3}{*}{2} 
     & 0.05 &277 (3.94)  & .8583(.0104) & .9280(.0132) & .8917(.0099) & .019(.010) & 0.023 & 1.00     \\
 &&  & 0.1   & 275 (3.90)&   .8631(.0100) & .9429(.0118) & .9011(.0088) & .014(.010) & 0.017 & 1.00   \\
 &&& 0.2     & 270 (3.70)    & .8679(.0101) & .9638(.0095) & .9132(.0075) & .010(.005) & 0.011         &1.00     \\
 &&  & 0.5    &280 (3.12)    & .8769(.0094) & .9822(.0066) & .9264(.0062) & .008(.004) & 0.009         &0.96      \\
 &&  & 1.0   & 273 (3.14)   & .8871(.0088) & .9841(.0064) & .9329(.0057) & .009(.005) & 0.010          &0.99      \\ \cline{3-11} 
 && \multirow{3}{*}{4} 
     & 0.05 &280 (3.92)   & .8565(.0114) & .9236(.0135) & .8876(.0104) & .027(.011) & 0.030 & 1.00     \\
 &&  & 0.1    &289 (4.95) &   .8578(.0134) & .9421(.0144) & .8882(.0103) & .020(.013) & 0.026 & 1.00      \\
 &&& 0.2      &282 (3.55)   & .8671(.0181) & .9613(.0122) & .9104(.0133) & .022(.015) & 0.024         &0.96       \\
 &&  & 0.5    &277 (3.12)   & .8797(.0094) & .9856(.0061) & .9295(.0061) & .013(.006) & 0.015         &0.98        \\
 &&  & 1.0   & 279 (3.70)   & .8922(.0086) & .9828(.0071) & .9352(.0056) & .013(.007) & 0.015         &0.98        \\ \hline
\multirow{10}{*}{0.4} & BRL& && 275 (2.99) & .5583(.0212)& .5665(.0642)& .5539(.0344) & .480(.047) & 0.483 & 0.57      \\ \cline{2-11} 
 &\multirow{15}{*}{BRLVOF}& \multirow{3}{*}{1} 
     & 0.05& 268 (5.10)   & .6591(.0145) & .7283(.0195) & .7201(0.0138) & .061(.014) & 0.068 & 0.97       \\
 &&  & 0.1    &267 (5.43) &   .6907(.0143) & .7750(.0202) & .7303(.0154) & .024(.014) & 0.028 & 1.00       \\
 &&& 0.2      &257 (6.31)   & .7091(.0134) & .8310(.0201) & .7651(.0139) & .009(.006) & 0.011          &1.00        \\
 &&  & 0.5    &241 (5.43)   & .7258(.0130) & .9142(.0159) & .8089(.0109) & .002(.001) & 0.002          &0.99         \\
 &&  & 1.0    &234 (5.03)   & .7391(.0128) & .9469(.0128) & .8299(.0097) & .000(.000) & 0.001          &1.00         \\ \cline{3-11} 
 && \multirow{3}{*}{2} 
     & 0.05&270 (5.26)  & .6903(.0146) & .7665(.0200) & .7263(.0156) & .034(.015) & 0.04 & 1.00        \\
 &&  & 0.1   &263 (5.65)&   .7047(.0138) & .8040(.0200) & .7510(.0146) & .026(.014) & 0.03 & 1.00        \\
 &&& 0.2     &249 (5.88)  & .7204(.0129) & .8695(.0191) & .7878(.0126) & .018(.010) & 0.020          &1.00         \\
 &&  & 0.5   &247 (4.97)  & .7353(.0129) & .9360(.0138) & .8233(.0101) & .014(.007) & 0.016          &0.96          \\
 &&  & 1.0   &238 (4.91)  & .7538(.0123) & .9513(.0123) & .8409(.0092) & .014(.007) & 0.016          &1.00          \\ \cline{3-11} 
 && \multirow{3}{*}{4} 
     & 0.05  &276 (5.22)  & .6623(.0140) & .7454(.0197) & .7665(.0101) & .054(.014) & 0.052 & 1.00      \\
 &&  & 0.1     &281 (5.26)&   .6360(.0125) & .7449(.0178) & .7724(.0113) & .043(.013) & 0.047 & 1.00      \\
 &&& 0.2       &256 (5.39)  & .7213(.0131) & .8926(.0172) & .7976(.0116) & .022(.012) & 0.026         &0.95      \\
 &&  & 0.5     &241 (4.94)  & .7399(.0128) & .9431(.0129) & .8290(.0098) & .021(.011) & 0.023         &0.98       \\
 &&  & 1.0     &242 (5.03)  & .7642(.0125) & .9451(.0130) & .8448(.0093) & .020(.011) & 0.023         &1.00       \\ \hline
\end{tabular}
\end{table}
\FloatBarrier

\FloatBarrier
\begin{table}[]
\caption {Results for estimation of $\rho$ under BRLVOF and BRL, when models for $\mathbf{X}_A$ given $\mathbf{X}_B$ include non-linear terms, and $\sigma=0.1$. Values in parentheses represent standard deviations of the corresponding quantities.} \label{tab:res3}
\centering
\begin{tabular}{ccccccccccc}
$\epsilon$ & Method &P & $\beta_M$ & $n_m$ & $\overline{TPR}$ & $\overline{PPV}$ & $\overline{F1}$ & $\overline{Bias}$ & $RMSE$&Coverage \\ \hline
\multirow{10}{*}{0.0} & BRL& &  &301 (0.17) &.9984(.0016) &.8845(.0418) &.9200(.0240) & .058(.023) & 0.062 & 0.91\\ \cline{2-11} 
 &\multirow{15}{*}{BRLVOF}& \multirow{3}{*}{1} & 0.05 &295 (1.03)& {\ul .9765(.0238)} & .9892(.0167) & .9927(.0208) & .061(.032) & 0.064 & 0.99 \\
 &&  & 0.1 &299 (1.97) & {\ul .9753(.0419)} & .9866(.0428) & .9808(.0412) & .065(.077) & 0.070 & 0.98 \\
 &&& 0.2   &  305 (1.84) & {\ul .9790(.0138)} & .9901(.0081) & .9844(.0102) & .051(.038) & 0.052&0.95 \\
 &&  & 0.5 &  301 (0.92) & {\ul .9802(.0122)} & .9904(.0071) & .9852(.0090) & .031(.027) & 0.032&0.97 \\
 &&  & 1.0 &  297 (1.02) & .9701(.0441) & .9768(.0441) & .9732(.0437) & .042(.087) & 0.046&0.99 \\ \cline{3-11} 
 && \multirow{3}{*}{2} & 0.05 &298 (0.74)& {\ul .9932(.0233)} & .9992(.0148) & .9962(.0194) & .031(.032) & 0.034 & 1.00 \\
 &&  & 0.1 &304 (0.82)& {\ul .9745(.0263)} & .9751(.0164) & .9938(.0015) & .048(.039) & 0.051 & 0.98 \\
 &&& 0.2   &  298 (0.74)& .9936(.0026) & .9994(.0010) & .9965(.0015) & .022(.009) & 0.024&1.00 \\
 &&  & 0.5 &  312 (0.73)& .9936(.0024) & .9989(.0014) & .9963(.0015) & .013(.006) & 0.015&0.93 \\
 &&  & 1.0 &  309 (1.01)& .9763(.0022) & .9776(.0031) & .9769(.0019) & .033(.004) & 0.034&0.95 \\ \cline{3-11} 
 && \multirow{3}{*}{4} & 0.05 &307 (0.57)& {\ul .9571(.0214)} & .9593(.0157) & .9980(.0185) & .032(.031) & 0.037 &1.00 \\
 &&  & 0.1 &319 (0.55) & {\ul .8975(.0030)} & .8997(.001) & .9981(.001) & .068(.077) & 0.076 &0.99 \\
 &&& 0.2   &  332 (1.51) & .9969(.0021) & .9992(.0011) & .9981(.0012) & .010(.007) & 0.013&0.92 \\
 &&  & 0.5 &  320 (0.73) & .9968(.0021) & .9981(.0020) & .9974(.0015) & .008(.006) & 0.010&0.94 \\
 &&  & 1.0 &  311 (2.01)& .9987(.0016) & .8814(.0427) & .9351(.0248) & .006(.005) & 0.009&0.96 \\ \hline
\multirow{10}{*}{0.2} &BRL&  &  & 288 (2.47)& .7780(.0151) &.7168(.0587)& .7424(.0326) & .248(.042) & 0.252&0.68 \\ \cline{2-11} 
 &\multirow{15}{*}{BRLVOF}& \multirow{3}{*}{1} & 0.05 &259 (4.38)& {\ul .8256(.0327)} & .9548(.0325) & .8854(.0305) & .081(.034) & 0.083 & 0.98 \\
 &&  & 0.1 &259 (4.43) & {\ul .8267(.0116)} & .9586(.0110) & .8876(.0086) & .072(.011) & 0.073 & 0.99 \\
 &&& 0.2   &  258 (4.24) & .8288(.0115) & .9662(.0099) & .8921(.0081) & .054(.007) & 0.055&0.99 \\
 &&  & 0.5 &  259 (3.94) & .8352(.0114) & .9787(.0051) & .9011(.0074) & .023(.002) & 0.024&0.96 \\
 &&  & 1.0 &  260 (4.56) & .8466(.0110) & .9790(.0077) & .9079(.0071) & .009(.001) & 0.009&0.97 \\ \cline{3-11} 
 && \multirow{3}{*}{2} & 0.05 &256 (4.23)& {\ul .8236(.0326)} & .9642(.0315) & .8887(.0300) & .054(.036) & 0.058&1.00 \\
 &&  & 0.1 &256 (4.25)  & {\ul .8248(.0114)} & .9668(.0098) & .8900(.0081) & .047(.017) & 0.051 &1.00 \\
 &&& 0.2   &  255 (4.11)  & .8269(.0113) & .9715(.0091) & .8932(.0078) & .038(.014) & 0.040&1.00 \\
 &&  & 0.5 &  259 (4.00)  & .8354(.0115) & .9776(.0079) & .9007(.0074) & .024(.010) & 0.026&0.99 \\
 &&  & 1.0 &  265 (4.22) & .8511(.0111) & .9720(.0092) & .9073(.0073) & .020(.009) & 0.022&0.99\\ \cline{3-11} 
 && \multirow{3}{*}{4} & 0.05 &258 (4.06)& {\ul .8082(.0326)} & .9538(.0306) & .8829(.0291) & .042(.0360) & 0.047 & 1.00 \\
 &&  & 0.1 &263 (3.99) & {\ul .7931(.0114)} & .9358(.0088) & .8841(.0079) & .051(.043) & 0.056 &0.99 \\
 &&& 0.2   &  268 (4.80) & .8207(.0115) & .9673(.0081) & .8871(.0078) & .032(.014) & 0.035&1.00 \\
 &&  & 0.5 &  264 (4.01) & .8322(.0113) & .9684(.0086) & .8943(.0074) & .025(.012) & 0.028&0.97 \\
 &&  & 1.0  & 265 (4.36)& .8485(.0109) & .9682(.0096) & .9043(.0072) & .020(.010) & 0.023&0.99\\ \hline
\multirow{10}{*}{0.4} & BRL& &  &275 (2.99)& .5583(.0212)& .5665(.0642)& .5539(.0344) & .424(.045) & 0.427&0.59 \\ \cline{2-11} 
 &\multirow{15}{*}{BRLVOF}& \multirow{3}{*}{1}& 0.05 &233 (6.44)& {\ul .6336(.0338)} & .8038(.0376) & .7379(.035) & .125(.091) & 0.128 &0.91 \\
 &&  & 0.1 &234 (6.46)& {\ul .6636(.0140)} & .8523(.0218) & .7459(.0139) & .081(.016) & 0.083 & 0.94 \\
 && & 0.2  &  228 (6.35)& .6683(.0143) & .8813(.0198) & .7598(.0112) & .063(.011) & 0.064&0.99 \\
 &&  & 0.5 &  219 (5.99)& .6763(.0144) & .9252(.0159) & .7811(.0113) & .030(.004) & 0.030&0.97 \\
 &&  & 1.0 &  222 (5.74)& .6937(.0143) & .9365(.0147) & .7967(.0108) & .011(.001) & 0.011&0.95 \\ \cline{3-11} 
 && \multirow{3}{*}{2} & 0.05 &223 (6.84)& {\ul .6467(.0348)} & .8709(.0381) & .7419(.0339) & .065(.040) & 0.072 & 1.00 \\
 &&  & 0.1 &224 (6.75) & {\ul .6436(.0239)} & .8732(.0236) & .7474(.0129) & .067(.038) & 0.073 & 0.99 \\
 &&& 0.2   &  218 (6.46) & .6532(.0147) & .9010(.0191) & .7570(.0121) & .047(.019) & 0.051&1.00  \\
 &&  & 0.5 &  222 (6.13) & .6535(.0145) & .9063(.0164) & .7591(.0112) & .052(.015) & 0.056&0.98 \\
 &&  & 1.0 &  227 (6.21) & .6907(.0143) & .9208(.0170) & .7890(.0110) & .027(.012) & 0.003&0.99 \\ \cline{3-11} 
 && \multirow{3}{*}{4} & 0.05 &216 (6.52)& {\ul .6242(.0344)} & .8774(.0372) & .7364(.0331) & .058(.039) & 0.064 & 1.00 \\
 &&  & 0.1 &234 (5.94) & {\ul .6392(.0137)} & .8291(.0184) & .7403(.0121) & .084(.065) & 0.092 & 0.98 \\
 &&& 0.2   &  230 (5.93)  & .6449(.0143) & .9112(.0183) & .7549(.0119) & .042(.020) & 0.047&0.96 \\
 &&  & 0.5 &  224 (5.97)  & .6614(.0144) & .9210(.0175) & .7695(.0112) & .034(.017) & 0.038&0.97 \\
 &&  & 1.0 &  225 (6.54)  & .6886(.0144) & .9165(.0182) & .7860(.0112) & .028(.016) & 0.032&0.99 \\ \hline
\end{tabular}
\end{table}
\FloatBarrier

\section{Application to Meals on Wheels Data} \label{data}
Home-delivered meals offered by Meals on Wheels (MOW) programs across the United States provide a crucial service to homebound older adults. These programs are associated with better nutritional outcomes, decreased rates of depression, and delayed onset of institutional care among its recipients \cite{Thomas2013}. 

In recent years, some MOW programs have implemented health and safety assessments with their services. These functional assessments can be used to identify patient populations who would benefit from the program, and experience reductions in healthcare utilization with meal delivery services \cite{Lloyd2015}. However, many MOW programs do not collect information on recipients' healthcare utilization. In order to evaluate the relationship between functional assessments and healthcare utilization, we link client records from MOW with Medicare claims data from the Centers for Medicare and Medicaid Services (CMS). This relationship can provide information on whether certain clients should be prioritized to receive services from MOW, and for targeting specific clients with additional services.

\subsection{Data} 
MOW client information was collected for all individuals who received home-delivered meals from participating programs in the state of Rhode Island, between January 1, 2010 and December 31, 2013. The client lists submitted by these programs contain information on the gender, date of birth (DOB), start date of service, and the 9-digit ZIP code for each client. The 9-digit zip code refers to a five-digit zip code plus a 4-digit add-on number. The 4-digit add on identifies a geographic segment within the 5-digit delivery zone, such as a city block, office building, a high-volume receiver of mail, or any other distinct mail unit \cite{grubesic,thomas}. In addition, two functional measures are provided, that inform the capacity of an individual to live independently in the community. The Activities of Daily Living (ADL) score evaluates the ability of an individual to perform daily self-care tasks necessary for independent living, including personal hygiene, dressing, eating, maintaining continence, and mobility. The Instrumental Activities of Daily Living (IADL) score measures the ability of an individual to perform more complex actions required for independent living. These tasks include basic communication, transportation, meal preparation, shopping, housework, managing medications, and managing personal finance. The ability to perform each of these tasks is assessed by the MOW intake coordinator, and each item is assigned a score ranging from 1 (no assistance required) to 3 (complete assistance required). The scores are summed to form an aggregate score for both ADL and IADL \cite{Mitty1988}. 

We restrict the analysis to individuals older than 65 at enrollment, as only those individuals are expected to be enrolled in Medicare. This results in a total of $n_A=3916$ MOW recipients, and $n_B=233,922$ Medicare beneficiaries. Table \ref{tab:mow0} summarizes the variables in the MOW and Medicare files.

\begin{table}[]
\centering
\caption{Description of variables in the MOW and Medicare files. Age is calculated as of the earliest year of MOW service (2010).} \label{tab:mow0}
\begin{tabular}{l|c}
\hline
\multicolumn{2}{c}{MOW file (n = 3916)}\\ \hline
Characteristic & Summary \\ \hline
Gender (n (\%)) \\
\hspace{3mm} Male & 1253 (32) \\
\hspace{3mm} Female  & 2663 (68)\\
Age (years, mean (SD)) & 82.0 (7.7) \\
IADL score (mean (SD)) & 13.6 (3.1)\\
ADL score (mean (SD)) & 4.6 (2.6)\\
\hline
\end{tabular}

\begin{tabular}{l|c}
\hline
\multicolumn{2}{c}{Medicare file (n = 233,922)}\\ \hline
Characteristic & Summary \\ \hline
Gender (n (\%)) \\
\hspace{3mm} Male & 102,350 (44) \\
\hspace{3mm} Female  & 131,572 (56)\\
Age (years, mean (SD)) & 82.9 (7.8) \\
\makecell[l]{Number of utilization \\events (n (\%)))} \\
\hspace{3mm} Inpatient & 34,620 (14.8)  \\
\hspace{3mm} SNF & 9143 (4.0) \\
\hspace{3mm} ER & 34,153 (14.6) \\
\makecell[l]{Prevalence of chronic \\conditions (n (\%))} \\
\hspace{3mm} Alzheimer's &           7858 (3.4)\\
\hspace{3mm} \makecell[l]{Acute Myocardial\\ Infarction}  &      5225 (2.2)\\
\hspace{3mm} Anaemia &              60,317 (25.8)\\
\hspace{3mm} Asthma &               17,827 (7.6)\\
\hspace{3mm} Atrial Fibrillation&   15,032 (6.4)\\
\hspace{3mm} Breast Cancer &         6335 (2.7)\\
\hspace{3mm} Colorectal Cancer &     3644 (1.6)\\
\hspace{3mm} Endometrial Cancer &     963 (0.4)\\
\hspace{3mm} Lung Cancer &           2091 (0.9)\\
\hspace{3mm} Prostate Cancer &       5700 (2.4)\\
\hspace{3mm} Cataract &             71,509 (30.6)\\
\hspace{3mm} COPD &                 28,557 (12.2)\\
\hspace{3mm} Depression &           44,627 (19.1)\\
\hspace{3mm} Diabetes &             42,504 (18.2)\\
\hspace{3mm} Hyperlipidemia & 89,850 (38.4) \\
\hspace{3mm} Hypertension & 96533 (41.3) \\
\hspace{3mm} Arthritis & 52,442 (22.4) \\
\hspace{3mm} Osteoporosis & 22,181 (9.5) \\
\hspace{3mm} Stroke & 13,398 (5.7)\\
\bottomrule
\end{tabular}
\end{table}

\subsection{Record linkage and analysis}
The comparison of all record pairs from the MOW and Medicare files involves over 900 million possible comparisons. To reduce the computational complexity, we  create blocks based on the 5-digit ZIP code and gender. This results in a total of 128 blocks, and $11,465,820$ record pairs. The minimum and maximum number of record-pairs per block are 50 and 785,017, respectively, with an average of 89,577 pairs.


\begin{table}[]
\centering
\caption{Linking variable description and disagreement levels.} \label{tab:mow1}
\begin{tabular}{lc}
\hline
Agreement Type & Level \\
\hline
Disagreement on DOB & $r_D=1$ \\
Agree on DOB Year only & $r_D=2$ \\
Agree on DOB Year and Month only & $r_D=3$ \\
Agree on DOB Year, Month, and Day & $r_D=4$ \\
Agree on first 5 digits of ZIP code only & $r_Z=1$ \\
Agree on first 6 digits of ZIP code only & $r_Z=2$ \\
Agree on first 7 digits of ZIP code only & $r_Z=3$ \\
Agree on first 8 digits of ZIP code only & $r_Z=4$ \\
Agree on all 9 digits of ZIP code & $r_Z=5$\\
\hline
\end{tabular}
\end{table}

We use the MOW recipients' DOB and 9-digit ZIP code as linking variables. Table \ref{tab:mow1} shows the levels of agreement for both linking variables. Following the recommendation in Winkler \cite{Winkler2002}, we model the interaction between agreement on all components of the DOB, and all digits of the ZIP code. The likelihood for the comparison data  under BRL is
\begin{align}
\begin{split}
    \mathcal{L}^{BRL_{Block}}(\mathbf{C},\theta_{M},\theta_{U}|\mathbf{Z}_A,\mathbf{Z}_B,\mathbf{Q})= \prod_{i=1}^{n_A} \prod_{j=1}^{n_B} &\bigg[ \prod_{l_D=1}^{3}\prod_{l_Z=1}^{4} \theta_{MDl_D}^{\mathbbm{1}(\gamma_{ijD}=l_{D})} \theta_{MZl_{Z}}^{\mathbbm{1}(\gamma_{ijZ}=l_{Z})} \theta_{MDZ}^{\mathbbm{1}(\gamma_{ijl_D}=4,\gamma_{ijl_Z}=5)}\bigg]^{C_{ij} Q_{ij}} \\ 
\times& \bigg[ \prod_{l_D=1}^{3}\prod_{l_Z=1}^{4} \theta_{UDl_D}^{\mathbbm{1}(\gamma_{ijD}=l_{D})} \theta_{UZl_{Z}}^{\mathbbm{1}(\gamma_{ijZ}=l_{Z})} \theta_{UDZ}^{\mathbbm{1}(\gamma_{ijl_D}=4,\gamma_{ijl_Z}=5)}\bigg]^{(1-C_{ij})Q_{ij}},
\end{split}
\end{align}
where $\theta_{MZ}$ and $\theta_{UZ}$ are parameters governing the distribution of ZIP code comparisons among the true links and non-links, respectively. Similarly, $\theta_{MD}$ and $\theta_{UD}$ denote parameters corresponding to the distributions of DOB comparisons among the true links and non-links, respectively.

We consider two specifications of the BRLVOF likelihood. Table \ref{tab:mow2} summarizes the variables, distributional forms, and the parameters for the models under both specifications. In the first specification, the likelihood includes models for relationships that are not of scientific interest. Specifically, we model the relationship between ADL and IADL scores in the MOW file, with the CCW indicators in the Medicare file ($f_{X2M}$, $f_{X2U}$ and $f_{X3M}$, $f_{X3U}$ in Table \ref{tab:mow2}). We call this specification BRLVOF$^{NS}$. 

For some record pairs, the date of death may occur before the recorded MOW enrollment date. If we assume that both dates are accurately recorded, then individuals with date of death prior to enrollment can be considered non-links. The likelihood can be restricted to linkage structures in which linked pairs have a date of death before MOW enrollment. If we assume that one of these dates is recorded erroneously, we can model the individuals' death status at enrollment for the links and non-links. A possible model for the death status is a Bernoulli distribution ($f_{X1M}$, $f_{X1U}$ in Table \ref{tab:mow2}). The likelihood under BRLVOF$^{NS}$ with possibly erroneous dates of death and enrollment is  
\begin{align} \label{eq:bvns}
\begin{split}
    \mathcal{L}^{BRLVOF^{NS}_{Block}}(\mathbf{C},\theta_{M},\theta_{U}, \beta_M, \beta_U&|\mathbf{X}_A,\mathbf{X}_B,\mathbf{Z}_A,\mathbf{Z}_B,\mathbf{Q}) \\
= &\prod_{i=1}^{n_A} \prod_{j=1}^{n_B} \bigg[f_{X1M} \times f_{X2M} \times f_{X3M} \times  \prod_{l_D=1}^{3}\prod_{l_Z=1}^{4} \theta_{MDl_D}^{\mathbbm{1}(\gamma_{ijD}=l_{D})} \theta_{MZl_{Z}}^{\mathbbm{1}(\gamma_{ijZ}=l_{Z})} \theta_{MDZ}^{\mathbbm{1}(\gamma_{ijl_D}=4,\gamma_{ijl_Z}=5)}\bigg]^{C_{ij}Q_{ij}} \\ 
&\times\bigg[f_{X1U} \times f_{X2U} \times f_{X3U} \times \prod_{l_D=1}^{3}\prod_{l_Z=1}^{4} \theta_{UDl_D}^{\mathbbm{1}(\gamma_{ijD}=l_{D})} \theta_{UZl_{Z}}^{\mathbbm{1}(\gamma_{ijZ}=l_{Z})} \theta_{UDZ}^{\mathbbm{1}(\gamma_{ijl_D}=4,\gamma_{ijl_Z}=5)}\bigg]^{(1-C_{ij})Q_{ij}}.
\end{split}
\end{align}

In the second specification, the BRLVOF likelihood also includes relationships of scientific interest in addition to the aforementioned models. We model the difference in acute inpatient, emergency room (ER), and 
skilled nursing facility (SNF) events in the 180 days before and after MOW enrollment as a function of the ADL and IADL scores, and the CCW indicators ($f_{Y1M}$, $f_{Y1U}$, $f_{Y2M}$, $f_{Y2U}$, and $f_{Y3M}$, $f_{Y3U}$ in Table \ref{tab:mow2}). We call this specification BRLVOF$^{S}$. The likelihood under BRLVOF$^{S}$ with possibly erroneous dates of death and enrollment is  
\begin{align} \label{eq:bvs}
\begin{split}
    &\mathcal{L}^{BRLVOF^{S}_{Block}}(\mathbf{C},\theta_{M},\theta_{U}, \beta_M, \beta_U|\mathbf{X}_A,\mathbf{X}_B,\mathbf{Z}_A,\mathbf{Z}_B,\mathbf{Q}) \\
= \prod_{i=1}^{n_A} \prod_{j=1}^{n_B} &\bigg[f_{X1M} \times f_{X2M} \times f_{X3M} \times f_{Y1M}\times f_{Y2M} \times f_{Y3M} \times \prod_{l_D=1}^{3}\prod_{l_Z=1}^{4} \theta_{MDl_D}^{\mathbbm{1}(\gamma_{ijD}=l_{D})} \theta_{MZl_{Z}}^{\mathbbm{1}(\gamma_{ijZ}=l_{Z})} \theta_{MDZ}^{\mathbbm{1}(\gamma_{ijl_D}=4,\gamma_{ijl_Z}=5)}\bigg]^{C_{ij}Q_{ij}} \\ 
\times& \bigg[f_{X1U} \times f_{X2U} \times f_{X3U} \times f_{Y1U}\times f_{Y2U} \times f_{Y3U} \times \prod_{l_D=1}^{3}\prod_{l_Z=1}^{4} \theta_{UDl_D}^{\mathbbm{1}(\gamma_{ijD}=l_{D})} \theta_{UZl_{Z}}^{\mathbbm{1}(\gamma_{ijZ}=l_{Z})} \theta_{UDZ}^{\mathbbm{1}(\gamma_{ijl_D}=4,\gamma_{ijl_Z}=5)}\bigg]^{(1-C_{ij})Q_{ij}}.
\end{split}
\end{align}

\begin{table}[]
\caption{Summary of linkage and analysis models. CCW indicators are disease indicators from the Chronic Conditions Data Warehouse under CMS.} \label{tab:mow2}
\centering
\resizebox{\textwidth}{!}{
\large{
\begin{tabular}{lllll}
\hline
Model & Outcome & Predictors & Model Form& Parameters \\ \hline
$f_Z$ & DOB, ZIP code & & Multinomial & $\theta_M$, $\theta_U$\\ 
$f_{X1M}$, $f_{X1U}$ & \makecell[l]{Death and MOW enrollment dates \\assumed erroneous} & & Bernoulli & $\beta_{M1}$, $\beta_{U1}$\\
$f_{X1M}$, $f_{X1U}$ & \makecell[l]{Death and MOW enrollment dates \\assumed correct} & & \makecell[l]{$f_{X1M} = \begin{cases}
    1, \text{ if, for pair $(i,j)$, enrollment of record $i$ occurs before death of record $j$}\\
    0, \text{ otherwise} \end{cases}$\\
$f_{X1U} \propto 1$} & {-} \\
$f_{X2M}$, $f_{X2U}$ & ADL score & CCW indicators & Linear Regression & $\beta_{M2}$, $\sigma_{M2}$, $\beta_{U2}$, $\sigma_{U2}$\\
$f_{X3M}$, $f_{X3U}$ & IADL score & CCW indicators & Linear Regression & $\beta_{M3}$, $\sigma_{M3}$, $\beta_{U3}$, $\sigma_{U3}$\\
$f_{Y1M}$, $f_{Y1U}$ & \makecell{180 Day pre-post difference in \\   Inpatient Acute Events} &  CCW indicators, ADL Score, IADL score& Linear Regression & $\beta_{Y1M}$, $\sigma_{Y1M}$, $\beta_{Y1U}$, $\sigma_{Y1U}$\\

$f_{Y2M}$, $f_{Y2U}$ &  \makecell{180 Day pre-post difference in \\   ER Events} &  CCW indicators, ADL Score, IADL score& Linear Regression & $\beta_{Y2M}$, $\sigma_{Y2M}$, $\beta_{Y2U}$, $\sigma_{Y2U}$\\

$f_{Y3M}$, $f_{Y3U}$ & \makecell{180 Day pre-post difference in \\  SNF Events} & CCW indicators, ADL Score, IADL score& Linear Regression & $\beta_{Y3M}$, $\sigma_{Y3M}$, $\beta_{Y3U}$, $\sigma_{Y3U}$ \\ \hline

\end{tabular} 
}
}
\end{table}

We use Dirichlet$(1, \dots, 1)$ priors for each of the parameters in $\theta_M$ and $\theta_U$ for BRL, BRLVOF$^{NS}$, and BRLVOF$^{S}$. We assume non-informative prior distributions on the linear model parameters \cite{gelmanbook} for both BRLVOF specifications.
For all algorithms, we generate 1000 samples of the linkage structure $\mathbf{C}$, and use the last 100 for analysis. The results assuming that dates of death and enrollment are correctly specified are similar to those assuming that they are erroneous. Thus, we only present results under the assumption that these dates are correctly recorded. 

To assess convergence of the MCMC chains, we perform Geweke's diagnostic test  \cite{geweke} for each parameter. Results from the test do not indicate convergence problems for any of the parameters. We present trace plots and autoassociation plots for selected parameters in the supplementary materials.

We also implement BRLVOF$_{ind}$, with our proposed adjustments for blocking (Section 3.2). We use the same likelihoods as in Equations \eqref{eq:bvns} and \eqref{eq:bvs}. However, $f_{X2U}, f_{X3U}, f_{Y1U}, f_{Y2U}, f_{Y3U}$ are modeled using independent normal distributions.

We examine the conditional association between differences in acute inpatient, emergency room (ER), and skilled nursing facility (SNF) events in the 180 days before and after enrollment, and the ADL and IADL scores. We confine this analysis to individuals who were fee-for-service 180 days prior to enrollment in MOW, and remain fee-for-service in the 180 days after enrollment. We estimate these associations separately for the ADL and the IADL scores. Each model includes pre-existing medical conditions, as listed in Table \ref{tab:mow2}. Each set of associations is estimated using a linear model. We calculate point and interval estimates of the estimated conditional associations using the formulae in Section \ref{sec:MI}.


\subsection{Linkage results}
Table \ref{linkres} provides a summary of the linkage performance for all the methods. The number of linked records under BRL ranges between 3134 and 3214, with a posterior mean of 3175, and a 95\% credible interval of [3139, 3213]. BRLVOF$^{NS}$ links an average of 3760 records, with a 95\% credible interval of [3289, 4230]. BRLVOF$^{NS}_{ind}$ links an average of 3802 records, with a 95\% credible interval of [3496, 4108].  When substantively important models are incorporated in the linkage, BRLVOF$^{S}$ links an average of 3788 records, with a 95\% credible interval of [3303, 4241]. Furthermore, BRLVOF$^{S}_{ind}$ links an average of 3849 individuals, with a 95\% credible interval of [3477, 4312].

Among the records linked using BRL, an average of 1524 individuals are Medicare fee-for-service in the six months prior to and following enrollment in MOW (95\% CI: [1503, 1546]). Using BRLVOF$^{NS}$, an average of 1835 individuals are Medicare fee-for-service in the six months prior to and following MOW services (95\% CI: [1586, 2085]).  BRLVOF$^{S}$ identifies an average of 1840 individuals are Medicare fee-for-service in the six months prior to and following MOW services (95\% CI: [1591, 2097]). The numbers of Medicare fee-for-service individuals under BRLVOF$^{NS}_{ind}$ and BRLVOF$^{S}_{ind}$ are higher, with an average of 1840 (95\% CI: [1591, 2097]) and 1830 (95\% CI: [1622, 2001]), respectively. 
\begin{table}[t]
\caption{Linkage results under all methods.} \label{linkres}
\centering
\begin{tabular}{c l c c}
 & Method & Estimate & 95\% Credible Interval \\ \hline
\multirow{5}{*}{$n$} & BRL & 3175 & (3139, 3213) \\
 & BRLVOF$^{NS}$ & 3760 & (3289, 4230) \\ 
 & BRLVOF$^{S}$ & 3788 & (3303, 4241) \\
 & BRLVOF$^{NS}_{ind}$ & 3802 & (3496, 4108) \\ 
 & BRLVOF$^{S}_{ind}$ & 3849 & (3477, 4312) \\ \hline
\multirow{5}{*}{$n_{FFS}$} & BRL & 1524 & (1503, 1546) \\
 & BRLVOF$^{NS}$ & 1835 & (1586, 2085) \\ 
  & BRLVOF$^{S}$ & 1840 & (1591, 2097) \\ 
  & BRLVOF$^{NS}_{ind}$ & 1802 & (1658, 1945) \\ 
 & BRLVOF$^{S}_{ind}$ & 1830 & (1622, 2001) \\ \hline

\end{tabular}
\end{table}

\subsection{Analysis results}

The estimates of the association between differences in six-month prior and post-enrollment utilization events, and  functional status measures, are depicted in Table \ref{anres}. Under all the methods, the ADL  and IADL scores are not significantly associated with the pre-post difference in inpatient, ER, or SNF events. BRL, BRLVOF$^{NS}$, and BRLVOF$^{S}$ estimate that patients with higher functional status would experience a greater reduction in the number of inpatient events after receiving MOW. BRLVOF$^{NS}$ estimates a smaller decrease in the number of inpatient events compared to BRL for a one unit increase in the ADL score (0.001 vs 0.003) and the IADL score (0.001 vs 0.008). BRLVOF$^{S}$ estimates a larger decrease in acute inpatient events than BRLVOF$^{NS}$ for a one unit increase in the ADL score (0.009 vs 0.001) and the IADL score (0.005 vs 0.001). BRLVOF$^{NS}_{ind}$ and BRLVOF$^{S}_{ind}$ estimate a decrease in the number of acute inpatient events for one unit increase in the ADL and IADL scores.  \par
BRLVOF$^{NS}$ estimates a smaller decrease in the number of ER events compared to BRL for a one point increase in the ADL score (0.002 vs 0.005). For the IADL score, BRLVOF$^{NS}$ estimates that individuals with higher functional status would experience a greater decrease in ER events after enrollment ($\hat{\beta}=-0.001)$, while BRL estimates that individuals with lower functional status would have fewer ER admissions ($\hat{\beta}=0.004)$. Results under BRLVOF$^{S}$ are qualitatively similar to those under BRLVOF$^{NS}$ ($\hat{\beta}=-0.013$ for ADL and $\hat{\beta}=-0.011$ for IADL). BRLVOF$^{NS}_{ind}$ and BRLVOF$^{S}_{ind}$ estimate a decrease in ER admissions per unit increase in the ADL score ($\hat{\beta}=-0.012$ and -0.002), and per unit increase in the IADL score ($\hat{\beta}=-0.011$ and -0.009).   \par

BRL estimates that patients with lower functional impairment according to the ADL score experience a greater reduction in the number of SNF events ($\hat{\beta}=-0.011)$. BRLVOF$^{NS}$ estimates a null effect of the ADL and IADL scores on the number of SNF events ($\hat{\beta}=0$).
BRLVOF$^{S}$ estimates patients with  a higher ADL score to have a larger number of SNF events ($\hat{\beta}=0.002$). Furthermore, BRLVOF$^{NS}_{ind}$ estimates an increase in SNF events with increase in ADL score, but a decrease in SNF events with an increase in the IADL score. BRLVOF$^{S}_{ind}$ estimates an increase in SNF events with increase in ADL score, but
no association between the IADL score and the number of SNF events.

\begin{table}[t]
\centering
\caption{Estimated associations under all methods.} \label{anres}
\begin{tabular}{cclcc}
Utilization Event & Variable & Method & $\hat{\beta}$ & 95\% Confidence Interval \\ \hline
\multirow{9}{*}{Inpatient} & \multirow{2}{*}{ADL} & BRL & -0.003 & (-0.017, 0.016) \\
 &  & BRLVOF$^{NS}$ & -0.001 & (-0.022, 0.021) \\ 
 &  & BRLVOF$^{S}$ & -0.009 & (-0.020, 0.019) \\ 
 & & BRLVOF$^{NS}_{ind}$ & -0.005 & (-0.026, 0.015) \\ 
 &  & BRLVOF$^{S}_{ind}$ & -0.001 & (-0.024, 0.020)  \\ \cline{2-5} 
 & \multirow{2}{*}{IADL} & BRL & -0.008 & (-0.022, 0.005) \\
 &  & BRLVOF$^{NS}$ & -0.001 & (-0.031, 0.033) \\ 
  &  & BRLVOF$^{S}$ & -0.005 & (-0.028, 0.026) \\ 
  & & BRLVOF$^{NS}_{ind}$ & -0.019 & (-0.036, 0.020) \\ 
 &  & BRLVOF$^{S}_{ind}$ & -0.014 & (-0.028, 0.025) \\ \hline
\multirow{9}{*}{ER} & \multirow{2}{*}{ADL} & BRL & -0.005 & (-0.023, 0.013) \\
 &  & BRLVOF$^{NS}$ & -0.002 & (-0.023, 0.020) \\
  &  & BRLVOF$^{S}$ & -0.013 & (-0.036, 0.020) \\ 
  & & BRLVOF$^{NS}_{ind}$ & -0.012 & (-0.031, 0.007) \\ 
 &  & BRLVOF$^{S}_{ind}$ & -0.002 & (-0.019, 0.020) \\ \cline{2-5} 
 & \multirow{2}{*}{IADL} & BRL & 0.004 & (-0.011, 0.018) \\
 &  & BRLVOF$^{NS}$ & -0.001 & (-0.031, 0.029) \\ 
 &  & BRLVOF$^{S}$ & -0.011 & (-0.028, 0.011) \\  
 & & BRLVOF$^{NS}_{ind}$ & -0.011 & ( -0.027, 0.005) \\ 
 &  & BRLVOF$^{S}_{ind}$ & -0.009 & ( -0.016, 0.019) \\ \hline
\multirow{9}{*}{SNF} & \multirow{2}{*}{ADL} & BRL & -0.011 & (-0.030, 0.008) \\
 &  & BRLVOF$^{NS}$ & 0.000 & (-0.020, 0.024) \\ 
 &  & BRLVOF$^{S}$ & 0.002 & (-0.010, 0.012) \\ 
 & & BRLVOF$^{NS}_{ind}$ & 0.016 & (-0.007, 0.039) \\ 
 &  & BRLVOF$^{S}_{ind}$ & 0.009 & (-0.012, 0.017) \\ \cline{2-5} 
 & \multirow{2}{*}{IADL} & BRL & -0.014 & (-0.030, 0.002) \\
 &  & BRLVOF$^{NS}$ & 0.000 & (-0.030, 0.030) \\ 
 &  & BRLVOF$^{S}$ & -0.006 & (-0.020, 0.012) \\  
 & & BRLVOF$^{NS}_{ind}$ & -0.008 & (-0.028, 0.011) \\ 
 &  & BRLVOF$^{S}_{ind}$ & 0.000 & (-0.011, 0.018) \\ \hline
\end{tabular}
\end{table}

\section{Discussion} \label{disc}
We propose an extension to the Bayesian Fellegi-Sunter methodology, that incorporates associations between variables in either file, and can utilize blocking. The proposed method, BRLVOF, models the associations within the linked and unlinked record pairs. This is in contrast to a recently proposed method \cite{Tang2020}, which assumes that $\mathbf{X}_A$ and $\mathbf{X}_B$ are independent among non-links. Incorporating associations among non-links can be beneficial when $\mathbf{X}_A$ or $\mathbf{X}_B$ is associated with the blocking variable, which can induce marginal associations between $\mathbf{X}_{A}$ and $\mathbf{X}_{B}$. A small simulation that illustrates this phenomenon is presented in Table S12 of the supplementary materials.

We show analytically and through simulations, that BRLVOF can result in improved linkage accuracy. The improvement in linkage accuracy results in inferences that are less biased, and with smaller RMSEs than BRL. The improvement is more noticeable when the information in the linking variables is limited, as observed in simulations with higher linkage error levels. The improvement is also stronger when the strength of the association between $\mathbf{X}_A$ and $\mathbf{X}_B$ is larger. \par


Using BRLVOF, we are able to link more MOW individuals to their Medicare enrollment records than BRL, providing a larger analytic sample of individuals. In our analysis, including the scientific model of interest in the BRLVOF likelihood did not substantially alter results compared to when it was not included. None of the methods identify significant associations between MOW recipients' ADL or IADL scores, and the change in acute inpatient, ER, or SNF events before and after receipt of the meals. The point estimates generally suggest that patients with less functional impairment may experience a greater reduction in healthcare utilization compared to patients with greater functional impairment. This trend is similar to BRL; however, for some events, the sign of the point estimate is reversed. 

An important extension of the BRLVOF method is allowing erroneous blocking fields, and accommodating complex blocking schemes like multiple passes or data-driven blocks.  In the supplementary material, we describe possible adjustments to the BRLVOF model that address erroneous blocking variables. Estimation of this model and examining its performance is an important area of future research.

\subsection*{Funding Information}
This work was supported by the National Institute on Aging (R21AG059120) and the Patient-Centered Outcomes Research Institute (ME-2017C3-10241). All statements in this report, including its findings and conclusions, are solely those of the authors and do not necessarily represent the views of the PCORI, its Board of Governors, or the Methodology Committee.

\printbibliography

@article{Winkler1993b,
  title={Improved decision rules in the Fellegi-Sunter model of record linkage},
  author={Winkler, W. E.},
  year={1993},
  journal={Proceedings of the Survey Research Methods Section, American Statistical Association},
  publisher={Bureau of the Census},
  pages={274-279},
}

@article{Winkler2002,
  title={Methods for Record Linkage and Bayesian Networks},
  author={Winkler, W. E.},
  year={2002},
  journal={Proceedings of the Survey Research Methods Section, American Statistical Association},
  publisher={Bureau of the Census},
  pages={274-279},
}

@article{binette22,
author = {O Binette and R C Steorts},
title = {(Almost) all of entity resolution},
journal = {Science Advances},
volume = {8},
number = {},
pages = {},
year  = {2022},
}

@book{Newcombe1988,
author={H. B. Newcombe},
title={Handbook of Record Linkage: Methods for Health and Statistical Studies, Administration, and Buisiness},
year={1988},
publisher={Oxford University Press},
address={Oxford}
}

@article{Gill1977,
author={L. E. Gill},
title={OX-LINK: The Oxford Medical Record Linkage System Demonstration of the PC Version},
year=1997,
journal={Proceedings of an International Workshop and Exposition},
pages={15-34}
}

@article{Sadinle2018,
author = "Sadinle, M.",
journal = "The Annals of Applied Statistics",
number = "2",
pages = "1013--1038",
title = "Bayesian propagation of record linkage uncertainty into population size estimation of human rights violations",
volume = "12",
year = "2018"
}

@article{Winkler1994,
author={W. E. Winkler},
title={Advanced methods for record linkage},
year=1994,
journal={Proceedings of the Survey Research Methods Section, American Statistical Association},
pages={467--472}
}

@incollection{Winkler1995,
author={W. E. Winkler},
title={Matching and record linkage},
year=1995,
booktitle={Buisiness Survey Methods },
editor={B.G. Cox and D.A. Binder and B.N. Chinnappa and A. Christianson and M.J. Colledge and P.S. Kott},
address={New York},
publisher={Wiley Publications},
pages={355--384}
}

@article{Gomatam2002,
author = {Gomatam, S. and Carter, R. and Ariet, M. and Mitchell, G.},
title = {An empirical comparison of record linkage procedures},
journal = {Statistics in Medicine},
volume = {21},
pages = {1485-1496},
year={2002}
}

@article{Campbell2008,
author = {K. M. Campbell and D. Deck and A. Krupski},
title ={Record linkage software in the public domain: A comparison of Link Plus, The Link King, and a `basic' deterministic algorithm},
journal = {Health Informatics Journal},
volume = {14},
pages = {5-15},
year = {2008}
}

@article{Fellegi1969,
author ={I. P. Fellegi and A. B. Sunter},
title={A theory for record linkage},
year= 1969,
journal = {Journal of the American Statistical Association},
number = {64},
pages = {1183--1210}
}

@article{Jaro1989,
author={M.A. Jaro},
title={Advances in record-linkage methodology as applied to matching the 1985 census of Tampa, Florida},
year=1989,
journal={Journal of the American Statistical Association},
number=84,
pages={414--420}
}

@article{Murray2015,
author = {J. S. Murray},
title = {Probabilistic record linkage and deduplication after indexing, blocking, and filtering},
journal = {Journal of Privacy and Confidentiality},
volume = {7},
number = {},
pages = {https://doi.org/10.29012/jpc.v7i1.643},
year  = {2015}
}

@article{Scheuren1993,
author = {Scheuren, F. and Winkler, W. E.},
year = {1993},
pages = {39-58},
title = {Regression analysis of data files that are computer matched - Part I},
volume = {19},
journal = {Survey Methodology}
}

@article{Scheuren1997,
author = {Scheuren, F. and Winkler, W. E.},
year = {1997},
pages = {157-165},
title = {Regression analysis of data files that are computer matched - Part II},
volume = {23},
journal= {Survey Methodology}
}

@article{Lahiri2005,
 author = {P. Lahiri and M. D. Larsen},
 journal = {Journal of the American Statistical Association},
 number = {469},
 pages = {222-230},
 title = {Regression analysis with linked data},
 volume = {100},
 year = {2005}
}

@article{wang2022,
author = {Wang, Z. and Ben-David, E. and Diao, G. and Slawski, M.},
title = {Regression with linked datasets subject to linkage error},
journal = {WIREs Computational Statistics},
volume = {14},
pages = {e1570},
year = {2022}
}

@article{kimchambers2012,
author = {G. Kim and R. Chambers},
title = {Regression analysis under probabilistic multi-linkage},
journal = {Statistica Neerlandica},
volume = {66},
pages = {64-79},
year = {2012}
}

@Article{chambersdasilva2020,
  author={R. Chambers and A. D. {da Silva}},
  title={{Improved secondary analysis of linked data: A framework and an illustration}},
  journal={Journal of the Royal Statistical Society, Series A},
  year=2020,
  volume={183},
  pages={37-59}
}

@article{Fortini2001,
author = {M. Fortini and B. Liseo and A. Nuccitelli},
year = {2001},
pages = {185-198},
title = {On Bayesian record linkage},
volume = {4},
journal = {Research in Official Statistics}
}

@article{Sadinle2017,
author = {M. Sadinle},
title = {Bayesian estimation of bipartite matchings for record linkage},
journal = {Journal of the American Statistical Association},
volume = {112},
pages = {600-612},
year  = {2017}
}

@article{Steorts2015,
author = {R. C. Steorts and R. Hall and S. E. Fienberg},
title = {A Bayesian approach to graphical record linkage and deduplication},
journal = {Journal of the American Statistical Association},
volume = {111},
pages = {1660-1672},
year = {2016}
}

@article{Gutman2013,
  title={A Bayesian procedure for file linking to analyze end-of-life medical costs},
  author={R. Gutman and C. C. Afendulis and A. M. Zaslavsky},
  journal={Journal of the American Statistical Association},
  year={2013},
  volume={108},
  pages={34--47}
}

@article{Dalzell2018,
author = {N. M. Dalzell and J. P. Reiter},
title = {Regression modeling and file matching using possibly erroneous matching variables},
journal = {Journal of Computational and Graphical Statistics},
volume = {27},
pages = {728-738},
year  = {2018}
}

@InProceedings{Tang2020,
author="J. Tang and  J. P. Reiter and R. C. Steorts",
editor="Domingo-Ferrer, Josep",
title="Bayesian modeling for simultaneous regression and record linkage",
booktitle="Privacy in Statistical Databases",
year="2020",
publisher="Springer International Publishing",
pages="209--223"
}

@article{Winkler1990,
author = {Winkler, W. E.},
year = {1990},
pages = {354–359},
title = {String comparator metrics and enhanced decision rules in the Fellegi-Sunter model of record linkage},
journal = {Proceedings of the Survey Research Methods Section, American Statistical Association}
}

@article{Winkler1989,
author = {W. E. Winkler},
year = {1989},
pages = {145-155},
title = {Near automatic weight computation in the Fellegi-Sunter model of record linkage},
journal={Proceedings of the Fifth Census Bureau Annual Research Conference}
}

@article{Larsen2005,
    author = {M. D. Larsen},
    title = {Hierarchical Bayesian record linkage theory},
   journal = {Proceedings of the
Section on Survey Research Methods},
     pages = {3277-3284},
    year = {2005}
}

@article{Guha2022,
author = "S. Guha and J. P. Reiter and A. Mercatanti",
journal = "Bayesian Analysis",
month = "",
number = "",
pages = "1--22",
publisher = "The Institute of Mathematical Statistics",
title = "Bayesian causal inference with bipartite record linkage",
volume = "",
year = "2022"
}

@article{tannerwong1987,
	Author = {M. Tanner and W. Wong},
	Journal = {Journal of the American Statistical Association},
	Number = {},
	Pages = {528-540},
	Title = {The calculation of posterior distributions by data augmentation},
	Volume = {82(398)},
	Year = {1987}}

@article{hastings,
author = {W. Hastings},
title = {Monte Carlo sampling methods using Markov chains and their application},
journal = {Biometrika},
volume = {57},
pages = {97–109},
year = {1970}
}

@article{Wu1995,
author={Y. Wu},
title={Random shuffling: A new approach to matching problem},
year={1995},
journal={Proceedings of the Statistical Computing Section, American Statistical Association},
pages={69-74}
}

@article{MardiaGreen2006,
    author = {Green, P. J. and Mardia, K. V.},
    title = "{Bayesian alignment using hierarchical models, with applications in protein bioinformatics}",
    journal = {Biometrika},
    volume = {93},
    pages = {235-254},
    year = {2006},
}

@book{Rubin1987,
author={D. B. Rubin},
year={1987},
title={Multiple Imputation for Nonresponse in Surveys},
publisher={Wiley},
location={New York}
}

@article{Meng1994,
author = {X. L. Meng},
title = {Multiple-imputation inferences with uncongenial sources of input},
journal = {Statistical Science},
volume = {9},
pages = {538--558},
year  = {1994}
}

@article{Rubin1996,
author = {D. B. Rubin},
title = {Multiple imputation after 18+ years},
journal = {Journal of the American Statistical Association},
volume = {91},
pages = {473–489},
year  = {1996}
}

@article{Thomas2013,
title = {Providing more home-delivered meals is one way to keep older adults with low care needs out of nursing homes},
author={K. S. Thomas and V. Mor},
journal = {Health Affairs},
volume = {32},
pages = {1796-1802},
year = {2013}
}

@article{Lloyd2015,
author = {J. L. Lloyd and N. S. Wellman},
title = {Older Americans Act nutrition programs: A community-based nutrition program helping older adults remain at home},
journal = {Journal of Nutrition in Gerontology and Geriatrics},
volume = {34},
pages = {90-109},
year  = {2015}
}

@article{Mitty1988,
author = {Mitty, L. E.},
year = {1988},
month = {},
pages = {539-557},
title = {Resource utilization groups. DRGs move to long-term care},
volume = {23},
journal = {The Nursing Clinics of North America}
}

@book{gelmanbook,
author={A. Gelman and J. Carlin and H. Stern and D. Dunson and A. Vehtari and D. B. Rubin.},
year={2013},
title={Bayesian Data Analysis},
publisher={CRC Press}
}

@INProceedings{geweke,
author="J. Geweke",
editor="J.M. Bernardo and J.O. Berger and A.P. Dawid and  A.F.M. Smith",
title="Evaluating the accuracy of sampling-based approaches to calculating posterior moments",
booktitle="Bayesian Statistics",
year="1992",
publisher="Clarendon Press: Oxford, UK.",
}

@article{Tancredi2011,
 author = {A. Tancredi and B. Liseo},
 journal = {The Annals of Applied Statistics},
 number = {2B},
 pages = {1553-1585},
 publisher = {Institute of Mathematical Statistics},
 title = {A hierarchical Bayesian approach to record linkage and population size problems},
 volume = {5},
 year = {2011}
}

@inproceedings{zanellabetancourt,
author = {Zanella, Giacomo and Betancourt, Brenda and Wallach, Hanna and Miller, Jeffrey and Zaidi, Abbas and Steorts, Rebecca C.},
title = {Flexible Models for Microclustering with Application to Entity Resolution},
year = {2016},
isbn = {9781510838819},
publisher = {Curran Associates Inc.},
address = {Red Hook, NY, USA},
booktitle = {Proceedings of the 30th International Conference on Neural Information Processing Systems},
pages = {1425–1433},
numpages = {9},
location = {Barcelona, Spain},
series = {NIPS'16}
}

@article{cochinwala2001,
author = {M Cochinwala and V Kurien and G Lalk and D Shasha},
title = {Efficient data reconciliation},
journal = {Information Sciences},
volume = {137},
number = {1},
pages = {1-15},
year = {2001}
}

@article{grubesic,
title = {Zip codes and spatial analysis: Problems and prospects},
journal = {Socio-Economic Planning Sciences},
volume = {42},
number = {2},
pages = {129-149},
year = {2008},
author = {Tony H. Grubesic}
}

@inproceedings{chen2009,
author = {Chen, Fei and Gao, Byron J. and Doan, AnHai and Yang, Jun and Ramakrishnan, Raghu},
title = {Optimizing complex extraction programs over evolving text data},
year = {2009},
isbn = {9781605585512},
publisher = {Association for Computing Machinery},
address = {New York, NY, USA},
booktitle = {Proceedings of the 2009 ACM SIGMOD International Conference on Management of Data},
pages = {321–334},
numpages = {14},
location = {Providence, Rhode Island, USA},
series = {SIGMOD '09}
}

@article{fisher1915,
 author = {R. A. Fisher},
 journal = {Biometrika},
 number = {4},
 pages = {507--521},
 title = {Frequency Distribution of the Values of the Correlation Coefficient in Samples from an Indefinitely Large Population},
 volume = {10},
 year = {1915}
}

@article{xiemeng, title={Dissecting multiple imputation from a multi-phase inference perspective: What happens when God’s, imputer’s and analyst’s models are uncongenial?}, ISSN={10170405}, url={http://www3.stat.sinica.edu.tw/statistica/J27N4/J27N41/J27N41-10.html}, DOI={10.5705/ss.2014.067}, journal={Statistica Sinica}, author={Xie, Xianchao and Meng, Xiao-Li}, year={2017},volume={27},pages={1485-1594}}

@article{thomas, title={A methodology to identify a cohort of Medicare beneficiaries residing in large assisted living facilities using administrative data}, journal={Medical Care}, author={Thomas, K. S. and Dosa, D.cand Gozalo, P. L. and Grabowski, D. C. and  Nazareno, J. and Makineni, R. and Mor, V.}, year={2018},volume={56},pages={e10–e15} }

@article{steorts16,
author = {Rebecca C. Steorts},
title = {{Entity resolution with empirically motivated priors}},
volume = {10},
journal = {Bayesian Analysis},
number = {4},
publisher = {International Society for Bayesian Analysis},
pages = {849 -- 875},
year = {2015},
doi = {10.1214/15-BA965SI},
URL = {https://doi.org/10.1214/15-BA965SI}
}

@article{brlvof,
author = {Gauri Kamat and Ming Shan and Roee Gutman},
title ={A Bayesian record linkage approach that adjusts for variables in one dataset},
journal = {Under Review.},
year = {2026}
}

\end{document}


\maketitle

\section{Theoretical insights}
\subsection{Proof of Proposition 1}
\begin{proof}
(a) We have
\begin{align}
    \frac{\mathcal{L}^{BRL}_{(i,j) \in \mathbf{M}}}{\mathcal{L}^{BRL}_{(i,j) \in \mathbf{U}}} = \frac{f_M(\Gamma(\mathbf{Z}_{Ai},\mathbf{Z}_{Bj})|\theta_M)}{f_U(\Gamma(\mathbf{Z}_{Ai},\mathbf{Z}_{Bj})|\theta_U)}= \frac{\prod_{k=1}^{K} \prod_{l_k=1}^{L_k} \theta_{Mkl_k}^{\mathbbm{1}(\gamma_{ijk}=l_k)}}{\prod_{k=1}^{K} \prod_{l_k=1}^{L_k} \theta_{Ukl_k}^{\mathbbm{1}(\gamma_{ijk}=l_k)}}. \label{first}
\end{align}
Under BRLVOF, the likelihood ratio will be 
\begin{align}
    \frac{\mathcal{L}^{BRLVOF}_{(i,j) \in \mathbf{M}}}{\mathcal{L}^{BRLVOF}_{(i,j) \in \mathbf{U}}} &= \frac{\prod_{k=1}^{K} \prod_{l_k=1}^{L_k} \theta_{Mkl_k}^{\mathbbm{1}(\gamma_{ijk}=l_k)}}{\prod_{k=1}^{K} \prod_{l_k=1}^{L_k} \theta_{Ukl_k}^{\mathbbm{1}(\gamma_{ijk}=l_k)}} \times \frac{f_M(\mathbf{X}_{Ai},\mathbf{X}_{Bj}|\mathbf{Z}_{Ai},\mathbf{Z}_{Bj},\beta_M)}{f_U(\mathbf{X}_{Ai},\mathbf{X}_{Bj}|\mathbf{Z}_{Ai},\mathbf{Z}_{Bj},\beta_U)} \nonumber \\
    & \times \frac{\prod_{(p,q)}f_M(\mathbf{X}_{Ap},\mathbf{X}_{Bq}|\mathbf{Z}_{Ap},\mathbf{Z}_{Bq},\beta_M) \mathbb{I}(C_{pq}=1)}{\prod_{(p,q)}f_U(\mathbf{X}_{Ap},\mathbf{X}_{Bq}|\mathbf{Z}_{Ap},\mathbf{Z}_{Bq},\beta_U) \mathbb{I}(C_{pq}=0)}
     \nonumber \\
    & \times \frac{\prod_{(p,q)}f_U(\mathbf{X}_{Ap},\mathbf{X}_{Bq}|\mathbf{Z}_{Ap},\mathbf{Z}_{Bq},\beta_U) \mathbb{I}(C_{pq}=0)}{\prod_{(p,q)}f_M(\mathbf{X}_{Ap},\mathbf{X}_{Bq}|\mathbf{Z}_{Ap},\mathbf{Z}_{Bq},\beta_M) \mathbb{I}(C_{pq}=1)}.
\end{align}
Thus, we have 
\begin{align}
   \text{log } \frac{\mathcal{L}^{BRLVOF}_{(i,j) \in \mathbf{M}}}{\mathcal{L}^{BRLVOF}_{(i,j) \in \mathbf{U}}} &= \text{log } \frac{\mathcal{L}^{BRL}_{(i,j) \in \mathbf{M}}}{\mathcal{L}^{BRL}_{(i,j) \in \mathbf{U}}} + \text{   log } \frac{f_M(\mathbf{X}_{Ai},\mathbf{X}_{Bj}|\mathbf{Z}_{Ai},\mathbf{Z}_{Bj},\beta_M)}{f_U(\mathbf{X}_{Ai},\mathbf{X}_{Bj}|\mathbf{Z}_{Ai},\mathbf{Z}_{Bj},\beta_U)}.\label{last}
\end{align}
Taking expectations with respect to $(\mathbf{\Gamma},\mathbf{X_A},\mathbf{X_B})$, we have 
\begin{align}
   &\E_{(\mathbf{\Gamma},\mathbf{X_A},\mathbf{X_B})} \text{log } \frac{\mathcal{L}^{BRLVOF}_{(i,j) \in \mathbf{M}}}{\mathcal{L}^{BRLVOF}_{(i,j) \in \mathbf{U}}} - \text{   } \E_{(\mathbf{\Gamma},\mathbf{X_A},\mathbf{X_B})} \text{log } \frac{\mathcal{L}^{BRL}_{(i,j) \in \mathbf{M}}}{\mathcal{L}^{BRL}_{(i,j) \in \mathbf{U}}} \\
   \\ \nonumber
   &= \sum_{\Gamma} \int_{(\mathbf{X_{Ai}},\mathbf{X_{Bj}})} \Big[\text{log } \frac{f_M(\mathbf{X}_{Ai},\mathbf{X}_{Bj}|\mathbf{Z}_{Ai},\mathbf{Z}_{Bj},\beta_M)}{f_U(\mathbf{X}_{Ai},\mathbf{X}_{Bj}|\mathbf{Z}_{Ai},\mathbf{Z}_{Bj},\beta_U)} \times f_M(\mathbf{X}_{Ai},\mathbf{X}_{Bj}|\mathbf{Z}_{Ai},\mathbf{Z}_{Bj},\beta_M)  \text{ }  d(\mathbf{X}_{Ai},\mathbf{X}_{Bj}) \Big ] \times f_M(\Gamma(\mathbf{Z}_{Ai},\mathbf{Z}_{Bj})|\theta_M) \\
   \\ \nonumber
   &=\sum_{\Gamma} \mathbb{K} f_M(\Gamma(\mathbf{Z}_{Ai},\mathbf{Z}_{Bj})|\theta_M)\\
   \\ \nonumber
   &=\mathbb{K} \geq 0,
\end{align}
where $\mathbb{K}$ is the Kullback-Leibler divergence between the densities $f_M(\mathbf{X}_{Ai},\mathbf{X}_{Bj}|\mathbf{Z}_{Ai},\mathbf{Z}_{Bj},\beta_M)$ and $f_U(\mathbf{X}_{Ai},\mathbf{X}_{Bj}|\mathbf{Z}_{Ai},\mathbf{Z}_{Bj},\beta_U)$. \\

\vspace{0.5cm}

\noindent (b) Following Equations(\eqref{first}) - (eq\ref{last}) above, we now have 
\begin{align}
   &\E_{(\mathbf{\Gamma},\mathbf{X_A},\mathbf{X_B})} \text{log } \frac{\mathcal{L}^{BRLVOF}_{(i,j) \in \mathbf{M}}}{\mathcal{L}^{BRLVOF}_{(i,j) \in \mathbf{U}}} - \E_{(\mathbf{\Gamma},\mathbf{X_A},\mathbf{X_B})} \text{log } \frac{\mathcal{L}^{BRL}_{(i,j) \in \mathbf{M}}}{\mathcal{L}^{BRL}_{(i,j) \in \mathbf{U}}} \\
   \\ \nonumber
   &= \sum_{\Gamma} \int_{(\mathbf{X_{Ai}},\mathbf{X_{Bj}})} \Big [\text{log } \frac{f_M(\mathbf{X}_{Ai},\mathbf{X}_{Bj}|\mathbf{Z}_{Ai},\mathbf{Z}_{Bj},\beta_M)}{f_U(\mathbf{X}_{Ai},\mathbf{X}_{Bj}|\mathbf{Z}_{Ai},\mathbf{Z}_{Bj},\beta_U)} \times f_U(\mathbf{X}_{Ai},\mathbf{X}_{Bj}|\mathbf{Z}_{Ai},\mathbf{Z}_{Bj},\beta_U) \text{ } d(\mathbf{X}_{Ai},\mathbf{X}_{Bj}) \Big ]\times f_U(\Gamma(\mathbf{Z}_{Ai},\mathbf{Z}_{Bj})|\theta_U)\\
   \\ \nonumber
   &=\sum_{\Gamma} -\mathbb{K} f_U(\Gamma(\mathbf{Z}_{Ai},\mathbf{Z}_{Bj})|\theta_U)\\
   \\ \nonumber
   &=-\mathbb{K} \leq 0,
\end{align}
where $\mathbb{K}$ is the Kullback-Leibler divergence between the densities $f_M(\mathbf{X}_{Ai},\mathbf{X}_{Bj}|\mathbf{Z}_{Ai},\mathbf{Z}_{Bj},\beta_M)$ and $f_U(\mathbf{X}_{Ai},\mathbf{X}_{Bj}|\mathbf{Z}_{Ai},\mathbf{Z}_{Bj},\beta_U)$. 
\end{proof}

\subsection{Illustration with bivariate normal data}
We let $f_{M}(\mathbf{X}_{Ai},\mathbf{X}_{Bj}|\mathbf{Z}_{Ai},\mathbf{Z}_{Bj},\beta_M)$ and $f_{U}(\mathbf{X}_{Ai},\mathbf{X}_{Bj}|\mathbf{Z}_{Ai},\mathbf{Z}_{Bj},\beta_U)$ be bivariate normal distributions $N_2(\boldsymbol{\mu_{M}},\boldsymbol{\Sigma_{M}})$ and $N_2(\boldsymbol{\mu_{U}},\boldsymbol{\Sigma_{U}})$, respectively. For simplicity, we assume that $\boldsymbol{\mu_M}=\boldsymbol{\mu_U}$, $\Sigma_{M}= \bigl( \begin{smallmatrix}1 & \rho_M \\\ \rho_M  & 1\end{smallmatrix}\bigr)
$, and  $\Sigma_{U}= \bigl( \begin{smallmatrix}1 & \rho_U\\\ \rho_U  & 1\end{smallmatrix}\bigr)
$.  The KL divergence between $f_M$ and $f_U$ will be 
\begin{align}
    \frac{1-\rho_M\rho_U}{1-\rho_U^2} -\frac{1}{2}\text{ log }\frac{1-\rho_M^2}{1-\rho_U^2}-1.
\end{align}

\noindent The table below displays the KL-divergence for different values of $\rho_M$ and $\rho_U$.

\begin{table}[ht]
\centering
\caption {KL divergence between $f_M$ and $f_U$ with varying $\rho_M$ and $\rho_U$.} 
\resizebox{\textwidth}{!}{
{\LARGE
\begin{tabular}{c|cccccccccccccccccccc}
  \hline
 \diagbox[]{$\rho_M$}{$\rho_U$}& -0.95 & -0.85 & -0.75 & -0.65 & -0.55 & -0.45 & -0.35 & -0.25 & -0.15 & 0.05 & 0.15 & 0.25 & 0.35 & 0.45 & 0.55 & 0.65 & 0.75 & 0.85 & 0.95 \\ 
  \hline
-0.95 & \cellcolor{lightgray}0.00 & 0.22 & 0.41 & 0.55 & 0.67 & 0.77 & 0.86 & 0.95 & 1.03 & 1.21 & 1.32 & 1.45 & 1.62 & 1.84 & 2.17 & 2.69 & 3.66 & 6.04 & 18.51 \\ 
  -0.85 & 0.45 & \cellcolor{lightgray}0.00 & 0.06 & 0.14 & 0.22 & 0.30 & 0.38 & 0.45 & 0.52 & 0.68 & 0.78 & 0.90 & 1.05 & 1.26 & 1.56 & 2.05 & 2.97 & 5.21 & 17.02 \\ 
  -0.75 & 1.20 & 0.08 & \cellcolor{lightgray}0.00 & 0.03 & 0.08 & 0.13 & 0.19 & 0.25 & 0.31 & 0.45 & 0.54 & 0.65 & 0.79 & 0.98 & 1.26 & 1.71 & 2.57 & 4.67 & 15.81 \\ 
  -0.65 & 2.03 & 0.25 & 0.03 & \cellcolor{lightgray}0.00 & 0.02 & 0.05 & 0.09 & 0.14 & 0.19 & 0.31 & 0.39 & 0.48 & 0.61 & 0.78 & 1.04 & 1.46 & 2.26 & 4.23 & 14.70 \\ 
  -0.55 & 2.91 & 0.46 & 0.11 & 0.02 & \cellcolor{lightgray}0.00 & 0.01 & 0.04 & 0.07 & 0.11 & 0.21 & 0.28 & 0.36 & 0.47 & 0.63 & 0.87 & 1.26 & 2.00 & 3.83 & 13.63 \\ 
  -0.45 & 3.82 & 0.70 & 0.21 & 0.06 & 0.01 & \cellcolor{lightgray}0.00 & 0.01 & 0.03 & 0.06 & 0.14 & 0.19 & 0.27 & 0.37 & 0.51 & 0.72 & 1.08 & 1.76 & 3.45 & 12.59 \\ 
  -0.35 & 4.75 & 0.96 & 0.34 & 0.13 & 0.04 & 0.01 & \cellcolor{lightgray}0.00 & 0.01 & 0.02 & 0.08 & 0.13 & 0.19 & 0.28 & 0.40 & 0.59 & 0.92 & 1.54 & 3.10 & 11.57 \\ 
  -0.25 & 5.69 & 1.23 & 0.48 & 0.21 & 0.09 & 0.03 & 0.01 & \cellcolor{lightgray}0.00 & 0.01 & 0.05 & 0.08 & 0.13 & 0.21 & 0.31 & 0.48 & 0.77 & 1.33 & 2.76 & 10.56 \\ 
  -0.15 & 6.64 & 1.51 & 0.63 & 0.30 & 0.15 & 0.07 & 0.03 & 0.01 & \cellcolor{lightgray}0.00 & 0.02 & 0.05 & 0.09 & 0.15 & 0.24 & 0.38 & 0.64 & 1.14 & 2.43 & 9.57 \\ 
  0.05 & 8.58 & 2.12 & 0.96 & 0.51 & 0.29 & 0.17 & 0.10 & 0.05 & 0.02 & \cellcolor{lightgray}0.00 & 0.01 & 0.02 & 0.06 & 0.11 & 0.22 & 0.40 & 0.79 & 1.81 & 7.61 \\ 
  0.15 & 9.57 & 2.43 & 1.14 & 0.64 & 0.38 & 0.24 & 0.15 & 0.09 & 0.05 & 0.01 & \cellcolor{lightgray}0.00 & 0.01 & 0.03 & 0.07 & 0.15 & 0.30 & 0.63 & 1.51 & 6.64 \\ 
  0.25 & 10.56 & 2.76 & 1.33 & 0.77 & 0.48 & 0.31 & 0.21 & 0.13 & 0.08 & 0.02 & 0.01 & \cellcolor{lightgray}0.00 & 0.01 & 0.03 & 0.09 & 0.21 & 0.48 & 1.23 & 5.69 \\ 
  0.35 & 11.57 & 3.10 & 1.54 & 0.92 & 0.59 & 0.40 & 0.28 & 0.19 & 0.13 & 0.05 & 0.02 & 0.01 & \cellcolor{lightgray}0.00 & 0.01 & 0.04 & 0.13 & 0.34 & 0.96 & 4.75 \\ 
  0.45 & 12.59 & 3.45 & 1.76 & 1.08 & 0.72 & 0.51 & 0.37 & 0.27 & 0.19 & 0.09 & 0.06 & 0.03 & 0.01 & \cellcolor{lightgray}0.00 & 0.01 & 0.06 & 0.21 & 0.70 & 3.82 \\ 
  0.55 & 13.63 & 3.83 & 2.00 & 1.26 & 0.87 & 0.63 & 0.47 & 0.36 & 0.28 & 0.15 & 0.11 & 0.07 & 0.04 & 0.01 & \cellcolor{lightgray}0.00 & 0.02 & 0.11 & 0.46 & 2.91 \\ 
  0.65 & 14.70 & 4.23 & 2.26 & 1.46 & 1.04 & 0.78 & 0.61 & 0.48 & 0.39 & 0.24 & 0.19 & 0.14 & 0.09 & 0.05 & 0.02 & \cellcolor{lightgray}0.00 & 0.03 & 0.25 & 2.03 \\ 
  0.75 & 15.81 & 4.67 & 2.57 & 1.71 & 1.26 & 0.98 & 0.79 & 0.65 & 0.54 & 0.38 & 0.31 & 0.25 & 0.19 & 0.13 & 0.08 & 0.03 & \cellcolor{lightgray}0.00 & 0.08 & 1.20 \\ 
  0.85 & 17.02 & 5.21 & 2.97 & 2.05 & 1.56 & 1.26 & 1.05 & 0.90 & 0.78 & 0.60 & 0.52 & 0.45 & 0.38 & 0.30 & 0.22 & 0.14 & 0.06 & \cellcolor{lightgray}0.00 & 0.45 \\ 
  0.95 & 18.51 & 6.04 & 3.66 & 2.69 & 2.17 & 1.84 & 1.62 & 1.45 & 1.32 & 1.12 & 1.03 & 0.95 & 0.86 & 0.77 & 0.67 & 0.55 & 0.41 & 0.22 & \cellcolor{lightgray}0.00 \\ \hline
\end{tabular}
}}
\end{table}

\section{Accounting for erroneous blocking variables}
Suppose that blocks are determined by $L$ faulty blocking variables, $\mathbf{\hat V}_A = \{\hat V_{Ai1},\dots,\hat V_{AiL}; i \in \mathbf{A}\}$ and $\mathbf{\hat V}_B = \{\hat V_{Bj1},\dots,\hat V_{BjL}; j \in \mathbf{B}\}$ in files $\mathbf{A}$ and $\mathbf{B}$, respectively. Let $\mathbf{ V}_A$ and $\mathbf{V}_B$ denote the true, latent, values of these blocking variables. A simplifying assumption is that $\mathbf{\hat V}_A$ is accurately recorded, and errors only occur in reporting $\mathbf{V}_B$. Thus, $\mathbf{\hat V}_A = \mathbf{ V}_A$. Define $\mathbf{E} = \{E_{jl}; j=1,\dots,n_B \text{ and } l=1,\dots,L\}$, where 
\begin{align}
    E_{jl} = \begin{cases}
      1, & \text{if}\ \hat V_{Bjl} \neq V_{Bjl} \\
      0, & \text{otherwise.}
    \end{cases}
\end{align}
The BRLVOF likelihood can be written as
\begin{align} \label{fblock}
    \mathcal{L}^{BRLVOF_{Block}} = f_1(\mathbf{X}_{A},\mathbf{X}_{B},\mathbf{Z}_{A},\mathbf{Z}_{B}| \mathbf{\hat V}_{B}, \mathbf{V}_{A},\mathbf{V}_{B}, \mathbf{E}) \times f_2(\mathbf{\hat V}_{B}|\mathbf{V}_{A},\mathbf{V}_{B},\mathbf{E}) \times f_3(\mathbf{E}|\mathbf{V}_{A},\mathbf{V}_{B}) \times f_4(\mathbf{V}_{A},\mathbf{V}_{B}).
\end{align}
We suppress the dependence of the likelihood on parameters for notational simplicity. Further,
\begin{align} \label{fblock}
    f_1(\mathbf{X}_{A},\mathbf{X}_{B},\mathbf{Z}_{A},\mathbf{Z}_{B}| \mathbf{\hat V}_{B}, \mathbf{V}_{A},\mathbf{V}_{B}, \mathbf{E}) = 
    \prod_{i=1}^{n_A} \prod_{j=1}^{n_B} &\bigg[f_M(\mathbf{X}_{Ai},\mathbf{X}_{Bj}|\mathbf{Z}_{Ai},\mathbf{Z}_{Bj},\beta_M) \prod_{k=1}^{K} \prod_{l_k=1}^{L_k} \theta_{Mkl_{k}}^{\mathbbm{1}(\gamma_{ijk}=l_k)} \bigg]^{C_{ij}\prod\limits_{l=1}^{L}\mathbb{I}(V_{Ail}=V_{Bjl})} \nonumber \\
     \times &\bigg[f_U(\mathbf{X}_{Ai},\mathbf{X}_{Bj}|\mathbf{Z}_{Ai},\mathbf{Z}_{Bj},\beta_U) \prod_{k=1}^{K} \prod_{l_k=1}^{L_k} \theta_{Ukl_{k}}^{\mathbbm{1}(\gamma_{ijk}=l_k)} \bigg]^{(1-C_{ij})\prod\limits_{l=1}^{L}\mathbb{I}(V_{Ail}=V_{Bjl})}.
\end{align}
The component $f_2(\mathbf{\hat V}_{B}|\mathbf{V}_{A},\mathbf{V}_{B},\mathbf{E})$ can be any measurement error model, such as Equation (5) in \cite{Dalzell2018}. Further, $f_3(\mathbf{E}|\mathbf{V}_{A},\mathbf{V}_{B}) = f_3(\mathbf{E}|\mathbf{V}_{B})$ can be a Bernoulli distribution, or a logistic regression that relates $\mathbf{E}$ to $\mathbf{V}_{B}$. Finally, $f_4(\mathbf{V}_{A},\mathbf{V}_{B})$ can be defined in multiple ways, for example, using latent class models as in \cite{Dalzell2018}, or a multinomial distribution, as in \cite{steorts16}. Posterior sampling would iterate between sampling $\mathbf{V}_B$, $\mathbf{E}$, and other model parameters. Prior distributions for the model parameters should be considered carefully, as the latent structures may induce weakly identifiable likelihoods.

\vspace{0.5cm}

\section{Updating $\mathbf{C}$: Metropolis Hastings algorithm}
We describe the Metropolis Hastings updates to the linking configuration $\mathbf{C}$ and the number of linked records $n_m$. Each Metropolis Hastings update will have an acceptance probability in the form
\begin{equation}
    A=\min\bigg(1,\frac{P(\mathbf{C}^{*},n_m^{*}|\mathbf{X}_A,\mathbf{X}_B,\mathbf{Z}_A,\mathbf{Z}_B,\boldsymbol{\theta}_M,\boldsymbol{\theta}_U,\boldsymbol{\beta}_M,\boldsymbol{\beta}_U)J(\mathbf{C},n_m|\mathbf{C}^{*},n_m^{*})}{P(\mathbf{C},n_m|\mathbf{X}_A,\mathbf{X}_B,\mathbf{Z}_A,\mathbf{Z}_B,\boldsymbol{\theta}_M,\boldsymbol{\theta}_U,\boldsymbol{\beta}_M,\boldsymbol{\beta}_U)J(\mathbf{C}^{*},n_m^{*}|\mathbf{C},n_m)}\bigg).
\end{equation}
$J(\mathbf{C}^{*},n_m^{*}|\mathbf{C},n_m)$ represents the transition probability for the proposed move and $J(\mathbf{C},n_m|\mathbf{C}^{*},n_m^{*})$ is the transition probability for the reverse move. It is useful to note that the posterior distribution of $(\mathbf{C},n_m)$ can be factorized into
\begin{align}
&P(\mathbf{C},n_m|\mathbf{X}_A,\mathbf{X}_B,\mathbf{Z}_A,\mathbf{Z}_B,\boldsymbol{\theta}_M,\boldsymbol{\theta}_U,\boldsymbol{\beta}_M,\boldsymbol{\beta}_U)\propto
P(\mathbf{C},n_m)\text{ }\mathcal{L}(\mathbf{C},n_m|\mathbf{X}_A,\mathbf{X}_B,\mathbf{Z}_A,\mathbf{Z}_B,\boldsymbol{\theta}_M,\boldsymbol{\theta}_U,\boldsymbol{\beta}_M,\boldsymbol{\beta}_U).
\end{align}
The following sections provide more detail about the form of the acceptance probability for the three types of updates.

\subsection{Unlinked Record Update Type 1}
The first update we consider is when record $i \in \mathbf{A}$ is not linked with any record from $\mathbf{B}$ at iteration $[t]$. We obtain an update of $\mathbf{C}$ by proposing a record $j \in \mathbf{B}$ that is not linked with any record in $\mathbf{A}$ at iteration $[t]$ to form the true link $C_{ij}^{[t+1]}=1$, which results in an increase in the number of linked records $n_m^{[t+1]}=n_m^{[t]}+1$. If record $j$ is selected with equal probability among the unlinked records in $\mathbf{B}$, the transition probability for adding record pair $(i,j)$ to $\mathbf{M}$ is $J(\mathbf{C}^{[t+1]},n_m^{[t+1]}|\mathbf{C}^{[t]},n_m^{[t]})=(n_B-n_m^{[t]})^{-1}$. The reverse move would assign $(i,j)$ to $\mathbf{U}$, which will have transition probability $J(\mathbf{C}^{[t]},n_m^{[t]}|\mathbf{C}^{[t+1]},n_m^{[t+1]})=(n_m^{[t]}+1)^{-1}$. 

The ratio of prior distributions will be
\begin{align}
\begin{split}
    \dfrac{p(\mathbf{C}^{[t+1]},n_m^{[t+1]})}{p(\mathbf{C}^{[t]},n_m^{[t]})}=&\dfrac{\dfrac{(\max(n_{A},n_{B})-(n_{m}^{[t]}+1))!}{\max(n_{A},n_{B})!} \dfrac{\Gamma(\alpha_{\pi}+\beta_{\pi})}{\Gamma(\alpha_{\pi})\Gamma(\beta_{\pi})}\dfrac{\Gamma(n_{m}^{[t]}+1+\alpha_{\pi})\Gamma(\min(n_{A},n_{B})-(n_{m}^{[t]}+1)+\beta_{\pi})} {\Gamma(\min(n_{A},n_{B})+\alpha_{\pi}+\beta_{\pi})}}{\dfrac{(\max(n_{A},n_{B})-n_{m}^{[t]})!}{\max(n_{A},n_{B})!} \dfrac{\Gamma(\alpha_{\pi}+\beta_{\pi})}{\Gamma(\alpha_{\pi})\Gamma(\beta_{\pi})}\dfrac{\Gamma(n_{m}^{t]}+\alpha_{\pi})\Gamma(\min(n_{A},n_{B})-n_{m}^{[t]}+\beta_{\pi})} {\Gamma(\min(n_{A},n_{B})+\alpha_{\pi}+\beta_{\pi})}}\\
    =&\dfrac{1}{\max(n_A,n_B)-n_m^{[t]}} \times \dfrac{n_m^{[t]}+\alpha_{\pi}}{\min(n_A,n_B)-n_m^{[t]}+\beta_{\pi}-1}.
\end{split}
\end{align}
The ratio of likelihoods will be reduced to the ratio of the true link versus non-link likelihoods for record pair $(i,j)$:
\begin{align}
    &\dfrac{\mathcal{L}(\mathbf{C}^{[t+1]},n_m^{[t+1]}|\mathbf{X}_A,\mathbf{X}_B,\mathbf{Z}_A,\mathbf{Z}_B,\boldsymbol{\theta}_M^{[t+1]},\boldsymbol{\theta}_U^{[t+1]},\boldsymbol{\beta}_M^{[t+1]},\boldsymbol{\beta}_U^{[t+1]})}{\mathcal{L}(\mathbf{C}^{[t]},n_m^{[t]}|\mathbf{X}_A,\mathbf{X}_B,\mathbf{Z}_A,\mathbf{Z}_B,\boldsymbol{\theta}_M^{[t+1]},\boldsymbol{\theta}_U^{[t+1]},\boldsymbol{\beta}_M^{[t+1]},\boldsymbol{\beta}_U^{[t+1]})}
    =\dfrac{f_M(\mathbf{X}_{Ai},\mathbf{X}_{Bj}|\mathbf{Z}_{Ai},\mathbf{Z}_{Bj},\boldsymbol{\beta}_M^{[t+1]})f_M(\mathbf{Z}_{Ai},\mathbf{Z}_{Bj}|\boldsymbol{\theta}_M^{[t+1]})}{f_U(\mathbf{X}_{Ai},\mathbf{X}_{Bj}|\mathbf{Z}_{Ai},\mathbf{Z}_{Bj},\boldsymbol{\beta}_U^{[t+1]})f_U(\mathbf{Z}_{Ai},\mathbf{Z}_{Bj}|\boldsymbol{\theta}_U^{[t+1]})}.
\end{align}

Thus, the acceptance probability for this Metropolis Hastings update is therefore the minimum between 1 and 
\begin{align}
\begin{split}
   \dfrac{n_B-n_m^{[t]} }{n_m^{[t]}+1} &\times\dfrac{1}{\max(n_A,n_B)-n_m^{[t]}} \times \dfrac{n_m^{[t]}+\alpha_{\pi}}{\min(n_A,n_B)-n_m^{[t]}+\beta_{\pi}-1} \\\\
    &\times \dfrac{f_M(\mathbf{X}_{Ai},\mathbf{X}_{Bj}|\mathbf{Z}_{Ai},\mathbf{Z}_{Bj},\boldsymbol{\beta}_M^{[t+1]})f_M(\mathbf{Z}_{Ai},\mathbf{Z}_{Bj}|\boldsymbol{\theta}_M^{[t+1]})}{f_U(\mathbf{X}_{Ai},\mathbf{X}_{Bj}|\mathbf{Z}_{Ai},\mathbf{Z}_{Bj},\boldsymbol{\beta}_U^{[t+1]})f_U(\mathbf{Z}_{Ai},\mathbf{Z}_{Bj}|\boldsymbol{\theta}_U^{[t+1]})}.  
\end{split}
\end{align}

\subsection{Linked Record Update Type 1}
The second move to consider is when record $i \in \mathbf{A}$ is linked with record $j \in \mathbf{B}$ at iteration $[t]$. One possible update to $\mathbf{C}$ is to unlink the record pair such that $C_{ij}^{[t+1]}=0$, which results in a decrease in the number of links $n_m^{[t+1]}=n_m^{[t]}-1$. The transition probability for removing record pair $(i,j)$ from the set of true links is $J(\mathbf{C}^{[t+1]},n_m^{[t+1]}|\mathbf{C}^{[t]},n_m^{[t]})=(n_m^{[t]})^{-1}$. The reverse move selects an unlinked record $j \in \mathbf{B}$ with equal probability to form a link with $i \in \mathbf{A}$, which will have transition probability equal to $J(\mathbf{C}^{[t]},n_m^{[t]}|\mathbf{C}^{[t+1]},n_m^{[t+1]})=(n_B-n_m^{[t]}+1)^{-1}$.

The ratio of prior distributions will be
\begin{align}
\begin{split}
\dfrac{p(\mathbf{C}^{[t+1]},n_m^{[t+1]})}{p(\mathbf{C}^{[t]},n_m^{[t]})}=&\dfrac{\dfrac{(\max(n_{A},n_{B})-(n_{m}^{[t]}-1))!}{\max(n_{A},n_{B})!} \dfrac{\Gamma(\alpha_{\pi}+\beta_{\pi})}{\Gamma(\alpha_{\pi})\Gamma(\beta_{\pi})}\dfrac{\Gamma(n_{m}^{[t]}-1+\alpha_{\pi})\Gamma(\min(n_{A},n_{B})-(n_{m}^{[t]}-1)+\beta_{\pi})} {\Gamma(\min(n_{A},n_{B})+\alpha_{\pi}+\beta_{\pi})}}{\dfrac{(\max(n_{A},n_{B})-n_{m}^{[t]})!}{\max(n_{A},n_{B})!} \dfrac{\Gamma(\alpha_{\pi}+\beta_{\pi})}{\Gamma(\alpha_{\pi})\Gamma(\beta_{\pi})}\dfrac{\Gamma(n_{m}^{t]}+\alpha_{\pi})\Gamma(\min(n_{A},n_{B})-n_{m}^{[t]}+\beta_{\pi})} {\Gamma(\min(n_{A},n_{B})+\alpha_{\pi}+\beta_{\pi})}}\\= &\dfrac{\max(n_A,n_B)-n_m^{[t]}+1}{1} \times \dfrac{\min(n_A,n_B)-n_m^{[t]}+\beta_{\pi})}{n_m^{[t]}+\alpha_{\pi}-1}.    
\end{split}
\end{align}
The ratio of likelihoods will be reduced to the ratio of the non-link versus true link likelihoods for record pair $(i,j)$:
\begin{align}
\begin{split}
     &\dfrac{\mathcal{L}(\mathbf{C}^{[t+1]},n_m^{[t+1]}|\mathbf{X}_A,\mathbf{X}_B,\mathbf{Z}_A,\mathbf{Z}_B,\boldsymbol{\theta}_M^{[t+1]},\boldsymbol{\theta}_U^{[t+1]},\boldsymbol{\beta}_M^{[t+1]},\boldsymbol{\beta}_U^{[t+1]})}{\mathcal{L}(\mathbf{C}^{[t]},n_m^{[t]}|\mathbf{X}_A,\mathbf{X}_B,\mathbf{Z}_A,\mathbf{Z}_B,\boldsymbol{\theta}_M^{[t+1]},\boldsymbol{\theta}_U^{[t+1]},\boldsymbol{\beta}_M^{[t+1]},\boldsymbol{\beta}_U^{[t+1]})}
    =\dfrac{f_U(\mathbf{X}_{Ai},\mathbf{X}_{Bj}|\mathbf{Z}_{Ai},\mathbf{Z}_{Bj},\boldsymbol{\beta}_U^{[t+1]})f_U(\mathbf{Z}_{Ai},\mathbf{Z}_{Bj}|\boldsymbol{\theta}_U^{[t+1]})}{f_M(\mathbf{X}_{Ai},\mathbf{X}_{Bj}|\mathbf{Z}_{Ai},\mathbf{Z}_{Bj},\boldsymbol{\beta}_M^{[t+1]})f_M(\mathbf{Z}_{Ai},\mathbf{Z}_{Bj}|\boldsymbol{\theta}_M^{[t+1]})}.
\end{split}
\end{align}
The Metropolis Hastings acceptance probability for this update is the minimum between 1 and
\begin{align}
\begin{split}
     \dfrac{n_m^{[t]}}{n_B-n_m^{[t]}+1} &\times \dfrac{\max(n_A,n_B)-n_m^{[t]}+1}{1} \times \dfrac{\min(n_A,n_B)-n_m^{[t]}+\beta_{\pi})}{n_m^{[t]}+\alpha_{\pi}-1}\\
    & \times \dfrac{f_U(\mathbf{X}_{Ai},\mathbf{X}_{Bj}|\mathbf{Z}_{Ai},\mathbf{Z}_{Bj},\boldsymbol{\beta}_U^{[t+1]})f_U(\mathbf{Z}_{Ai},\mathbf{Z}_{Bj}|\boldsymbol{\theta}_U^{[t+1]})}{f_M(\mathbf{X}_{Ai},\mathbf{X}_{Bj}|\mathbf{Z}_{Ai},\mathbf{Z}_{Bj},\boldsymbol{\beta}_M^{[t+1]})f_M(\mathbf{Z}_{Ai},\mathbf{Z}_{Bj}|\boldsymbol{\theta}_M^{[t+1]})}.
\end{split}
\end{align}

\subsection{Linked Record Update Type 2}
An alternative update to $\mathbf{C}$ when record $i \in \mathbf{A}$ is linked with record $j \in \mathbf{B}$ at iteration $[t]$ is to swap linkage with another true link pair, $(r,s):r \in \mathbf{A}, s \in \mathbf{B}$. This update will unlink records $(i,j)$ and $(r,s)$ and will form the new true links $(i,s):i \in \mathbf{A}, s \in \mathbf{B}, C_{ir}^{[t+1]}=1$ and $(r,j):r \in \mathbf{A}, j \in \mathbf{B}, C_{rj}^{[t+1]}=1$. The total number of linked records does not change with this update. The transition probability of selecting $(r,s)$ with equal likelihood among the true links at iteration $[t]$ is equal to $J(\mathbf{C}^{[t+1]},n_m^{[t+1]}|\mathbf{C}^{[t]},n_m^{[t]})=(n_m^{[t]}-1)^{-1}$. The reverse move would involve selecting record pair $(r,j)$ to swap linkage with $(i,s)$, which will have transition probability equal to $J(\mathbf{C}^{[t]},n_m^{[t]}|\mathbf{C}^{[t+1]},n_m^{[t+1]})=(n_m^{[t]}-1)^{-1}$. Because $n_m^{[t+1]}=n_m^{[t]}$ for this update, the ratio of prior distributions will be equal to 1. The Metropolis Hastings acceptance probability for swapping the linkage of $(i,j)$ and $(k,l)$ will only depend on the ratio of likelihoods, and will be equal to the minimum of 1 and
\begin{align}
\begin{split}
    &\dfrac{f_U(\mathbf{X}_{Ai},\mathbf{X}_{Bj}|\mathbf{Z}_{Ai},\mathbf{Z}_{Bj},\boldsymbol{\beta}_U^{[t+1]})f_U(\mathbf{Z}_{Ai},\mathbf{Z}_{Bj}|\boldsymbol{\theta}_U^{[t+1]})}{f_M(\mathbf{X}_{Ai},\mathbf{X}_{Bj}|\mathbf{Z}_{Ai},\mathbf{Z}_{Bj},\boldsymbol{\beta}_M^{[t+1]})f_M(\mathbf{Z}_{Ai},\mathbf{Z}_{Bj}|\boldsymbol{\theta}_M^{[t+1]})}\dfrac{f_U(\mathbf{X}_{Ar},\mathbf{Y}_{Bs}|\mathbf{Z}_{Ar},\mathbf{Z}_{Bs},\boldsymbol{\beta}_U^{[t+1]})f_U(\mathbf{Z}_{Ar},\mathbf{Z}_{Bs}|\boldsymbol{\theta}_U^{[t+1]})}{f_M(\mathbf{X}_{Ar},\mathbf{Y}_{Bs}|\mathbf{Z}_{Ar},\mathbf{Z}_{Bs},\boldsymbol{\beta}_M^{[t+1]})f_M(\mathbf{Z}_{Ar},\mathbf{Z}_{Bs}|\boldsymbol{\theta}_M^{[t+1]})} \\
    &\times \dfrac{f_M(\mathbf{X}_{Ai},\mathbf{Y}_{Bs}|\mathbf{Z}_{Ai},\mathbf{Z}_{Bs},\boldsymbol{\beta}_M^{[t+1]})f_M(\mathbf{Z}_{Ai},\mathbf{Z}_{Bs}|\boldsymbol{\theta}_M^{[t+1]})}{f_U(\mathbf{X}_{Ai},\mathbf{Y}_{Bs}|\mathbf{Z}_{Ai},\mathbf{Z}_{Bs},\boldsymbol{\beta}_U^{[t+1]})f_U(\mathbf{Z}_{Ai},\mathbf{Z}_{Bs}|\boldsymbol{\theta}_U^{[t+1]})}\dfrac{f_M(\mathbf{X}_{Ar},\mathbf{X}_{Bj}|\mathbf{Z}_{Ar},\mathbf{Z}_{Bj},\boldsymbol{\beta}_M^{[t+1]})f_M(\mathbf{Z}_{Ar},\mathbf{Z}_{Bj}|\boldsymbol{\theta}_M^{[t+1]})}{f_U(\mathbf{X}_{Ar},\mathbf{X}_{Bj}|\mathbf{Z}_{Ar},\mathbf{Z}_{Bj},\boldsymbol{\beta}_U^{[t+1]})f_U(\mathbf{Z}_{Ar},\mathbf{Z}_{Bj}|\boldsymbol{\theta}_U^{[t+1]})}.  
\end{split}
\end{align}

\vspace{0.5cm}

\section{Updating C: Adaptive multinomial method}
Following the Gibbs sampling algorithm proposed in \textcite{Sadinle2017}, one can update the configuration of $\mathbf{C}$ by proposing new link designations for each record $i \in \mathbf{A}$. At each iteration, there are two sets of options: record $i \in \mathbf{A}$ forms a link with an unlinked record $j \in \mathbf{B}$, or record $i$ is not linked with any record in $\mathbf{B}$. Let $\mathbf{C}_{-i,*}^{[t+1]} = (\mathbf{C}_{1,*}^{[t+1]},\dots,\mathbf{C}_{i-1,*}^{[t+1]},\mathbf{C}_{i+1,*}^{[t]},\dots,\mathbf{C}_{n_A,*}^{[t]})^T$ represent the linkage structure excluding the designations for $i \in \mathbf{A}$ that is about to be sampled. Let $n_{m(-i)}=\sum_{i=1}^{n_A} \sum_{j=1}^{n_B} \mathbf{C}_{-i,*}^{[t+1]}$ be the number of true links excluding the designations belonging to record $i \in \mathbf{A}$. 

The posterior probability for record $i \in \mathbf{A}$ to link with any record $j \in \mathbf{B}$ is
\begin{align}
\begin{split}
    &P(C_{ij}^{[t+1]}=1,n_m^{[t+1]}=n_{m(-i)}^{[t+1]}+1|\mathbf{X}_A, \mathbf{X}_B, \mathbf{Z}_A,\mathbf{Z}_B, \boldsymbol{\theta}_M, \boldsymbol{\theta}_U, \boldsymbol{\beta}_M, \boldsymbol{\beta}_U) \propto\\
&p(\mathbf{C},n_m^{[t+1]}=n_{m(-i)}^{[t+1]}+1) f_M(\mathbf{X}_{Ai},\mathbf{X}_{Bj}|\mathbf{Z}_{Ai},\mathbf{Z}_{Bj},\boldsymbol{\beta}_M^{[t+1]}) f_M(\mathbf{Z}_{Ai},\mathbf{Z}_{Bj}|\boldsymbol{\theta}_M^{[t+1]})\mathbbm{1}(C_{*,j}=0) \times \\
&
\prod\limits_{j' \neq j}f_U(\mathbf{X}_{Ai},\mathbf{Y}_{Bj'}|\mathbf{Z}_{Ai},\mathbf{Z}_{Bj'},\boldsymbol{\beta}_U^{[t+1]})f_U(\mathbf{Z}_{Ai},\mathbf{Z}_{Bj'}|\boldsymbol{\theta}_U^{[t+1]}),
\end{split}
\end{align}
where $\mathbbm{I}(C_{*,j}=0)$ is an indicator that $j \in \mathbf{B}$ is not linked with any record in $\mathbf{A}$. The posterior probability of record $i \in \mathbf{A}$ not linking with any record in $\mathbf{B}$ is
\begin{align}
\begin{split}
 &P(\mathbf{C}_{i,*}^{[t+1]}=0,n_m^{[t+1]}=n_{m(-i)}^{[t+1]}|\mathbf{X}_A, \mathbf{X}_B, \mathbf{Z}_A,\mathbf{Z}_B,\boldsymbol{\theta}_M^{[t+1]},\boldsymbol{\theta}_U^{[t+1]},\boldsymbol{\beta}_M^{[t+1]},\boldsymbol{\beta}_U^{[t+1]}) \propto\\
&p(\mathbf{C},n_m^{[t+1]}=n_{m(-i)}^{[t+1]}) \prod_{j=1}^{n_B} f_U(\mathbf{X}_{Ai},\mathbf{X}_{Bj}|\mathbf{Z}_{Ai},\mathbf{Z}_{Bj},\boldsymbol{\beta}_U^{[t+1]})f_U(\mathbf{Z}_{Ai},\mathbf{Z}_{Bj}|\boldsymbol{\theta}_U^{[t+1]}).   
\end{split}
\end{align}

Updating the link designation for record $i \in \mathbf{A}$ is equivalent to sampling from a multinomial distribution containing the unlinked records in $\mathbf{B}$ and a non-link option. Upon rearranging the terms in the posterior probabilities above and marginalizing over $n_m$, the probability of record $i \in \mathbf{A}$ pairing with the unlinked record $j \in \mathbf{B}$ is
\begin{align}
\begin{split}
  &P(C^{[t+]}_{ij}=1|\mathbf{X}_A,\mathbf{X}_B,\mathbf{Z}_A,\mathbf{Z}_B,\boldsymbol{\theta}_M^{[t+1]},\boldsymbol{\theta}_U^{[t+1]},\boldsymbol{\beta}_M^{[t+1]},\boldsymbol{\beta}_U^{[t+1]})=\\
    &\dfrac{
    \dfrac{f_M(\mathbf{X}_{Ai},\mathbf{X}_{Bj}|\mathbf{Z}_{Ai},\mathbf{Z}_{Bj},\boldsymbol{\beta}_M^{[t+1]}) f_U(\mathbf{Z}_{Ai},\mathbf{Z}_{Bj}|\boldsymbol{\theta}_M^{[t+1]})}{f_U(\mathbf{X}_{Ai},\mathbf{X}_{Bj}|\mathbf{Z}_{Ai},\mathbf{Z}_{Bj},\boldsymbol{\beta}_U^{[t+1]})f_U(\mathbf{Z}_{Ai},\mathbf{Z}_{Bj}|\boldsymbol{\theta}_U^{[t+1]})}\mathbbm{I}(C_{*,j}=0)
    }{\sum\limits_{j'=1}^{n_B} \dfrac{f_M(\mathbf{X}_{Ai},\mathbf{X}_{Bj}|\mathbf{Z}_{Ai},\mathbf{Z}_{Bj},\boldsymbol{\beta}_M^{[t+1]}) f_M(\mathbf{Z}_{Ai},\mathbf{Z}_{Bj}|\boldsymbol{\theta}_M^{[t+1]})}{f_U(\mathbf{X}_{Ai},\mathbf{X}_{Bj}|\mathbf{Z}_{Ai},\mathbf{Z}_{Bj},\boldsymbol{\beta}_U^{[t+1]})f_U(\mathbf{Z}_{Ai},\mathbf{Z}_{Bj}|\boldsymbol{\theta}_U^{[t+1]})}\mathbbm{I}(C_{*,j}=0)+\dfrac{(n_B-n_{m(-i)}^{[t+1]})(n_A-n_{m(-i)}^{[t+1]}+\beta_{\pi}-1)}{n_{m(-i)}^{[t+1]}+\alpha_{\pi}}},   
\end{split}
\end{align}
and the probability for record $i$ to not link with any record from $\mathbf{B}$ is
\begin{align}
\begin{split}
      &P(C^{[t+]}_{ij}=0|\mathbf{X}_A,\mathbf{X}_B,\mathbf{Z}_A,\mathbf{Z}_B,\boldsymbol{\theta}_M^{[t+1]},\boldsymbol{\theta}_U^{[t+1]},\boldsymbol{\beta}_M^{[t+1]},\boldsymbol{\beta}_U^{[t+1]})=\\
    &\dfrac{
    \dfrac{(n_B-n_{m(-i)}^{[t+1]})(n_A-n_{m(-i)}^{[t+1]}+\beta_{\pi}-1)}{n_{m(-i)}^{[t+1]}+\alpha_{\pi}}
    }{\sum\limits_{j'=1}^{n_B} \dfrac{f_M(\mathbf{X}_{Ai},\mathbf{X}_{Bj}|\mathbf{Z}_{Ai},\mathbf{Z}_{Bj},\boldsymbol{\beta}_M^{[t+1]}) f_M(\mathbf{Z}_{Ai},\mathbf{Z}_{Bj}|\boldsymbol{\theta}_M^{[t+1]})}{f_U(\mathbf{X}_{Ai},\mathbf{X}_{Bj}|\mathbf{Z}_{Ai},\mathbf{Z}_{Bj},\boldsymbol{\beta}_U^{[t+1]})f_U(\mathbf{Z}_{Ai},\mathbf{Z}_{Bj}|\boldsymbol{\theta}_U^{[t+1]})}\mathbbm{I}(C_{*,j}=0)+\dfrac{(n_B-n_{m(-i)}^{[t+1]})(n_A-n_{m(-i)}^{[t+1]}+\beta_{\pi}-1)}{n_{m(-i)}^{[t+1]}+\alpha_{\pi}}}. 
\end{split}
\end{align}

\vspace{0.7cm}
\section{Additional simulation results} \label{apsim}

\subsection{Results under BRLVOF with $\sigma=0.5$}
Table \ref{tablin} depicts results for $\rho$, when the association model between $\mathbf{X_A}$ and $\mathbf{X_B}$ is linear. Table \ref{tablinW} displays results when this relationship includes $W$. Table \ref{tabnonlin} displays results when the relationship between $\mathbf{X}_A$ and $\mathbf{X}_B$ includes non linear terms. In all three cases, the value of $\sigma=0.5$.
\FloatBarrier 
\begin{table}[t]
\caption {Results for estimation of $\rho$ under BRLVOF and BRL, when conditional models for $\mathbf{X}_A$ given $\mathbf{X}_B$ are linear, and $\sigma=0.5$. Values in parentheses represent standard deviations of the corresponding quantities.} \label{tablin}
\centering
\resizebox{\textwidth}{!}{
{\LARGE
\begin{tabular}{cccc|ccccccc}
$\epsilon$ & Method& P & $\beta_M$ & $n$ & $\overline{TPR}$ & 
$\overline{PPV}$ & $\overline{F1}$ & $\overline{Bias}$ & $RMSE$&Coverage \\ \hline
\multirow{10}{*}{0.0} & BRL& &  &301 (0.17) &.9984(.0016) &.8845(.0418) &.9200(.0240) & .058(.023) & 0.063 & 0.90 \\ \cline{2-11} 
 & \multirow{15}{*}{BRLVOF} &\multirow{3}{*}{1} & 0.05 &300 (0.27)& .9990(.0012) & .9993(.0006) & .9991(.0009) & .001(.001) & 0.002 & 1.00 \\
 &&  & 0.1 &302 (0.27) & .9902(.0027) & .9900(.0015) & .9901(.0020) & .009(.003) & 0.011 & 1.00\\
 &&& 0.2   &  302 (1.01) & { .9991(.0012)} & .9994(.0008) & .9992(.0009) & .001(.001) & 0.001&1.00 \\
 &&  & 0.5 &  307 (1.21) & { .9993(.0070)} & .9993(.0130) & .9993(.0101) & .001(.030) & 0.004&0.97 \\
 &&  & 1.0 &  306 (0.43) & .9937(.0456) & .9927(.0444) & .9933(.0447) & .010(.076) & 0.013&0.97\\ \cline{3-11} 
 && \multirow{3}{*}{2} & 0.05 &300 (0.27)& .9991(.0011) & .9994(.0005) & .9992(.0008) & .001(.001) & 0.002 & 1.00 \\
 &&  & 0.1 &300 (0.27)& .9991(.0011) & .9994(.0005) & .9993(.0007) & .001(.001) & 0.002 & 1.00 \\
 &&& 0.2   &  300 (0.27)& .9992(.0011) & .9995(.0006) & .9993(.0008) & .001(.002) & 0.002&1.00 \\
 &&  & 0.5 &  318 (0.25)& .9993(.0010) & .9994(.0006) & .9993(.0007) & .000(.001) & 0.001&0.91 \\
 &&  & 1.0 &  308 (0.59)& .9995(.0008) & .9979(.0021) & .9987(.0012) & .001(.002) & 0.003&0.96 \\ \cline{3-11} 
 && \multirow{3}{*}{4} & 0.05 &308 (0.52)& .9566(.0058) & .9565(.0050) & .9994(.0008) & .020(.006) & 0.025 & 1.00 \\
 &&  & 0.1 &333 (1.29) & .8398(.0017) & .8399(.0008) & .9993(.0008) & .092(.002) & 0.101 & 0.99 \\
 &&& 0.2   &  330 (1.35) & .9991(.0123) & .9991(.0172) & .9991(.0147) & .001(.013) & 0.003&0.95 \\
 &&  & 0.5 &  320 (0.71) & .9996(.0009) & .9949(.0009) & .9967(.0007) & .003(.004) & 0.005&0.94 \\
 &&  & 1.0 &  312 (2.02) & .9997(.0008) & .9971(.0026) & .9984(.0014) & .003(.004) & 0.005&0.94\\ \hline
\multirow{10}{*}{0.2} &BRL&  &  & 288 (2.47)& .7780(.0151) &.7168(.0587)& .7424(.0326) & .248(.042) & 0.251&0.68 \\ \cline{2-11} 
 & \multirow{15}{*}{BRLVOF}& \multirow{3}{*}{1} & 0.05 &265 (4.36)& .8368(.0111) & .9476(.0120) & .8887(.0090) & .022(.011) & 0.026 & 1.00 \\
 &&  & 0.1    &267 (5.49) & .8317(.0146) & .9411(.0157) & .8893(.0096) & .027(.015) & 0.032 & 1.00 \\
 &&& 0.2      &  263 (4.26) & .8388(.0110) & .9561(.0109) & .8935(.0082) & .015(.008) & 0.017&1.00 \\
 &&  & 0.5    &  263 (4.01) & .8366(.0113) & .9629(.0087) & .8952(.0076) & .020(.038) & 0.021&0.99 \\
 &&  & 1.0    &  261 (3.90) & .8539(.0111) & .9800(.0074) & .9124(.0071) & .001(.001) & 0.002&0.99 \\ \cline{3-11} 
 && \multirow{3}{*}{2} & 0.05 &265 (4.39)& .8364(.0111) & .9475(.0121) & .8884(.0090) & .025(.011) & 0.002 & 1.00 \\
 &&  & 0.1 &266 (5.36) & .8325(.0150) & .9451(.0160) & .8902(.0087) & .028(.015) & 0.032 & 1.00 \\
 &&& 0.2   &  265 (5.01) & .8399(.0112) & .9608(.0103) & .8961(.0082) & .017(.009) & 0.019&1.00 \\
 &&  & 0.5 &  266 (3.84) & .8484(.0110) & .9765(.0078) & .9078(.0072) & .010(.005) & 0.012&0.98 \\
 &&  & 1.0 &  265 (3.88) & .8628(.0107) & .9778(.0078) & .9166(.0069) & .011(.006) & 0.011&1.00 \\ \cline{3-11} 
 && \multirow{3}{*}{4} & 0.05 &257 (4.13)& .7918(.0109) & .8954(.0113) & .8837(.0089) & .045(.011) & 0.025 & 1.00  \\
 &&  & 0.1 &270 (4.21) & .7948(.0109) & .9015(.0107) & .8883(.0085) & .050(.011) & 0.056 & 1.00 \\
 &&& 0.2    & 275 (3.77) & .8402(.0110) & .9660(.0094) & .8985(.0077) & .017(.009) & 0.020&0.97 \\
 &&  & 0.5  & 261 (4.15) & .8528(.0108) & .9785(.0075) & .9112(.0071) & .016(.008) & 0.018&0.98 \\
 &&  & 1.0  & 265 (4.46) & .8706(.0101) & .9760(.0078) & .9202(.0066) & .017(.009) & 0.019&1.00 \\ \hline
\multirow{10}{*}{0.4} & BRL & &  &275 (2.99)& .5583(.0212)& .5665(.0642)& .5539(.0344) & .459(.046) & 0.427&0.59 \\ \cline{2-11} 
 & \multirow{15}{*}{BRLVOF}& \multirow{3}{*}{1} & 0.05 &245 (6.69)&.6584(.0146) & .8045(.0226) & .7313(.0154) & .053(.015) & 0.026 & 0.99 \\
 &&  & 0.1 &243 (6.59) & .6675(.0147) & .8205(.0221) & .7359(.0150) & .037(.015) & 0.043 & 1.00 \\
 &&& 0.2   &  237 (7.52) & .6725(.0145) & .8427(.0216) & .7478(.0142) & .025(.015) & 0.029&1.00 \\
 &&  & 0.5 &  231 (5.94) & .6833(.0145) & .9019(.0180) & .7772(.0120) & .008(.004) & 0.009&1.00 \\
 &&  & 1.0 &  229 (5.82) & .6986(.0144) & .9326(.0151) & .7985(.0109) & .003(.001) & 0.003&1.00 \\ \cline{3-11} 
 && \multirow{3}{*}{2} & 0.05 &245 (6.69)& .6614(.0149) & .8100(.0229) & .7280(.0155) & .040(.015) & 0.002 & 1.00 \\
 &&  & 0.1 &243 (6.59) & .6653(.0148) & .8222(.0221) & .7353(.0150) & .036(.015) & 0.042 & 1.00 \\
 &&& 0.2   &  237 (7.52) & .6720(.0146) & .8572(.0208) & .7531(.0137) & .026(.015) & 0.030&1.00\\
 &&  & 0.5 &  231 (5.94) & .6865(.0141) & .9151(.0163) & .7842(.0112) & .016(.010) & 0.019&0.98 \\
 &&  & 1.0 &  229 (5.82) & .7105(.0139) & .9305(.0151) & .8055(.0106) & .017(.003) & 0.019&1.00 \\ \cline{3-11} 
 && \multirow{3}{*}{4} & 0.05 &212 (6.48)& .6624(.0129) & .8102(.0229) & .7319(.0145) & .043(.012) & 0.040 & 1.00 \\
 &&  & 0.1 &232 (6.08) & .6649(.0149) & .8219(.0201) & .7399(.0151) & .032(.015) & 0.030 & 0.99 \\
 &&& 0.2   &  226 (6.39) & .6763(.0139) & .8818(.0195) & .7652(.0124) & .029(.016) & 0.034&0.97 \\
 &&  & 0.5 &  223 (6.15) & .6960(.0141) & .9285(.0195) & .7952(.0127) & .023(.013) & 0.027&0.99 \\
 &&  & 1.0 &  227 (6.61) & .7258(.0134) & .9336(.0148) & .8160(.0102) & .023(.012) & 0.026&0.99\\ \hline
\end{tabular}
}}
\end{table}
\FloatBarrier

\begin{table}[]
\caption {Results for estimation of $\rho$ under BRLVOF and BRL, when conditional models for $\mathbf{X}_A$ given $\mathbf{X}_B$ include $W$, and $\sigma=0.5$. Values in parentheses represent standard deviations of the corresponding quantities.} \label{tablinW}
\centering
\resizebox{\textwidth}{!}{
{\LARGE
\begin{tabular}{cccc|ccccccc}
$\epsilon$ & Method& P & $\beta_M$ & $n$ & $\overline{TPR}$ & 
$\overline{PPV}$ & $\overline{F1}$ & $\overline{Bias}$ & $RMSE$&Coverage \\ \hline
\multirow{10}{*}{0.0} & BRL& &  &301 (0.17) &.9984(.0016) &.8845(.0418) &.9200(.0240) & .058(.023) & 0.065 & 0.90\\ \cline{2-11} 
 & \multirow{15}{*}{BRLVOF} &\multirow{3}{*}{1} 
 & 0.05    & 300 (0.29) & .9992(.0012) & .9995(.0005) & .9993(.0008) & .001(.001) & 0.002 & 1.00 \\ 
 &&  & 0.1 &   300 (0.29) & .9991(.0012) & .9994(.0004) & .9993(.0008) & .001(.001) & 0.002 & 1.00 \\ 
 &&& 0.2   &   300 (0.29) & .9992(.0012) & .9996(.0004) & .9994(.0007) & .001(.001) & 0.002 & 1.00 \\ 
 &&  & 0.5 &   300 (0.27) & .9994(.0010) & .9997(.0003) & .9995(.0006) & .000(.001) & 0.001 & 1.00 \\ 
 &&  & 1.0 &   300 (0.44) & .9995(.0009) & .9991(.0013) & .9993(.0009) & .000(.001) & 0.000 & 1.00 \\
 \cline{3-11} 
 && \multirow{3}{*}{2} 
      & 0.05& 304 (0.28) & .9796(.0016) & .9797(.0007) & .9994(.0007) & .013(.002) & 0.015 & 1.00 \\ 
 &&  & 0.1 &    300 (0.28) & .9992(.0011) & .9995(.0004) & .9993(.0007) & .001(.001) & 0.002 & 1.00 \\ 
 &&& 0.2   &    300 (0.22) & .0200(.0000) & .0200(.0000) & .0200(.0000) & .000(.000) & 0.000 & 0.02 \\ 
 &&  & 0.5 &    316 (0.26) & .9214(.0011) & .9208(.0006) & .9210(.0007) & .102(.001) & 0.103 & 0.92 \\ 
 &&  & 1.0 &    312 (0.61) & .9448(.0009) & .9415(.0020) & .9429(.0011) & .068(.001) & 0.069 & 0.94 \\ 
 \cline{3-11} 
 && \multirow{3}{*}{4} 
 & 0.05    & 310 (0.28) & .9593(.0014) & .9556(.0007) & .9966(.0008) & .018(.001) & 0.021 & 1.00 \\ 
 &&  & 0.1 &   314 (0.28) & .9393(.0014) & .9355(.0007) & .9965(.0008) & .035(.001) & 0.039 & 0.99 \\ 
 &&& 0.2   &   320 (0.73) & .9096(.0015) & .9054(.0017) & .9964(.0014) & .062(.001) & 0.067 & 0.96 \\ 
 &&  & 0.5 &   318 (0.32) & .9201(.0023) & .9158(.0016) & .9968(.0007) & .061(.002) & 0.065 & 0.92 \\ 
 &&  & 1.0 &   309 (0.76) & .9598(.0008) & .9576(.0025) & .9986(.0013) & .030(.001) & 0.033 & 0.96 \\ 
 \hline
\multirow{10}{*}{0.2} &BRL&  &  & 288 (2.47)& .7780(.0151) &.7168(.0587)& .7424(.0326) & .248(.042) & 0.253&0.68 \\ \cline{2-11} 
 & \multirow{15}{*}{BRLVOF}& \multirow{3}{*}{1} 
 & 0.05       &  264 (4.46) & .8337(.0111) & .9484(.0122) & .8872(.0089) & .026(.011) & 0.029 & 1.00 \\ 
 &&  & 0.1    &    263 (4.43) & .8345(.0112) & .9506(.0119) & .8886(.0088) & .023(.011) & 0.026 & 1.00 \\ 
 &&& 0.2      &    262 (4.3) & .8360(.0110) & .9561(.0111) & .8919(.0084) & .016(.011) & 0.019 & 1.00 \\ 
 &&  & 0.5    &    260 (4.03) & .8419(.0111) & .9713(.0087) & .9018(.0076) & .006(.011) & 0.006 & 1.00 \\ 
 &&  & 1.0    &    261 (3.9) & .8512(.0110) & .9790(.0074) & .9105(.0071) & .002(.011) & 0.002 & 1.00 \\  \cline{3-11} 
 && \multirow{3}{*}{2} 
 & 0.05     & 264 (4.31) & .8350(.0111) & .9503(.0118) & .8888(.0089) & .025(.011) & 0.029 & 1.00 \\ 
 &&  & 0.1  &   263 (4.26) & .8362(.0111) & .9533(.0113) & .8908(.0087) & .022(.011) & 0.026 & 1.00 \\ 
 &&& 0.2   &    264 (4.08) & .8302(.0107) & .9518(.0101) & .8957(.0080) & .029(.011) & 0.032 & 0.99 \\ 
 &&  & 0.5 &    268 (3.78) & .8232(.0107) & .9474(.0076) & .8807(.0071) & .044(.011) & 0.046 & 0.97 \\ 
 &&  & 1.0 &    265 (3.88) & .8620(.0106) & .9774(.0076) & .9159(.0068) & .011(.011) & 0.013 & 1.00 \\ 
 \cline{3-11} 
 && \multirow{3}{*}{4} 
     & 0.05 &265 (4.39) & .8315(.0114) & .9466(.0117) & .8845(.0089) & .026(.011) & 0.03 & 1.00 \\ 
 &&  & 0.1  &  281 (4.1) & .7838(.0125) & .8896(.0115) & .8845(.0085) & .059(.012) & 0.065 & 0.99 \\ 
 &&& 0.2    &  284 (3.74) & .7714(.0105) & .8801(.0089) & .8919(.0078) & .074(.010) & 0.08 & 0.97 \\ 
 &&  & 0.5  &  268 (3.75) & .8342(.0108) & .9556(.0072) & .9081(.0071) & .033(.011) & 0.035 & 0.98 \\ 
 &&  & 1.0  &  267 (3.87) & .8689(.0102) & .9776(.0080) & .9200(.0066) & .016(.010) & 0.018 & 1.00 \\ 
 \hline
\multirow{10}{*}{0.4} & BRL& &  &275 (2.99)& .5583(.0212)& .5665(.0642)& .5539(.0344) & .424(.045) & 0.431&0.57\\ \cline{2-11} 
 & \multirow{15}{*}{BRLVOF}& \multirow{3}{*}{1} 
 & 0.05    &  241 (6.96) & .6510(.0147) & .8060(.0233) & .7273(.0155) & .052(.015) & 0.059 & 0.99 \\ 
 &&  & 0.1 &    247 (6.67) & .6468(.0154) & .8038(.0231) & .7236(.0155) & .052(.015) & 0.059 & 1.00 \\ 
 &&& 0.2   &    238 (6.75) & .6650(.0145) & .8403(.0217) & .7422(.0141) & .026(.015) & 0.031 & 1.00 \\ 
 &&  & 0.5 &    229 (6.43) & .6690(.0148) & .8883(.0185) & .7704(.0122) & .021(.015) & 0.023 & 0.99 \\ 
 &&  & 1.0 &    222 (5.95) & .6897(.0144) & .9307(.0154) & .7920(.0110) & .003(.014) & 0.003 & 1.00 \\ 
 \cline{3-11} 
 && \multirow{3}{*}{2} 
 & 0.05    &  265 (4.39) & .8315(.0114) & .9466(.0117) & .8845(.0089) & .026(.011) & 0.03 & 1.00 \\ 
 &&  & 0.1 &    281 (4.1) & .7838(.0125) & .8896(.0115) & .8845(.0085) & .059(.012) & 0.065 & 0.99 \\ 
 &&& 0.2   &    245 (6.22) & .6590(.0137) & .8368(.0198) & .7357(.0127) & .058(.014) & 0.062 & 0.97 \\ 
 &&  & 0.5 &    230 (5.98) & .6741(.0138) & .9018(.0161) & .7711(.0108) & .038(.014) & 0.041 & 0.98 \\ 
 &&  & 1.0 &    227 (5.77) & .7090(.0139) & .9353(.0147) & .8063(.0105) & .017(.014) & 0.019 & 1.00 \\ 
 \cline{3-11} 
 && \multirow{3}{*}{4} 
 & 0.05    &  265 (4.39) & .8315(.0114) & .9466(.0117) & .8845(.0089) & .026(.011) & 0.03 & 1.00 \\ 
 &&  & 0.1 &    281 (4.1) & .7838(.0125) & .8896(.0115) & .8845(.0085) & .059(.012) & 0.065 & 0.99 \\ 
 &&& 0.2   &    244 (6.31) & .6390(.0143) & .8329(.0192) & .7607(.0127) & .060(.014) & 0.067 & 0.99 \\ 
 &&  & 0.5 &    224 (5.96) & .6922(.0144) & .9265(.0157) & .7921(.0111) & .025(.014) & 0.028 & 1.00 \\ 
 &&  & 1.0 &    233 (5.83) & .7212(.0138) & .9304(.0152) & .8122(.0105) & .023(.014) & 0.027 & 1.00 \\ 
 \hline
\end{tabular}
}
}
\end{table}
\FloatBarrier

\FloatBarrier
\begin{table}[]
\caption {Results for estimation of $\rho$ under BRLVOF and BRL, when conditional models for $\mathbf{X}_A$ given $\mathbf{X}_B$ include non-linear terms, and $\sigma=0.5$. Values in parentheses represent standard deviations of the corresponding quantities.} \label{tabnonlin}
\centering
\resizebox{\textwidth}{!}{
{\LARGE
\begin{tabular}{cccc|ccccccc}
$\epsilon$ & Method &P & $\beta_M$ & $n$ &$\overline{TPR}$ & $\overline{PPV}$ & $\overline{F1}$ & $\overline{Bias}$ & $RMSE$&Coverage \\ \hline
\multirow{10}{*}{0.0} & BRL& &  &301 (0.17) &.9984(.0016) &.8845(.0418) &.9200(.0240) & .058(.023) & 0.069 & 0.88 \\ \cline{2-11} 
 &\multirow{15}{*}{BRLVOF}& \multirow{3}{*}{1} & 0.05 &299 (0.59)& .9954(.0021) & .9993(.0006) & .9974(.0012) & .029(.002) & 0.032 & 1.00 \\
 &&  & 0.1 &301 (0.58) & .9862(.0029) & .9898(.0011) & .9880(.0018) & .035(.003) & 0.039 & 1.00 \\
 &&& 0.2   &  305 (0.57) & { .9917(.0422)} & .9949(.0422) & .9933(.0433) & .025(.021) & 0.029&0.98 \\
 &&  & 0.5 &  303 (0.58) & { .9876(.0123)} & .9901(.0073) & .9888(.0092) & .023(.039) & 0.023&0.98 \\
 &&  & 1.0 &  300 (0.79) & .9966(.0017) & .9980(.0023) & .9973(.0015) & .003(.009) & 0.004&1.00 \\ \cline{3-11} 
 && \multirow{3}{*}{2} & 0.05 &301 (0.57)& .9869(.0021) & .9895(.0006) & .9981(.0012) & .024(.002) & 0.028 & 1.00 \\
 &&  & 0.1 &307 (0.60) & .9678(.0040) & .9659(.0019) & .9954(.0012) & .047(.004) & 0.052 & 0.97 \\
 &&& 0.2   &  307 (0.55) & .9945(.0017) & .9923(.0047) & .9916(.0026) & .014(.009) & 0.016&0.96 \\
 &&  & 0.5 &  311 (0.63) & .9966(.0017) & .9987(.0016) & .9976(.0013) & .008(.006) & 0.006&0.94 \\
 &&  & 1.0 &  306 (1.04) & .9969(.0018) & .9961(.0018) & .9965(.0019) & .006(.004) & 0.008&0.97 \\ \cline{3-11} 
 && \multirow{3}{*}{4} & 0.05 &308 (0.52)& .9681(.0019) & .9654(.0009) & .9960(.0011) & .027(.002) & 0.032 & 1.00 \\
 &&  & 0.1 &333 (1.29) & .8317(.0067) & .8323(.0053) & .9985(.0012) & .102(.007) & 0.114 & 0.97 \\
 &&& 0.2   &  330 (1.35) & .9978(.0020) & .9990(.0014) & .9984(.0013) & .008(.008) & 0.011&0.97 \\
 &&  & 0.5 &  320 (0.71) & .9977(.0019) & .9979(.0022) & .9978(.0015) & .006(.006) & 0.008&0.93 \\
 &&  & 1.0 &  312 (2.02) & .9980(.0017) & .9956(.0035) & .9968(.0019) & .006(.005) & 0.008&0.95 \\ \hline
\multirow{10}{*}{0.2} &BRL&  &  & 288 (2.47)& .7780(.0151) &.7168(.0587)& .7424(.0326) & .248(.042) & 0.266&0.62\\ \cline{2-11} 
 &\multirow{15}{*}{BRLVOF}& \multirow{3}{*}{1} & 0.05 &257 (4.29)& .8241(.0115) & .9615(.0104) & .8874(.0084) & .045(.012) & 0.032 & 1.00 \\
 &&  & 0.1 &257 (4.31)& .8246(.0114) & .9636(.0103) & .8885(.0082) & .040(.011) & 0.043 & 1.00 \\
 &&& 0.2   &  261 (4.14)& .8261(.0114) & .9675(.0096) & .8911(.0079) & .031(.009) & 0.032&1.00 \\
 &&  & 0.5 &  256 (4.03)& .8316(.0114) & .9761(.0081) & .8979(.0075) & .015(.004) & 0.015&1.00 \\
 &&  & 1.0 &  259 (4.21)& .8427(.0112) & .9756(.0085) & .9042(.0072) & .007(.002) & 0.007&1.00 \\ \cline{3-11} 
 && \multirow{3}{*}{2} & 0.05 &255 (4.23)& .8199(.0115) & .9659(.0098) & .8868(.0082) & .043(.012) & 0.028 & 1.00 \\
 &&  & 0.1 &259 (4.16)& .8046(.0114) & .9484(.0093) & .8881(.0081) & .056(.011) & 0.06 & 0.98 \\
 &&& 0.2   &  257 (4.05)& .8141(.0264) & .9597(.0323) & .8818(.0084) & .046(.039) & 0.051&0.99 \\
 &&  & 0.5 &  258 (4.12)& .8307(.0113) & .9755(.0084) & .8971(.0074) & .021(.010) & 0.024&0.99 \\
 &&  & 1.0 &  264 (4.46)& .8463(.0112) & .9694(.0097) & .9035(.0074) & .018(.008) & 0.020&0.99 \\ \cline{3-11} 
 && \multirow{3}{*}{4} & 0.05 &257 (4.13)& .8058(.0118) & .9539(.0091) & .8815(.0083) & .041(.012) & 0.032 & 1.00 \\
 &&  & 0.1 &270 (4.21)& .7674(.0130) & .9070(.0095) & .8830(.0082) & .066(.013) & 0.072 & 0.99 \\
 &&& 0.2   &  275 (3.77)& .8193(.0117) & .9717(.0087) & .8889(.0079) & .029(.014) & 0.033&0.97 \\
 &&  & 0.5 &  261 (4.15)& .8303(.0113) & .9721(.0086) & .8955(.0075) & .023(.012) & 0.026&0.99 \\
 &&  & 1.0 &  265 (4.46)& .8467(.0110) & .9670(.0100) & .9028(.0074) & .021(.011) & 0.024&0.99 \\ \hline
\multirow{10}{*}{0.4} & BRL& &  &275 (2.99)& .5583(.0212)& .5665(.0642)& .5539(.0344) & .424(.045) & 0.444&0.52 \\ \cline{2-11} 
 &\multirow{15}{*}{BRLVOF}& \multirow{3}{*}{1} & 0.05 &226 (6.81)& .6456(.0146) & .8546(.0213) & .7427(.0136) & .066(.015) & 0.032 & 1.00 \\
 &&  & 0.1 &226 (6.85)& .6542(.0146) & .8703(.0211) & .7466(.0132) & .050(.015) & 0.055 & 1.00 \\
 &&& 0.2   &  224 (7.86)& .6565(.0146) & .8863(.0201) & .7540(.0126) & .037(.014) & 0.041&1.00 \\
 &&  & 0.5 &  223 (6.1)& .6646(.0147) & .9182(.0174) & .7708(.0116) & .019(.006) & 0.020&0.98 \\
 &&  & 1.0 &  221 (6.28& .6820(.0145) & .9259(.0168) & .7851(.0111) & .009(.002) & 0.009&1.00 \\ \cline{3-11} 
 && \multirow{3}{*}{2} & 0.05 &218 (6.86)& .6388(.0148) & .8804(.0210) & .7401(.0131) & .059(.015) & 0.028 & 1.00 \\
 &&  & 0.1 &220 (6.70)& .6343(.0147) & .8779(.0205) & .7435(.0130) & .061(.015) & 0.067 & 0.99 \\
 &&& 0.2   &  221 (6.49)& .6442(.0148) & .9010(.0191) & .7509(.0123) & .043(.019) & 0.048&0.98 \\
 &&  & 0.5 &  217 (6.31)& .6512(.0147) & .9098(.0173) & .7588(.0114) & .041(.015) & 0.045&0.99 \\
 &&  & 1.0 &  224 (6.64)& .6828(.0146) & .9138(.0178) & .7812(.0113) & .027(.012) & 0.030&1.00 \\ \cline{3-11} 
 && \multirow{3}{*}{4} & 0.05 &212 (6.48) & .6263(.0145) & .8886(.0196) & .7344(.0124) & .050(.014) & 0.032 & 1.00 \\
 &&  & 0.1 &232 (6.08)& .5921(.0149) & .8390(.0183) & .7457(.0123) & .085(.015) & 0.094 & 0.98 \\
 &&& 0.2   &  226 (6.39)& .6403(.0144) & .9093(.0185) & .7511(.0119) & .041(.020) & 0.046&0.96 \\
 &&  & 0.5 &  223 (6.15)& .6586(.0144) & .9172(.0177) & .7663(.0114) & .034(.018) & 0.038&0.98 \\
 &&  & 1.0 &  227 (6.61)& .6841(.0143) & .9130(.0181) & .7818(.0110) & .028(.016) & 0.033&0.99 \\ \hline
\end{tabular}
}
}
\end{table}
\FloatBarrier

\subsection {Results under BRLVOF for regression slope $\beta$}
We display the results for inference on $\beta$, the slope of the regression of $\mathbf{X}_A$ on $\mathbf{X}_B$. Since the results are qualitatively very similar to those for $\rho$, we only present them under selected scenarios. Specifically, we consider the scenario when $\mathbf{X}_B$ comprises four covariates ($P=4$), $\beta_M \in \{0.05,0.2\}$, $\sigma=0.1$, and $\epsilon \in \{0.0,0.2,0.4\}$.
\begin{figure} [H]
\centering
\begin{subfigure}[t]{1.06\textwidth}
  \centering
  \includegraphics[scale=.16]{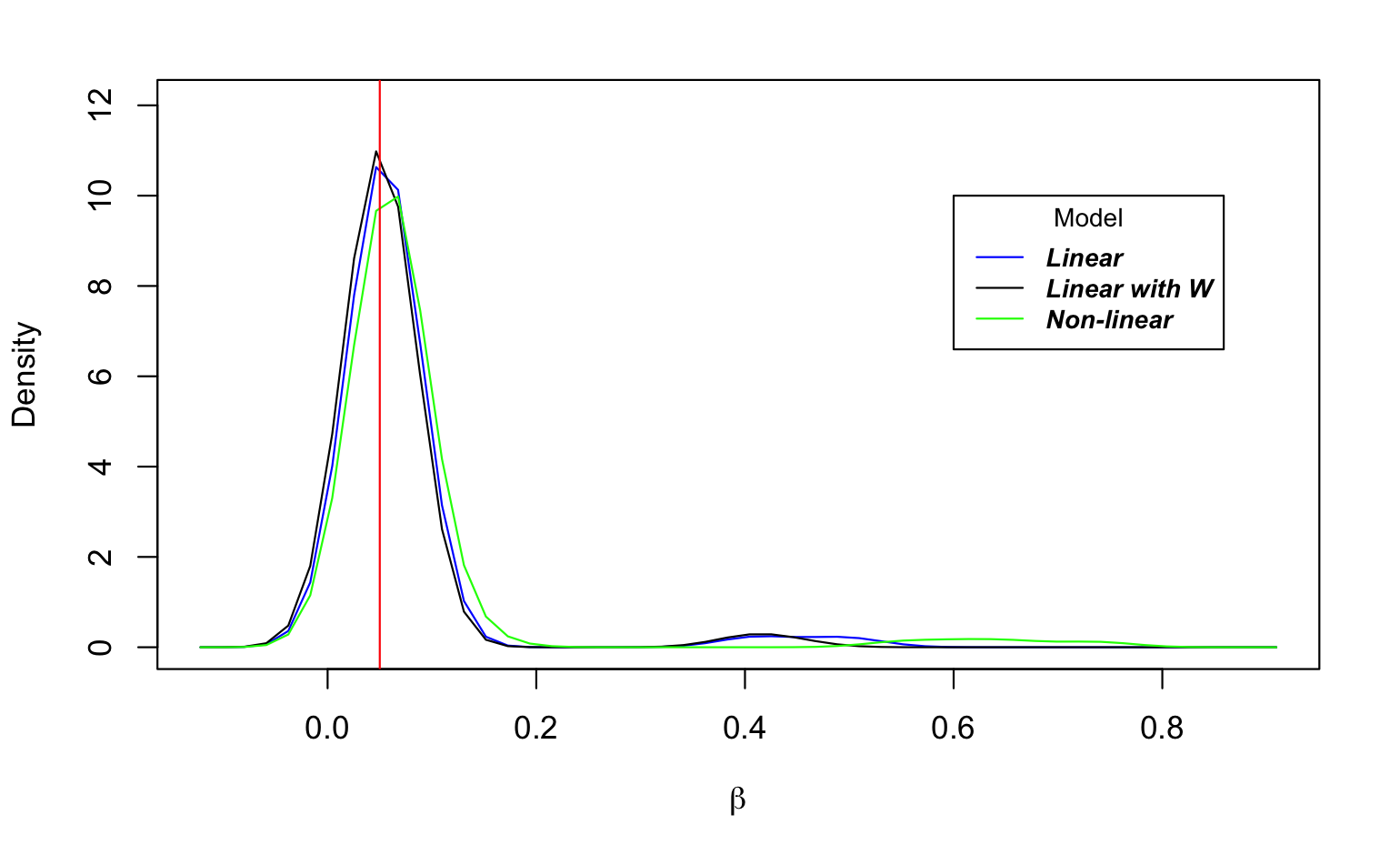}
  \includegraphics[scale=.16]{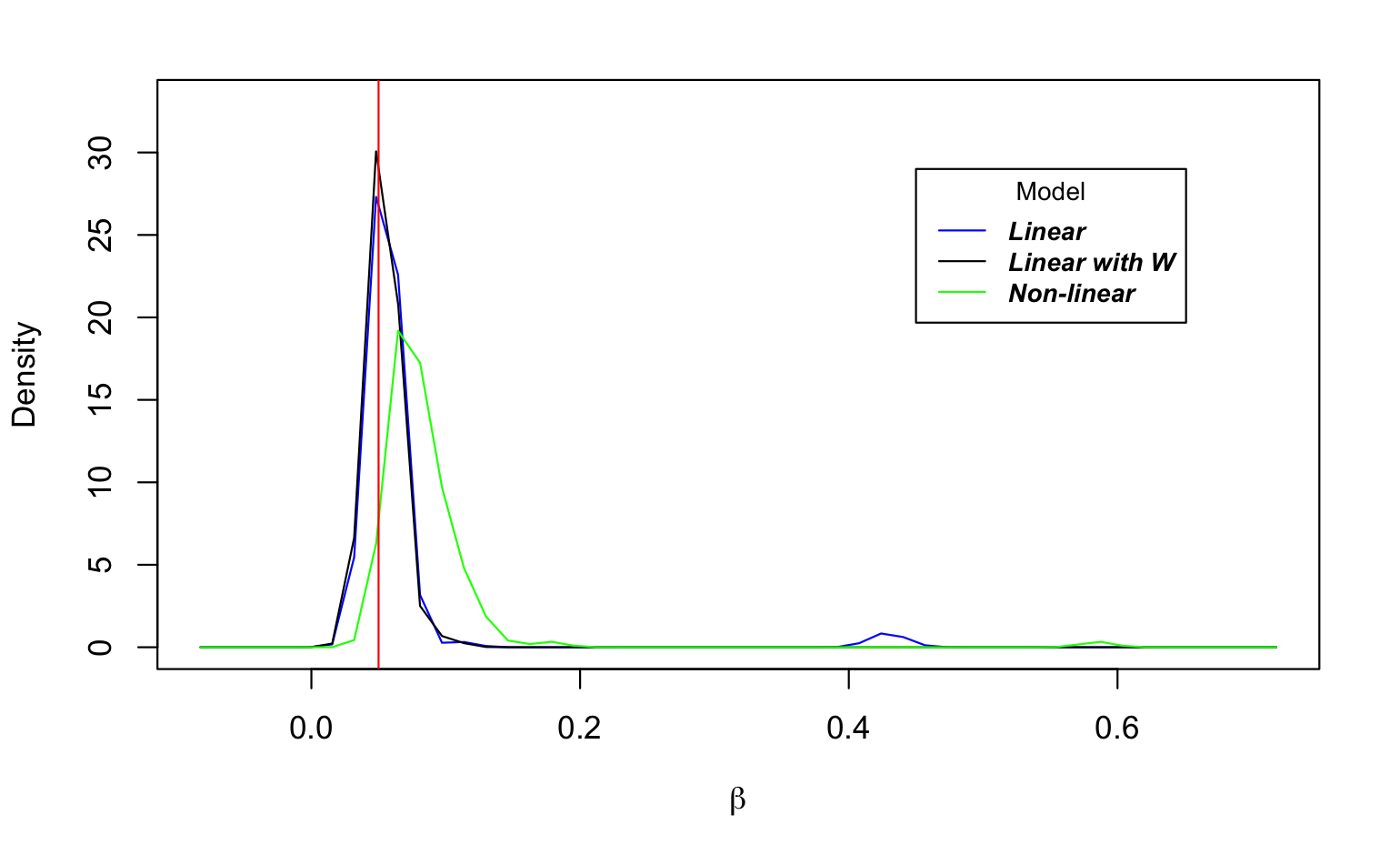}
  \\
  \includegraphics[scale=.16]{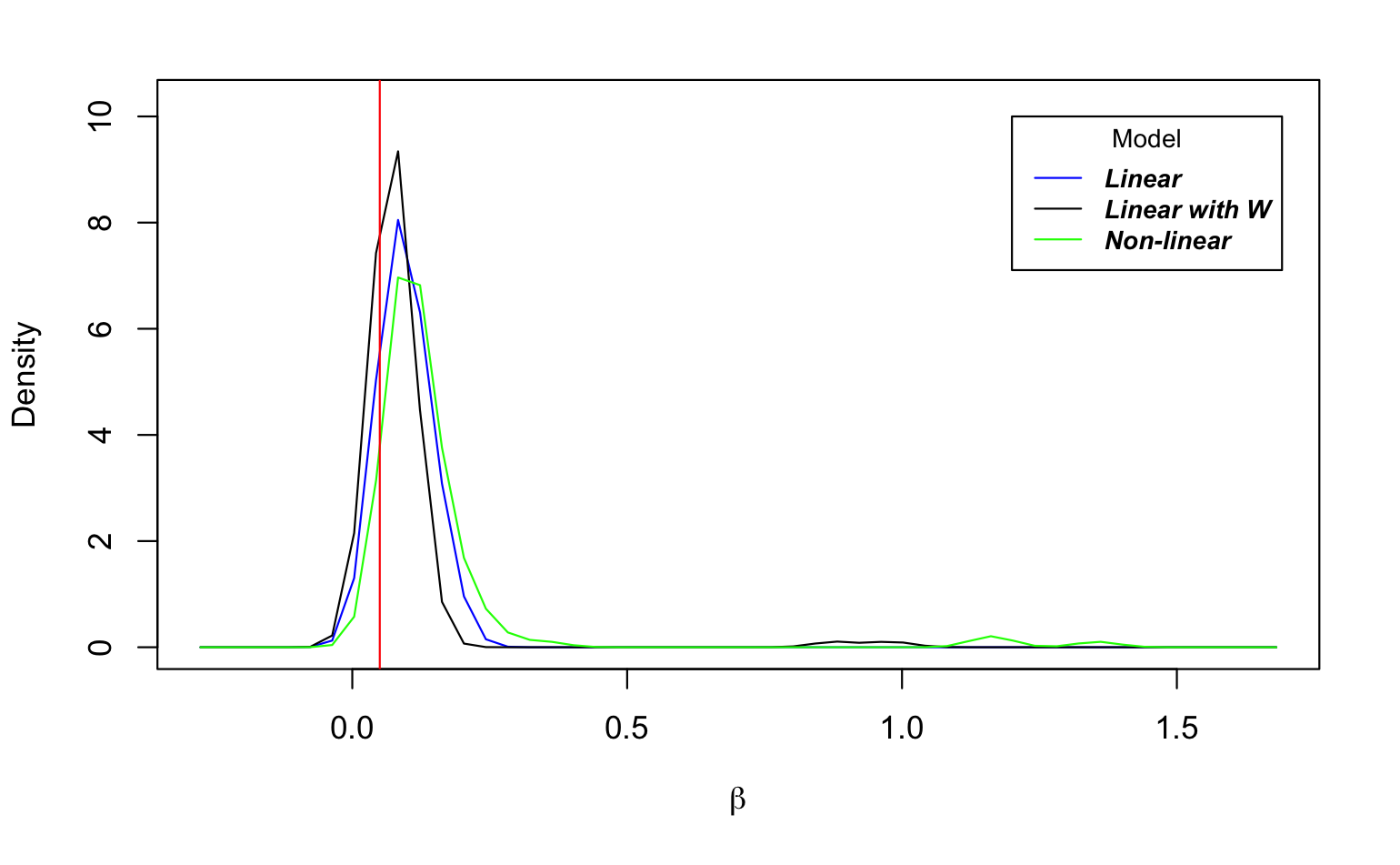} 
\end{subfigure}
\caption{Distribution of point estimates of $\beta$, when true value of $\beta_M=0.05$. The upper panel depicts the case when the error levels are $\epsilon=0$ (left) and $\epsilon=0.2$ (right). The lower panel depicts the case when $\epsilon=0.4$.}

\end{figure}
\FloatBarrier

\begin{figure} [H]
\centering
\begin{subfigure}[t]{1.06\textwidth}
  \centering
  \includegraphics[scale=.16]{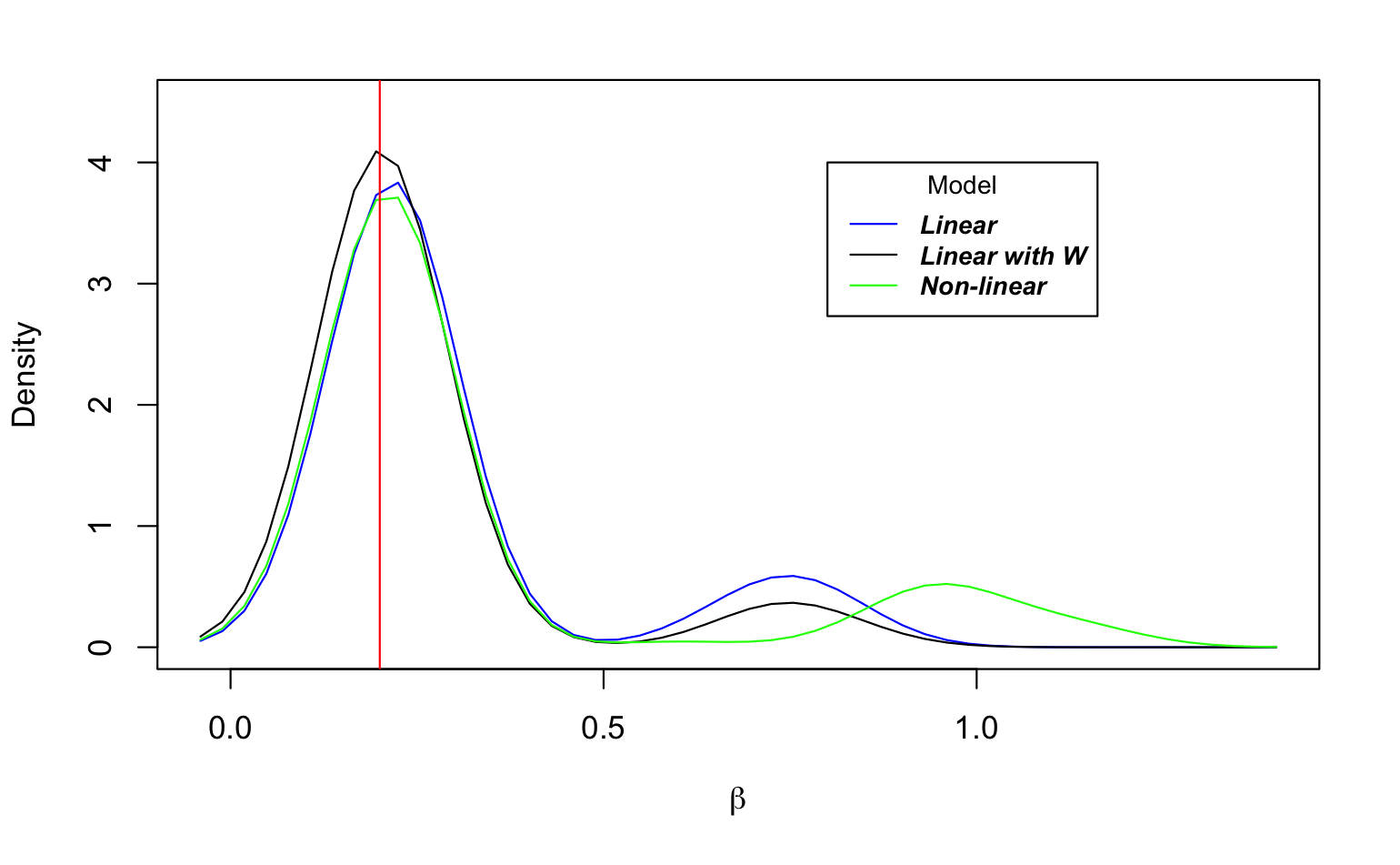}
  \includegraphics[scale=.16]{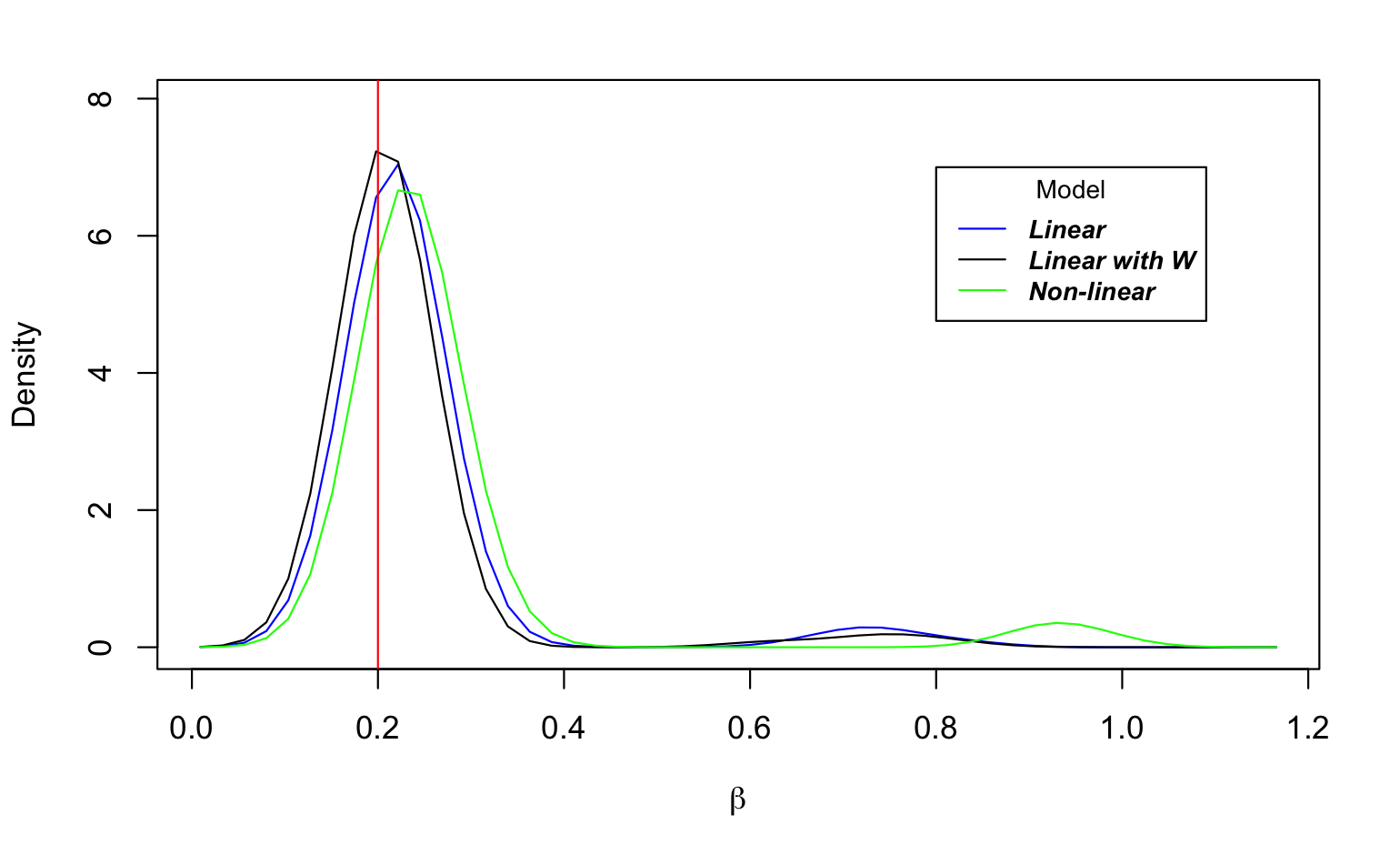}
  \\
  \includegraphics[scale=.16]{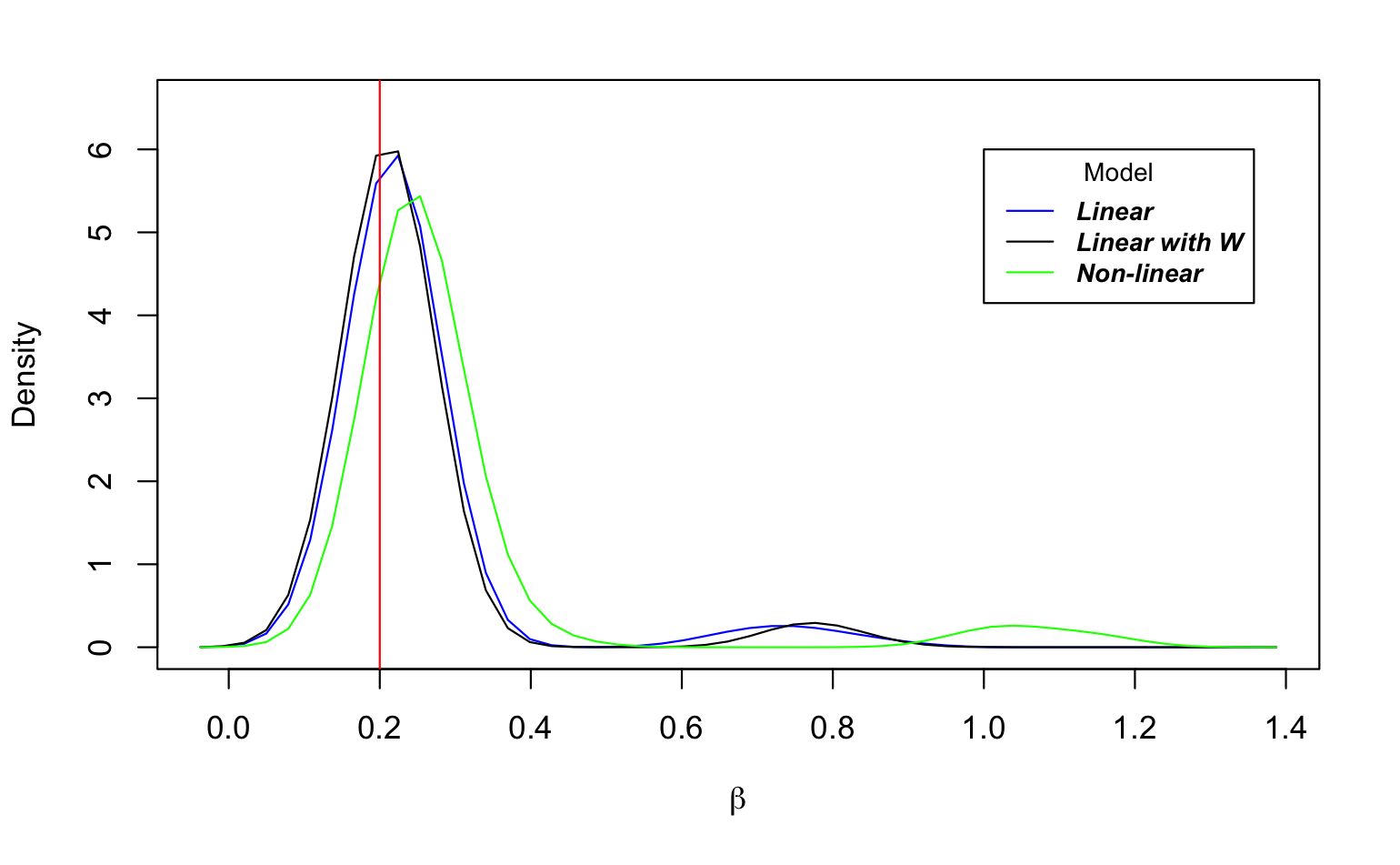} 
\end{subfigure}
\caption{Distribution of point estimates of $\beta$, when true value of $\beta_M=0.2$. The upper panel depicts the case when the error levels are $\epsilon=0$ (left) and $\epsilon=0.2$ (right). The lower panel depicts the case when $\epsilon=0.4$.}

\end{figure}
\FloatBarrier

\subsection{Results under BRLVOF for $\beta_M$ and $\beta_U$}
We display the results for inference on $\beta_M$ and $\beta_U$. We compute estimates $\hat\beta_M$ and $\hat\beta_U$ using the posterior mean from the last 900 iterations of the Gibbs sampler. Since the results are qualitatively similar across the various simulation scenarios, we only display them under selected settings. Specifically, we consider the scenario when $\mathbf{X}_B$ comprises four covariates ($P=4$), $\beta_M \in \{0.05,0.2\}$, $\sigma=0.1$, and $\epsilon \in \{0.0,0.2,0.4\}$.
\FloatBarrier
\begin{table}[H]
\caption {Bias and RMSE in estimating $\beta_M$ and $\beta_U$ across varying error rates when $\mathbf{X}_B$ comprises four covariates, and $\sigma=0.1$. Values in parentheses represent standard deviations of the corresponding quantities.}
\centering
\newsavebox\mybox
\sbox{\mybox}{%
\begin{tabular}{cccc|cc|cc}
Model for $\mathbf{X}_A$ given $\mathbf{X}_B$ &$\epsilon$ & $\beta_M$ &$\beta_U$ & \multicolumn{2}{c}{$\hat\beta_M$}&\multicolumn{2}{c}{$\hat\beta_U$} \\ \hline
&&&& $\overline{Bias}$ & $RMSE$ & $\overline{Bias}$ & $RMSE$\\ \hline
\multirow{6}{*}{Linear}&\multirow{2}{*}{0.0}&0.05& 0.05.  &   .014(.062) & 0.021 & .068(.092) & 0.095 \\ 
                    && 0.2 &0.05   &     .074(.069) & 0.095 & .062(.056) & 0.085 \\ 
  
\cline{2-8}
&\multirow{2}{*}{0.2}  &0.05& 0.05     & .011(.046) & 0.017 & .061(.065) & 0.081 \\ 
 
                     && 0.2 &  0.05   &    .027(.031) & 0.037 & .071(.095) & 0.086 \\
   
\cline{2-8}
&\multirow{2}{*}{0.4}  &0.05& 0.05   & .050(.003) & 0.005 & .051(.005) & 0.051 \\ 
  &&  0.2&  0.05   &                   .032(.018) & 0.039 & .075(.004) & 0.073 \\  

 \hline
 
 \multirow{6}{*}{Linear with W}&\multirow{2}{*}{0.0}&0.05& 0.05  &  .014(.050) & 0.021 & .062(.067) & 0.083 \\ 
                    && 0.2 &0.05                                  &   .052(.038) & 0.068 & .086(.118) & 0.085 \\ 
  
\cline{2-8}
&\multirow{2}{*}{0.2}  &0.05& 0.05     & .006(.004) & 0.011 & .050(.000) & 0.051 \\ 
 
                     && 0.2 &  0.05   &    .024(.088) & 0.034 & .066(.081) & 0.092 \\ 
   
\cline{2-8}
&\multirow{2}{*}{0.4}  &0.05& 0.05   & .018(.055) & 0.027 & .065(.077) & 0.090 \\ 
  &&  0.2&  0.05   &                     .031(.019) & 0.041 & .074(.050) & 0.093 \\ 

 \hline
 
 \multirow{6}{*}{Non-linear}&\multirow{2}{*}{0.0}&0.05& 0.05.  &   .014(.062) & 0.021 & .068(.092) & 0.095 \\ 
                    && 0.2 &0.05   &     .074(.069) & 0.095 & .062(.056) & 0.085 \\ 
  
\cline{2-8}
&\multirow{2}{*}{0.2}  &0.05& 0.05     & .011(.046) & 0.017 & .061(.065) & 0.081 \\ 
 
                     && 0.2 &  0.05   &    .027(.031) & 0.037 & .071(.095) & 0.086 \\
   
\cline{2-8}
&\multirow{2}{*}{0.4}  &0.05& 0.05   & .091(.053) & 0.101 & .065(.072) & 0.091 \\ 
  &&  0.2&  0.05   &                     .086(.028) & 0.098 & .085(.050) & 0.099 \\  

 \hline
\end{tabular}
}
{\centering\scalebox{.9}{\usebox{\mybox}}\par}
\end{table}
\FloatBarrier 

\subsection{Results under $BRLVOF_{ind}$ }
We implement the approach of \cite{Tang2020} (abbreviated as BRLVOF$_{ind}$) using the algorithm outlined in Section 3.3.2 on page 215 of the article. Specifically, we model $f_M(X_{Ai},\mathbf{X}_{Bj}|\mathbf{Z}_{Ai},\mathbf{Z}_{Bj},\beta_M)$ using a linear model $X_{Ai}=\mathbf{X}_{Bj} \beta_M + {\varepsilon}_{M(i,j)}$ for $(i,j) \in \mathbf{M}$. For $(i,j) \in \mathbf{U}$, we assume $X_{Ai} \independent \mathbf{X}_{Bj}$, and model $f_U(X_{Ai}|\mathbf{Z}_{Ai},\mathbf{Z}_{Bj},\beta_U)$ as a normal distribution with mean $\beta_U$ and variance $\sigma^2_U$. Under this algorithm, the full conditional distributions for $\beta_M$ and $\sigma^2_M$ remain the same as in Section 4.2 of the main text. The full conditional distributions for $\beta_U$ and $\sigma^2_U$ will be 
\begin{align} \label{rlmod}
\begin{split}
    \beta_U|\mathbf{X}_A,\mathbf{C},\sigma_U & \sim N \big(\overline{\mathbf{X}}_{AU}, \frac{\sigma_U^2}{n_An_B-n_m} \big)\\
    \sigma_U^2|\mathbf{X}_A,\mathbf{C},\beta_U&\sim Inv-Gamma \bigg(\dfrac{n_An_B-n_m}{2}, \dfrac{1}{2}\big(\mathbf{X}_{AU}-\beta_U \big)^{T}\big(\mathbf{X}_{AU}-\beta_U \big) \bigg).
\end{split}
\end{align}

\noindent Tables S6-S11 display results under all simulation scenarios.

\FloatBarrier
\begin{table}[]
\caption {Results for estimation of $\rho$ under BRLVOF$_{ind}$, when conditional models for $\mathbf{X}_A$ given $\mathbf{X}_B$ are linear, and $\sigma=0.1$.Values in parentheses represent standard deviations of the corresponding quantities.} \label{tab:res2}
\centering
\resizebox{\textwidth}{!}{
{\LARGE
\begin{tabular}{cccc|ccccccc}
$\epsilon$ & Method &P & $\beta_M$ &$n$ & $\overline{TPR}$ & $\overline{PPV}$ & $\overline{F1}$ & $\overline{Bias}$ & $RMSE$ &Coverage\\ \hline
\multirow{10}{*}{0.0} 
 &\multirow{15}{*}{BRLVOF$_{ind}$}& \multirow{3}{*}{1} 
     &  0.05  &  300 (0.12) & .9994(.0007) & .9995(.0005) & .9995(.0006) & .000(.001) & 0.001 & 1.00 \\  
 &&  &  0.1   &    300 (0.11) & .9996(.0006) & .9996(.0004) & .9996(.0005) & .000(.001) & 0.000 & 1.00 \\ 
 &&  &  0.2   &    300 (0.11) & .9997(.0005) & .9998(.0002) & .9997(.0003) & .000(.000) & 0.000 & 1.00 \\ 
 &&  &  0.5   &    300 (0.1) & .9999(.0004) & 1.0000(.0001) & .9999(.0002) & .000(.000) & 0.000 & 1.00 \\ 
 &&  &  1.0   &    300 (0.22) & .9999(.0004) & .9998(.0006) & .9999(.0004) & .000(.000) & 0.000 & 1.00 \\  \cline{3-11} 
 && \multirow{3}{*}{2} 
     &0.05   &300 (0.12) & .9994(.0006) & .9995(.0003) & .9994(.0005) & .000(.001) & 0.001 & 1.00 \\ 
 &&  &0.1    &  300 (0.12) & .9996(.0005) & .9997(.0002) & .9996(.0004) & .000(.001) & 0.000 & 1.00 \\ 
 &&. &0.2    &  300 (0.12) & .9997(.0005) & .9998(.0002) & .9997(.0003) & .000(.000) & 0.000 & 1.00 \\ 
 &&  &0.5    &  300 (0.11) & .9999(.0004) & .9999(.0001) & .9999(.0002) & .000(.000) & 0.000 & 1.00 \\ 
 &&  &1.0    &  300 (0.41) & 1.0000(.0003) & .9989(.0013) & .9995(.0007) & .001(.000) & 0.002 & 1.00 \\ \cline{3-11} 
 && \multirow{3}{*}{4} 
& 0.05        &300 (0.12) & .9997(.0006) & .9997(.0003) & .9997(.0004) & .000(.001) & 0.001 & 1.00 \\ 
 &&  & 0.1    &  300 (0.11) & .9999(.0005) & .9999(.0002) & .9999(.0003) & .000(.000) & 0.00 & 1.00 \\ 
 &&& 0.2      &  300 (0.1) & .9999(.0004) & .9999(.0001) & .9999(.0003) & .000(.000) & 0.00 & 1.00 \\ 
 &&  & 0.5    &  300 (0.14) & .9999(.0003) & .9999(.0003) & .9999(.0003) & .000(.000) & 0.001 & 1.00 \\ 
 &&  & 1.0    &  300 (0.51) & 1.0000(.0003) & .9987(.0017) & .9993(.0009) & .002(.000) & 0.003 & 1.00 \\ \hline
\multirow{10}{*}{0.2} 
 &\multirow{15}{*}{BRLVOF$_{ind}$}& \multirow{3}{*}{1} 
    & 0.05  &287 (3.32) & .8716(.0100) & .9127(.0129) & .8917(.0102) & .012(.010) & 0.014 & 1.00 \\  
 &&  & 0.1  &  282 (3.46) & .8805(.0091) & .9356(.0118) & .9072(.0088) & .005(.009) & 0.005 & 1.00 \\  
 &&& 0.2    &  277 (3.41) & .8878(.0086) & .9623(.0095) & .9234(.0070) & .001(.009) & 0.001 & 1.00 \\  
 &&  & 0.5  &  273 (2.92) & .8933(.0081) & .9833(.0063) & .9361(.0055) & .000(.008) & 0.000 & 1.00 \\ 
 &&  & 1.0  &  273 (2.83) & .8987(.0082) & .9893(.0053) & .9417(.0052) & .000(.008) & 0.000 & 1.00 \\ \cline{3-11}
 
 && \multirow{3}{*}{2} 
     & 0.05  &285 (3.41) & .8753(.0096) & .9227(.0124) & .8983(.0096) & .014(.010) & 0.016 & 1.00 \\ 
 &&  & 0.1   &  280 (3.49) & .8843(.0088) & .9490(.0109) & .9154(.0080) & .010(.009) & 0.011 & 1.00 \\ 
 &&& 0.2     &  275 (3.23) & .8896(.0085) & .9713(.0084) & .9286(.0065) & .008(.009) & 0.010 & 1.00 \\ 
 &&  & 0.5   &  273 (2.79) & .8963(.0080) & .9866(.0057) & .9392(.0053) & .008(.008) & 0.009 & 1.00 \\ 
 &&  & 1.0   &  275 (2.97) & .9050(.0082) & .9879(.0055) & .9445(.0051) & .009(.008) & 0.010 & 1.00 \\  \cline{3-11} 
 && \multirow{3}{*}{4} 
     & 0.05   &288 (5.24) & .8760(.0099) & .9221(.0157) & .8972(.0115) & .023(.010) & 0.028 & 1.00 \\ 
 &&  & 0.1    &  282 (3.36) & .8841(.0089) & .9496(.0098) & .9146(.0075) & .019(.009) & 0.021 & 1.00 \\ 
 &&& 0.2      &  276 (3.9) & .8893(.0089) & .9711(.0094) & .9277(.0075) & .016(.009) & 0.018 & 1.00 \\ 
 &&  & 0.5    &  273 (2.81) & .8983(.0081) & .9881(.0053) & .9410(.0052) & .012(.008) & 0.014 & 1.00 \\ 
 &&  & 1.0    &  277 (3.24) & .9096(.0088) & .9868(.0058) & .9465(.0053) & .013(.009) & 0.015 & 1.00 \\  \hline
\multirow{10}{*}{0.4} 
 &\multirow{15}{*}{BRLVOF$_{ind}$}& \multirow{3}{*}{1} 
     & 0.05  &282 (3.85) & .6899(.0142) & .7297(.0172) & 7237(.0152)  & .048(.014) & 0.052 & 0.98 \\ 
 &&  & 0.1    &  279 (4.58) & .7277(.0136) & .7871(.0180) & .7558(.0144) & .016(.014) & 0.017 & 0.99 \\ 
 &&& 0.2      &  263 (5.4) & .7529(.0123) & .8593(.0178) & .8025(.0124) & .002(.012) & 0.002 & 1.00 \\ 
 &&  & 0.5    &  273 (2.81) & .7671(.0115) & .9351(.0137) & .8426(.0095) & .000(.012) & 0.00 & 1.00 \\ 
 &&  & 1.0    &  244 (4.53) & .7790(.0118) & .9596(.0108) & .8598(.0085) & .000(.012) & 0.00 & 1.00 \\  \cline{3-11} 
 && \multirow{3}{*}{2} 
     & 0.05  &287 (4.07) & .7095(.0142) & .7491(.0175) & .7275(.0148) & .040(.014) & 0.044 & 0.97 \\ 
 &&  & 0.1   &  271 (5.75) & .7377(.0133) & .8174(.0197) & .7754(.0146) & .018(.013) & 0.022 & 1.00 \\ 
 &&& 0.2     &  256 (5.4) & .7572(.0121) & .8878(.0171) & .8172(.0115) & .014(.012) & 0.016 & 1.00 \\ 
 &&  & 0.5   &  244 (4.73) & .7715(.0118) & .9485(.0123) & .8507(.0090) & .012(.012) & 0.014 & 1.00 \\ 
 &&  & 1.0   &  246 (4.71) & .7883(.0122) & .9605(.0108) & .8657(.0085) & .014(.012) & 0.015 & 1.00 \\ \cline{3-11} 
 && \multirow{3}{*}{4} 
     & 0.05    &276 (5.22)  & .6623(.0140) & .7454(.0197) & .7665(.0101) & .054(.014) & 0.052 & 1.00      \\
 &&  & 0.1     &281 (5.26)&   .6360(.0125) & .7449(.0178) & .7724(.0113) & .043(.013) & 0.047 & 1.00      \\
 &&& 0.2       &256 (5.39)  & .7213(.0131) & .8926(.0172) & .7976(.0116) & .022(.012) & 0.026  &0.95      \\
 &&  & 0.5     &241 (4.94)  & .7399(.0128) & .9431(.0129) & .8290(.0098) & .021(.011) & 0.023  &0.98       \\
 &&  & 1.0     &242 (5.03)  & .7642(.0125) & .9451(.0130) & .8448(.0093) & .020(.011) & 0.023  &1.00       \\ \hline
\end{tabular}
}
}
\end{table}
\FloatBarrier

\FloatBarrier
\begin{table}[]
\caption {Results for estimation of $\rho$ under BRLVOF$_{ind}$, when conditional models for $\mathbf{X}_A$ given $\mathbf{X}_B$ include $W$, and $\sigma=0.1$.Values in parentheses represent standard deviations of the corresponding quantities.} \label{tab:res2}
\centering
\resizebox{\textwidth}{!}{
{\LARGE
\begin{tabular}{cccc|ccccccc}
$\epsilon$ & Method &P & $\beta_M$ &$n$ & $\overline{TPR}$ & $\overline{PPV}$ & $\overline{F1}$ & $\overline{Bias}$ & $RMSE$ &Coverage\\ \hline
\multirow{10}{*}{0.0} 
 &\multirow{15}{*}{BRLVOF$_{ind}$}& \multirow{3}{*}{1} 
     &  0.05  &    300 (0.19) & .9993(.0009) & .9995(.0004) & .9994(.0007) & .001(.001) & 0.001 & 1.00 \\   
 &&  &  0.1   &      300 (0.19) & .9992(.0009) & .9993(.0004) & .9993(.0006) & .000(.001) & 0.001 & 1.00 \\  
 &&  &  0.2   &      300 (0.19) & .9995(.0008) & .9997(.0003) & .9996(.0005) & .000(.001) & 0.001 & 1.00 \\  
 &&  &  0.5   &      300 (0.17) & .9997(.0007) & .9998(.0002) & .9998(.0004) & .000(.001) & 0.000 & 1.00 \\  
 &&  &  1.0   &      300 (0.30) & .9998(.0006) & .9996(.0007) & .9997(.0005) & .000(.001) & 0.000 & 1.00 \\   \cline{3-11} 
 && \multirow{3}{*}{2} 
     &0.05   & 300 (0.18) & .9993(.0009) & .9995(.0004) & .9994(.0006) & .001(.001) & 0.002 & 1.00 \\ 
 &&  &0.1    &   300 (0.19) & .9996(.0008) & .9998(.0003) & .9997(.0005) & .001(.001) & 0.001 & 1.00 \\ 
 &&. &0.2    &   302 (0.18) & .9996(.0007) & .9958(.0002) & .9972(.0005) & .005(.001) & 0.005 & 0.99 \\ 
 &&  &0.5    &   300 (0.17) & .9997(.0006) & .9999(.0002) & .9998(.0004) & .000(.001) & 0.001 & 1.00 \\ 
 &&  &1.0    &   300 (0.49) & .9998(.0005) & .9990(.0015) & .9994(.0008) & .001(.000) & 0.002 & 1.00 \\ \cline{3-11} 
 && \multirow{3}{*}{4} 
& 0.05        &300 (0.20) & .9993(.0009) & .9994(.0005) & .9994(.0007) & .001(.001) & 0.002 & 1.00 \\ 
 &&  & 0.1    &  301 (1.18) & .9995(.0008) & .9976(.0024) & .9983(.0018) & .003(.001) & 0.004 & 1.00 \\ 
 &&& 0.2      &  302 (0.18) & .9996(.0007) & .9958(.0002) & .9972(.0004) & .004(.001) & 0.004 & 1.00 \\ 
 &&  & 0.5    &  300 (0.19) & .9998(.0006) & .9999(.0004) & .9998(.0004) & .000(.001) & 0.001 & 1.00 \\ 
 &&  & 1.0    &  300 (0.60) & .9998(.0004) & .9983(.0019) & .9991(.0010) & .002(.000) & 0.003 & 1.00 \\  \hline
\multirow{10}{*}{0.2} 
 &\multirow{15}{*}{BRLVOF$_{ind}$}& \multirow{3}{*}{1} 
    & 0.05  &  279 (3.86) & .8565(.0106) & .9223(.0136) & .8881(.0104) & .020(.011) & 0.023 & 1.00 \\   
 &&  & 0.1  &    276 (3.95) & .8600(.0103) & .9335(.0127) & .8952(.0095) & .013(.010) & 0.015 & 1.00 \\   
 &&& 0.2    &    269 (0.18) & .8656(.0100) & .9541(.0107) & .9076(.0082) & .005(.010) & 0.006 & 1.00 \\   
 &&  & 0.5  &    268 (3.46) & .8724(.0099) & .9774(.0074) & .9218(.0067) & .001(.010) & 0.001 & 1.00 \\  
 &&  & 1.0  &    268 (3.19) & .8804(.0094) & .9856(.0059) & .9299(.0060) & .000(.009) & 0.000 & 1.00 \\  \cline{3-11}
 
 && \multirow{3}{*}{2} 
     & 0.05  &  278 (3.92) & .8581(.0103) & .9274(.0131) & .8913(.0098) & .019(.010) & 0.022 & 1.00 \\  
 &&  & 0.1   &    275 (3.99) & .8633(.0101) & .9431(.0121) & .9014(.0090) & .015(.010) & 0.017 & 1.00 \\  
 &&& 0.2     &    271 (3.68) & .8688(.0098) & .9635(.0096) & .9136(.0076) & .010(.010) & 0.012 & 1.00 \\  
 &&  & 0.5   &    268 (3.27) & .8772(.0095) & .9821(.0066) & .9266(.0064) & .008(.010) & 0.009 & 1.00 \\  
 &&  & 1.0   &    270 (3.14) & .8876(.0088) & .9845(.0061) & .9334(.0057) & .009(.009) & 0.010 & 1.00 \\  \cline{3-11} 
 && \multirow{3}{*}{4} 
     & 0.05   &278 (3.92) & .8581(.0103) & .9274(.0131) & .8913(.0098) & .019(.010) & 0.022 & 1.00 \\ 
 &&  & 0.1    &  275 (3.99) & .8633(.0101) & .9431(.0121) & .9014(.0090) & .015(.010) & 0.017 & 1.00 \\ 
 &&& 0.2      &  269 (3.56) & .8699(.0098) & .9716(.0085) & .9178(.0071) & .014(.010) & 0.016 & 1.00 \\ 
 &&  & 0.5    &  268 (3.16) & .8802(.0092) & .9857(.0060) & .9299(.0060) & .013(.009) & 0.015 & 1.00 \\ 
 &&  & 1.0    &  272 (3.16) & .8922(.0086) & .9839(.0064) & .9357(.0055) & .014(.009) & 0.015 & 1.00 \\  \hline
\multirow{10}{*}{0.4} 
 &\multirow{15}{*}{BRLVOF$_{ind}$}& \multirow{3}{*}{1} 
     & 0.05   &  272 (5.05) & .6792(.0149) & .7501(.0198) & .7128(.0159) & .038(.015) & 0.044 & 1.00 \\  
 &&  & 0.1    &    272 (5.28) & .6892(.0145) & .7673(.0200) & .7251(.0154) & .035(.015) & 0.039 & 0.98 \\  
 &&& 0.2      &    259 (5.89) & .7080(.0137) & .8250(.0201) & .7614(.0140) & .016(.014) & 0.018 & 0.99 \\ 
 &&  & 0.5    &    239 (5.49) & .7264(.0129) & .9138(.0159) & .8092(.0109) & .002(.013) & 0.002 & 1.00 \\ 
 &&  & 1.0    &    234 (4.98) & .7396(.0127) & .9469(.0129) & .8303(.0097) & .000(.013) & 0.001 & 1.00 \\  \cline{3-11} 
 && \multirow{3}{*}{2} 
     & 0.05  & 272 (5.2) & .6888(.0145) & .7619(.0201) & .7229(.0157) & .037(.015) & 0.044 & 1.00 \\ 
 &&  & 0.1   &   263 (5.67) & .7042(.0138) & .8041(.0204) & .7507(.0148) & .026(.014) & 0.031 & 1.00 \\ 
 &&& 0.2     &   249 (5.82) & .7208(.0129) & .8676(.0191) & .7872(.0127) & .018(.013) & 0.021 & 1.00 \\ 
 &&  & 0.5   &   236 (5.15) & .7360(.0129) & .9356(.0139) & .8236(.0102) & .014(.013) & 0.016 & 1.00 \\ 
 &&  & 1.0   &   238 (4.89) & .7538(.0124) & .9522(.0122) & .8412(.0092) & .015(.012) & 0.017 & 1.00 \\ \cline{3-11} 
 && \multirow{3}{*}{4} 
     & 0.05    &268 (5.20) & .6888(.0145) & .7619(.0201) & .7229(.0157) & .037(.015) & 0.044 & 1.00 \\ 
 &&  & 0.1     &  255 (5.67) & .7042(.0138) & .8041(.0204) & .7507(.0148) & .026(.014) & 0.031 & 1.00 \\ 
 &&& 0.2       &  243 (5.62) & .7227(.0130) & .8909(.0172) & .7979(.0116) & .022(.013) & 0.026 & 1.00 \\ 
 &&  & 0.5     &  236 (4.99) & .7413(.0128) & .9429(.0128) & .8298(.0097) & .021(.013) & 0.024 & 1.00 \\ 
 &&  & 1.0     &  242 (5.01) & .7643(.0123) & .9477(.0126) & .8459(.0091) & .021(.012) & 0.024 & 1.00 \\  \hline
\end{tabular}
}
}
\end{table}
\FloatBarrier

\FloatBarrier
\begin{table}[]
\caption {Results for estimation of $\rho$ under BRLVOF$_{ind}$, when conditional models for $\mathbf{X}_A$ given $\mathbf{X}_B$ are non-linear, and $\sigma=0.1$.Values in parentheses represent standard deviations of the corresponding quantities.} \label{tab:res2}
\centering
\resizebox{\textwidth}{!}{
{\LARGE
\begin{tabular}{cccc|ccccccc}
$\epsilon$ & Method &P & $\beta_M$ &$n$ & $\overline{TPR}$ & $\overline{PPV}$ & $\overline{F1}$ & $\overline{Bias}$ & $RMSE$ &Coverage\\ \hline
\multirow{10}{*}{0.0}  
 &\multirow{15}{*}{BRLVOF$_{ind}$}& \multirow{3}{*}{1} 
     &  0.05  &    296 (1.01) & .9868(.0035) & .9994(.0010) & .9930(.0020) & .057(.003) & 0.059 & 0.99 \\ 
 &&  &  0.1   &      296 (1.01) & .9869(.0035) & .9993(.0011) & .9930(.0020) & .050(.003) & 0.051 & 0.99 \\  
 &&  &  0.2   &      296 (0.99) & .9872(.0033) & .9991(.0011) & .9931(.0019) & .037(.003) & 0.038 & 0.99 \\  
 &&  &  0.5   &      297 (0.93) & .9884(.0031) & .9995(.0008) & .9939(.0017) & .017(.003) & 0.017 & 0.99 \\  
 &&  &  1.0   &      297 (1.01) & .9899(.0027) & .9984(.0020) & .9941(.0018) & .006(.003) & 0.006 & 0.99 \\   \cline{3-11} 
 && \multirow{3}{*}{2} 
     &0.05   & 298 (0.75) & .9933(.0026) & .9993(.0007) & .9963(.0014) & .030(.003) & 0.034 & 1.00 \\ 
 &&  &0.1    &   298 (0.76) & .9934(.0026) & .9993(.0007) & .9963(.0015) & .027(.003) & 0.03 & 1.00 \\  
 &&. &0.2    &   298 (0.76) & .9934(.0026) & .9993(.0008) & .9963(.0015) & .022(.003) & 0.024 & 1.00 \\  
 &&  &0.5    &   298 (0.79) & .9937(.0024) & .9991(.0012) & .9964(.0014) & .013(.002) & 0.014 & 1.00 \\  
 &&  &1.0    &   299 (1.08) & .9943(.0021) & .9968(.0029) & .9955(.0019) & .009(.002) & 0.01 & 1.00 \\  \cline{3-11} 
 && \multirow{3}{*}{4} 
& 0.05        &  299 (0.6) & .9968(.0021) & .9993(.0007) & .9981(.0012) & .014(.002) & 0.018 & 1.00 \\ 
 &&  & 0.1    &    301 (0.6) & .9968(.0020) & .9953(.0007) & .9955(.0012) & .015(.002) & 0.018 & 1.00 \\ 
 &&& 0.2      &    299 (0.64) & .9967(.0021) & .9991(.0009) & .9979(.0012) & .011(.002) & 0.013 & 1.00 \\ 
 &&  & 0.5    &    300 (0.79) & .9970(.0020) & .9984(.0018) & .9977(.0014) & .008(.002) & 0.01 & 1.00 \\ 
 &&  & 1.0    &    300 (1.09) & .9973(.0018) & .9962(.0031) & .9967(.0019) & .007(.002) & 0.009 & 1.00 \\  \hline
\multirow{10}{*}{0.2} 
 &\multirow{15}{*}{BRLVOF$_{ind}$}& \multirow{3}{*}{1} 
    & 0.05  &   259 (4.40) & .8254(.0114) & .9549(.0115) & .8853(.0087) & .082(.011) & 0.084 & 0.98 \\  
 &&  & 0.1  &     259 (4.30) & .8272(.0113) & .9589(.0109) & .8880(.0086) & .072(.011) & 0.073 & 0.98 \\   
 &&& 0.2    &     258 (4.30) & .8293(.0116) & .9662(.0098) & .8924(.0082) & .054(.012) & 0.055 & 0.99 \\   
 &&  & 0.5  &     256 (4.07) & .8352(.0114) & .9788(.0077) & .9012(.0074) & .024(.011) & 0.024 & 0.97 \\  
 &&  & 1.0  &     259 (3.98) & .8467(.0109) & .9795(.0076) & .9081(.0070) & .009(.011) & 0.009 & 0.97 \\  \cline{3-11}
 
 && \multirow{3}{*}{2} 
     & 0.05  &  256 (4.3) & .8229(.0115) & .9642(.0101) & .8878(.0082) & .054(.011) & 0.058 & 1.00 \\ 
 &&  & 0.1   &    258 (4.19) & .8242(.0114) & .9621(.0097) & .8870(.0081) & .050(.011) & 0.053 & 1.00 \\   
 &&& 0.2     &    256 (4.16) & .8275(.0114) & .9714(.0090) & .8936(.0078) & .038(.011) & 0.041 & 1.00 \\   
 &&  & 0.5   &    256 (4.04) & .8356(.0113) & .9781(.0078) & .9011(.0074) & .025(.011) & 0.027 & 1.00 \\   
 &&  & 1.0   &    262 (4.24) & .8508(.0111) & .9729(.0088) & .9076(.0073) & .020(.011) & 0.022 & 1.00 \\  \cline{3-11} 
 && \multirow{3}{*}{4} 
     & 0.05   & 261 (5.19) & .8158(.0117) & .9514(.0114) & .8757(.0095) & .042(.012) & 0.047 & 1.00 \\ 
 &&  & 0.1    &   263 (3.98) & .8176(.0117) & .9504(.0088) & .8760(.0081) & .037(.012) & 0.041 & 1.00 \\  
 &&& 0.2      &   257 (5.23) & .8214(.0114) & .9648(.0108) & .8861(.0090) & .031(.011) & 0.035 & 1.00 \\  
 &&  & 0.5    &   256 (4.09) & .8323(.0111) & .9742(.0083) & .8975(.0074) & .024(.011) & 0.026 & 1.00 \\  
 &&  & 1.0    &   263 (4.39) & .8486(.0110) & .9695(.0094) & .9049(.0073) & .020(.011) & 0.023 & 1.00 \\  \hline
\multirow{10}{*}{0.4} 
 &\multirow{15}{*}{BRLVOF$_{ind}$}& \multirow{3}{*}{1} 
     & 0.05   &  235 (6.53) & .6460(.0142) & .8195(.0218) & .7370(.0146) & .104(.014) & 0.107 & 0.98 \\ 
 &&  & 0.1    &    234 (6.37) & .6635(.0140) & .8524(.0213) & .7459(.0138) & .082(.014) & 0.084 & 0.99 \\   
 &&& 0.2      &    228 (6.35) & .6689(.0140) & .8805(.0194) & .7600(.0125) & .063(.014) & 0.064 & 0.99 \\  
 &&  & 0.5    &    220 (5.97) & .6768(.0144) & .9251(.0159) & .7814(.0112) & .030(.014) & 0.03 & 0.97 \\  
 &&  & 1.0    &    222 (5.76) & .6936(.0143) & .9373(.0144) & .7969(.0108) & .012(.014) & 0.012 & 0.96 \\  \cline{3-11} 
 && \multirow{3}{*}{2} 
     & 0.05  & 223 (6.93) & .6465(.0146) & .8710(.0215) & .7418(.0131) & .066(.015) & 0.072 & 1.00 \\ 
 &&  & 0.1   &   221 (6.69) & .6496(.0145) & .8807(.0201) & .7474(.0126) & .058(.014) & 0.064 & 1.00 \\  
 &&& 0.2     &   218 (6.62) & .6537(.0147) & .8994(.0191) & .7568(.0122) & .046(.015) & 0.051 & 1.00 \\  
 &&  & 0.5   &   216 (6.25) & .6667(.0148) & .9248(.0164) & .7745(.0114) & .031(.015) & 0.034 & 1.00 \\  
 &&  & 1.0   &   225 (6.3) & .6902(.0144) & .9225(.0166) & .7893(.0111) & .027(.014) & 0.03 & 1.00 \\ \cline{3-11} 
 && \multirow{3}{*}{4} 
     & 0.05    & 216 (6.48) & .6363(.0146) & .8906(.0198) & .7410(.0125) & .056(.015) & 0.061 & 1.00 \\ 
 &&  & 0.1     &   216 (6.4) & .6392(.0144) & .8958(.0190) & .7448(.0120) & .051(.014) & 0.057 & 1.00 \\ 
 &&& 0.2       &   212 (6.28) & .6447(.0144) & .9111(.0182) & .7548(.0119) & .043(.014) & 0.048 & 1.00 \\ 
 &&  & 0.5     &   215 (6.24) & .6622(.0145) & .9224(.0169) & .7706(.0113) & .033(.015) & 0.037 & 1.00 \\ 
 &&  & 1.0     &   225 (6.41) & .6880(.0143) & .9190(.0170) & .7865(.0110) & .028(.014) & 0.033 & 1.00 \\  \hline
\end{tabular}
}
}
\end{table}
\FloatBarrier

\FloatBarrier
\begin{table}[]
\caption {Results for estimation of $\rho$ under BRLVOF$_{ind}$, when conditional models for $\mathbf{X}_A$ given $\mathbf{X}_B$ are linear, and $\sigma=0.5$.Values in parentheses represent standard deviations of the corresponding quantities.} 
\centering
\resizebox{\textwidth}{!}{
{\LARGE
\begin{tabular}{cccc|ccccccc}
$\epsilon$ & Method &P & $\beta_M$ &$n$ & $\overline{TPR}$ & $\overline{PPV}$ & $\overline{F1}$ & $\overline{Bias}$ & $RMSE$ &Coverage\\ \hline
\multirow{10}{*}{0.0} 
 &\multirow{15}{*}{BRLVOF$_{ind}$}& \multirow{3}{*}{1} 
     &  0.05  &  300 (0.28) & .9990(.0012) & .9993(.0006) & .9992(.0009) & .001(.001) & 0.002 & 1.00 \\ 
 &&  &  0.1   &    300 (0.28) & .9991(.0012) & .9994(.0006) & .9993(.0009) & .001(.001) & 0.002 & 1.00 \\  
 &&  &  0.2   &    300 (0.27) & .9991(.0012) & .9994(.0005) & .9993(.0008) & .001(.001) & 0.001 & 1.00 \\  
 &&  &  0.5   &    300 (0.26) & .9995(.0010) & .9997(.0004) & .9996(.0007) & .000(.001) & 0 & 1.00 \\  
 &&  &  1.0   &    300 (0.43) & .9996(.0009) & .9993(.0012) & .9995(.0008) & .000(.001) & 0 & 1.00 \\   \cline{3-11} 
 && \multirow{3}{*}{2} 
     &0.05   & 300 (0.28) & .9990(.0012) & .9993(.0005) & .9992(.0008) & .001(.001) & 0.003 & 1.00 \\ 
 &&  &0.1    &   300 (0.27) & .9992(.0011) & .9995(.0004) & .9993(.0007) & .001(.001) & 0.002 & 1.00 \\ 
 &&. &0.2    &   300 (0.27) & .9992(.0011) & .9995(.0004) & .9994(.0007) & .001(.001) & 0.002 & 1.00 \\ 
 &&  &0.5    &   300 (0.26) & .9995(.0009) & .9996(.0004) & .9995(.0006) & .000(.001) & 0.001 & 1.00 \\ 
 &&  &1.0    &   300 (0.63) & .9996(.0008) & .9982(.0020) & .9989(.0011) & .002(.001) & 0.003 & 1.00 \\  \cline{3-11} 
 && \multirow{3}{*}{4} 
& 0.05        &300 (0.29) & .9992(.0012) & .9995(.0005) & .9994(.0008) & .001(.001) & 0.002 & 1.00 \\ 
 &&  & 0.1    &  300 (0.28) & .9993(.0011) & .9996(.0004) & .9994(.0007) & .001(.001) & 0.002 & 1.00 \\ 
 &&& 0.2      &  300 (0.27) & .9995(.0010) & .9998(.0003) & .9996(.0006) & .000(.001) & 0.001 & 1.00 \\ 
 &&  & 0.5    &  300 (0.31) & .9997(.0008) & .9996(.0007) & .9997(.0006) & .001(.001) & 0.002 & 1.00 \\ 
 &&  & 1.0    &  301 (0.73) & .9998(.0007) & .9974(.0024) & .9986(.0013) & .003(.001) & 0.004 & 1.00 \\ \hline
\multirow{10}{*}{0.2} 
 &\multirow{15}{*}{BRLVOF$_{ind}$}& \multirow{3}{*}{1} 
    & 0.05  & 265 (4.38) & .8368(.0113) & .9477(.0121) & .8887(.0091) & .022(.011) & 0.026 & 1.00 \\  
 &&  & 0.1  &   265 (4.36) & .8376(.0112) & .9492(.0120) & .8898(.0090) & .020(.011) & 0.024 & 1.00 \\   
 &&& 0.2    &   263 (4.26) & .8395(.0110) & .9563(.0108) & .8940(.0083) & .015(.011) & 0.017 & 1.00 \\   
 &&  & 0.5  &   261 (4.06) & .8452(.0110) & .9718(.0088) & .9039(.0075) & .005(.011) & 0.006 & 1.00 \\  
 &&  & 1.0  &   261 (3.86) & .8539(.0108) & .9803(.0072) & .9126(.0069) & .002(.011) & 0.002 & 1.00 \\  \cline{3-11}
 
 && \multirow{3}{*}{2} 
     & 0.05  & 265 (4.38) & .8362(.0112) & .9473(.0121) & .8882(.0091) & .026(.011) & 0.029 & 1.00 \\ 
 &&  & 0.1   &   264 (4.35) & .8378(.0113) & .9509(.0116) & .8907(.0088) & .023(.011) & 0.026 & 1.00 \\  
 &&& 0.2     &   263 (4.16) & .8405(.0109) & .9604(.0103) & .8964(.0081) & .017(.011) & 0.019 & 1.00 \\  
 &&  & 0.5   &   261 (3.87) & .8494(.0110) & .9765(.0077) & .9084(.0072) & .011(.011) & 0.012 & 1.00 \\  
 &&  & 1.0   &   265 (3.93) & .8631(.0106) & .9785(.0076) & .9170(.0068) & .011(.011) & 0.012 & 1.00 \\  \cline{3-11} 
 && \multirow{3}{*}{4} 
     & 0.05   &  267 (4.31) & .8342(.0112) & .9427(.0118) & .8844(.0090) & .024(.011) & 0.027 & 1.00 \\ 
 &&  & 0.1    &    271 (4.14) & .8359(.0113) & .9391(.0110) & .8825(.0087) & .025(.011) & 0.028 & 1.00 \\ 
 &&& 0.2      &    264 (4.07) & .8413(.0110) & .9604(.0095) & .8961(.0078) & .020(.011) & 0.022 & 1.00 \\ 
 &&  & 0.5    &    262 (3.78) & .8538(.0109) & .9786(.0072) & .9118(.0071) & .016(.011) & 0.018 & 1.00 \\ 
 &&  & 1.0    &    267 (3.75) & .8709(.0101) & .9770(.0076) & .9208(.0066) & .017(.010) & 0.019 & 1.00 \\ 
 \hline
\multirow{10}{*}{0.4} 
 &\multirow{15}{*}{BRLVOF$_{ind}$}& \multirow{3}{*}{1} 
     & 0.05   &  243 (6.7) & .6516(.0145) & .7966(.0221) & .7313(.0153) & .054(.015) & 0.061 & 0.99 \\  
 &&  & 0.1    &    244 (6.6) & .6671(.0145) & .8194(.0223) & .7352(.0149) & .036(.014) & 0.042 & 1.00 \\  
 &&& 0.2      &    242 (6.53) & .6720(.0143) & .8378(.0213) & .7449(.0140) & .031(.014) & 0.035 & 0.99 \\ 
 &&  & 0.5    &    228 (6.18) & .6836(.0142) & .9015(.0176) & .7773(.0118) & .008(.014) & 0.009 & 1.00 \\ 
 &&  & 1.0    &    225 (5.76) & .6987(.0142) & .9329(.0149) & .7987(.0108) & .003(.014) & 0.003 & 1.00 \\  \cline{3-11} 
 && \multirow{3}{*}{2} 
     & 0.05  &  245 (6.62) & .6610(.0150) & .8098(.0228) & .7277(.0156) & .041(.015) & 0.048 & 1.00 \\ 
 &&  & 0.1   &    243 (6.73) & .6650(.0146) & .8215(.0227) & .7348(.0151) & .036(.015) & 0.042 & 1.00 \\ 
 &&& 0.2     &    236 (6.63) & .6729(.0143) & .8551(.0206) & .7529(.0134) & .026(.014) & 0.031 & 1.00 \\ 
 &&  & 0.5   &    225 (5.99) & .6872(.0142) & .9148(.0163) & .7846(.0113) & .017(.014) & 0.02 & 1.00 \\ 
 &&  & 1.0   &    229 (5.81) & .7110(.0140) & .9317(.0147) & .8062(.0106) & .017(.014) & 0.02 & 1.00 \\  \cline{3-11} 
 && \multirow{3}{*}{4} 
     & 0.05    &245 (6.62) & .6610(.0150) & .8098(.0228) & .7277(.0156) & .041(.015) & 0.048 & 1.00 \\ 
 &&  & 0.1     &  243 (6.73) & .6650(.0146) & .8215(.0227) & .7348(.0151) & .036(.015) & 0.042 & 1.00 \\ 
 &&& 0.2       &  232 (6.4) & .6779(.0139) & .8787(.0195) & .7651(.0124) & .030(.014) & 0.034 & 1.00 \\ 
 &&  & 0.5     &  225 (5.79) & .6976(.0140) & .9294(.0151) & .7967(.0107) & .024(.014) & 0.027 & 1.00 \\ 
 &&  & 1.0     &  233 (5.63) & .7258(.0134) & .9353(.0144) & .8170(.0101) & .022(.013) & 0.026 & 1.00 \\  \hline
\end{tabular}
}
}
\end{table}
\FloatBarrier

\FloatBarrier
\begin{table}[]
\caption {Results for estimation of $\rho$ under BRLVOF$_{ind}$, when conditional models for $\mathbf{X}_A$ given $\mathbf{X}_B$ include $W$, and $\sigma=0.5$.Values in parentheses represent standard deviations of the corresponding quantities.} 
\centering
\resizebox{\textwidth}{!}{
{\LARGE
\begin{tabular}{cccc|ccccccc}
$\epsilon$ & Method &P & $\beta_M$ &$n$ & $\overline{TPR}$ & $\overline{PPV}$ & $\overline{F1}$ & $\overline{Bias}$ & $RMSE$ &Coverage\\ \hline
\multirow{10}{*}{0.0}  
 &\multirow{15}{*}{BRLVOF$_{ind}$}& \multirow{3}{*}{1} 
     &  0.05  &    300 (0.29) & .9992(.0012) & .9995(.0005) & .9993(.0008) & .001(.001) & 0.002 & 1.00 \\  
 &&  &  0.1   &      300 (0.29) & .9990(.0012) & .9993(.0005) & .9992(.0008) & .001(.001) & 0.002 & 1.00 \\   
 &&  &  0.2   &      300 (0.29) & .9992(.0012) & .9995(.0004) & .9993(.0008) & .001(.001) & 0.002 & 1.00 \\   
 &&  &  0.5   &      300 (0.28) & .9994(.0011) & .9997(.0003) & .9995(.0006) & .000(.001) & 0.001 & 1.00 \\   
 &&  &  1.0   &      300 (0.44) & .9996(.0010) & .9992(.0012) & .9994(.0008) & .000(.001) & 0 & 1.00 \\   \cline{3-11} 
 && \multirow{3}{*}{2} 
     &0.05   & 300 (0.28) & .9992(.0011) & .9995(.0004) & .9994(.0007) & .001(.001) & 0.002 & 1.00 \\ 
 &&  &0.1    &   300 (0.28) & .9993(.0011) & .9996(.0004) & .9995(.0007) & .001(.001) & 0.002 & 1.00 \\  
 &&. &0.2    &   300 (0.27) & .9994(.0011) & .9996(.0004) & .9995(.0007) & .001(.001) & 0.002 & 1.00 \\  
 &&  &0.5    &   300 (0.28) & .9996(.0009) & .9998(.0004) & .9997(.0006) & .001(.001) & 0.001 & 1.00 \\  
 &&  &1.0    &   300 (0.63) & .9997(.0008) & .9983(.0020) & .9990(.0011) & .002(.001) & 0.003 & 1.00 \\ \cline{3-11} 
 && \multirow{3}{*}{4} 
& 0.05        & 300 (0.28) & .9990(.0012) & .9993(.0005) & .9991(.0008) & .001(.001) & 0.002 & 1.00 \\ 
 &&  & 0.1    &   300 (0.28) & .9991(.0012) & .9994(.0005) & .9993(.0008) & .001(.001) & 0.002 & 1.00 \\ 
 &&& 0.2      &   300 (0.27) & .9994(.0011) & .9996(.0004) & .9995(.0007) & .001(.001) & 0.001 & 1.00 \\ 
 &&  & 0.5    &   300 (0.32) & .9995(.0009) & .9995(.0008) & .9995(.0007) & .001(.001) & 0.002 & 1.00 \\ 
 &&  & 1.0    &   301 (0.77) & .9997(.0007) & .9975(.0025) & .9986(.0013) & .003(.001) & 0.004 & 1.00 \\  \hline
\multirow{10}{*}{0.2} 
 &\multirow{15}{*}{BRLVOF$_{ind}$}& \multirow{3}{*}{1} 
    & 0.05  &  264 (4.36) & .8339(.0110) & .9488(.0121) & .8875(.0089) & .026(.011) & 0.029 & 1.00 \\  
 &&  & 0.1  &    264 (4.41) & .8350(.0112) & .9505(.0118) & .8889(.0088) & .023(.011) & 0.026 & 1.00 \\    
 &&& 0.2    &    262 (4.35) & .8364(.0112) & .9563(.0110) & .8922(.0085) & .017(.011) & 0.019 & 1.00 \\    
 &&  & 0.5  &    260 (4.04) & .8418(.0112) & .9716(.0088) & .9019(.0077) & .006(.011) & 0.007 & 1.00 \\   
 &&  & 1.0  &    261 (3.86) & .8516(.0110) & .9790(.0074) & .9107(.0071) & .002(.011) & 0.002 & 1.00 \\   \cline{3-11}
 
 && \multirow{3}{*}{2} 
     & 0.05  &  266 (4.36) & .8347(.0112) & .9459(.0120) & .8861(.0089) & .027(.011) & 0.03 & 1.00 \\ 
 &&  & 0.1   &    263 (4.27) & .8360(.0110) & .9533(.0114) & .8907(.0086) & .022(.011) & 0.025 & 1.00 \\ 
 &&& 0.2     &    262 (4.19) & .8392(.0110) & .9613(.0101) & .8960(.0080) & .017(.011) & 0.019 & 1.00 \\ 
 &&  & 0.5   &    260 (3.94) & .8473(.0111) & .9762(.0079) & .9071(.0074) & .011(.011) & 0.013 & 1.00 \\ 
 &&  & 1.0   &    265 (3.88) & .8625(.0106) & .9771(.0076) & .9161(.0069) & .011(.011) & 0.013 & 1.00 \\  \cline{3-11} 
 && \multirow{3}{*}{4} 
     & 0.05   &266 (4.36) & .8347(.0112) & .9459(.0120) & .8861(.0089) & .027(.011) & 0.03 & 1.00 \\ 
 &&  & 0.1    &  263 (4.27) & .8360(.0110) & .9533(.0114) & .8907(.0086) & .022(.011) & 0.025 & 1.00 \\  
 &&& 0.2      &  260 (4.18) & .8384(.0111) & .9663(.0098) & .8977(.0079) & .018(.011) & 0.021 & 1.00 \\  
 &&  & 0.5    &  261 (3.87) & .8517(.0109) & .9798(.0073) & .9112(.0070) & .015(.011) & 0.017 & 1.00 \\  
 &&  & 1.0    &  267 (3.85) & .8688(.0102) & .9775(.0080) & .9198(.0067) & .016(.010) & 0.018 & 1.00 \\  \hline
\multirow{10}{*}{0.4} 
 &\multirow{15}{*}{BRLVOF$_{ind}$}& \multirow{3}{*}{1} 
     & 0.05   &  243 (6.77) & .6575(.0150) & .8131(.0233) & .7269(.0156) & .040(.015) & 0.047 & 1.00 \\ 
 &&  & 0.1    &    244 (6.76) & .6592(.0150) & .8146(.0230) & .7278(.0154) & .038(.015) & 0.044 & 1.00 \\  
 &&& 0.2      &    238 (6.76) & .6645(.0147) & .8393(.0217) & .7415(.0143) & .027(.015) & 0.031 & 1.00 \\ 
 &&  & 0.5    &    226 (6.31) & .6756(.0144) & .8966(.0182) & .7703(.0121) & .009(.014) & 0.01 & 1.00 \\ 
 &&  & 1.0    &    223 (6.06) & .6905(.0147) & .9305(.0154) & .7924(.0111) & .003(.015) & 0.003 & 1.00 \\  \cline{3-11} 
 && \multirow{3}{*}{2} 
     & 0.05  & 243 (6.77) & .6575(.0150) & .8131(.0233) & .7269(.0156) & .040(.015) & 0.047 & 1.00 \\ 
 &&  & 0.1   &   244 (6.76) & .6592(.0150) & .8146(.0230) & .7278(.0154) & .038(.015) & 0.044 & 1.00 \\ 
 &&& 0.2     &   234 (6.65) & .6747(.0141) & .8645(.0209) & .7576(.0132) & .026(.014) & 0.031 & 1.00 \\ 
 &&  & 0.5   &   224 (6.02) & .6869(.0142) & .9200(.0164) & .7862(.0113) & .017(.014) & 0.02 & 1.00 \\ 
 &&  & 1.0   &   228 (5.79) & .7095(.0138) & .9351(.0150) & .8065(.0105) & .017(.014) & 0.019 & 1.00 \\  \cline{3-11} 
 && \multirow{3}{*}{4} 
     & 0.05    &243 (6.77) & .6575(.0150) & .8131(.0233) & .7269(.0156) & .040(.015) & 0.047 & 1.00 \\ 
 &&  & 0.1     &  244 (6.76) & .6592(.0150) & .8146(.0230) & .7278(.0154) & .038(.015) & 0.044 & 1.00 \\ 
 &&& 0.2       &  230 (6.57) & .6723(.0144) & .8766(.0196) & .7607(.0126) & .028(.014) & 0.032 & 1.00 \\ 
 &&  & 0.5     &  224 (5.89) & .6924(.0144) & .9263(.0157) & .7921(.0113) & .025(.014) & 0.028 & 1.00 \\ 
 &&  & 1.0     &  233 (5.83) & .7214(.0138) & .9306(.0150) & .8124(.0105) & .023(.014) & 0.027 & 1.00 \\ \hline
\end{tabular}
}
}
\end{table}
\FloatBarrier

\FloatBarrier
\begin{table}[]
\caption {Results for estimation of $\rho$ under BRLVOF$_{ind}$, when conditional models for $\mathbf{X}_A$ given $\mathbf{X}_B$ are non-linear, and $\sigma=0.5$.Values in parentheses represent standard deviations of the corresponding quantities.}
\centering
\resizebox{\textwidth}{!}{
{\LARGE
\begin{tabular}{cccc|ccccccc}
$\epsilon$ & Method &P & $\beta_M$ &$n$ & $\overline{TPR}$ & $\overline{PPV}$ & $\overline{F1}$ & $\overline{Bias}$ & $RMSE$ &Coverage\\ \hline
\multirow{10}{*}{0.0} 
 &\multirow{15}{*}{BRLVOF$_{ind}$}& \multirow{3}{*}{1} 
     &  0.05  &   299 (0.59) & .9956(.0021) & .9995(.0006) & .9975(.0013) & .029(.002) & 0.032 & 1.00 \\ 
 &&  &  0.1   &     299 (0.59) & .9956(.0021) & .9994(.0007) & .9975(.0012) & .025(.002) & 0.027 & 1.00 \\  
 &&  &  0.2   &     299 (0.6) & .9957(.0021) & .9993(.0007) & .9975(.0013) & .018(.002) & 0.02 & 1.00 \\  
 &&  &  0.5   &     299 (0.59) & .9962(.0020) & .9994(.0009) & .9978(.0012) & .008(.002) & 0.009 & 1.00 \\  
 &&  &  1.0   &     300 (0.81) & .9967(.0017) & .9981(.0021) & .9974(.0015) & .003(.002) & 0.004 & 1.00 \\  \cline{3-11} 
 && \multirow{3}{*}{2} 
     &0.05   & 299 (0.57) & .9968(.0020) & .9994(.0006) & .9981(.0012) & .016(.002) & 0.02 & 1.00 \\ 
 &&  &0.1    &   299 (0.56) & .9968(.0020) & .9994(.0006) & .9981(.0011) & .015(.002) & 0.019 & 1.00 \\  
 &&. &0.2    &   299 (0.58) & .9966(.0020) & .9993(.0008) & .9979(.0012) & .013(.002) & 0.016 & 1.00 \\   
 &&  &0.5    &   299 (0.68) & .9967(.0019) & .9988(.0014) & .9978(.0013) & .008(.002) & 0.01 & 1.00 \\   
 &&  &1.0    &   300 (1.07) & .9971(.0017) & .9965(.0031) & .9968(.0018) & .006(.002) & 0.008 & 1.00 \\ \cline{3-11} 
 && \multirow{3}{*}{4} 
& 0.05        &  300 (0.56) & .9978(.0019) & .9993(.0008) & .9985(.0011) & .010(.002) & 0.013 & 1.00 \\ 
 &&  & 0.1    &    300 (0.56) & .9977(.0018) & .9992(.0009) & .9985(.0011) & .009(.002) & 0.012 & 1.00 \\ 
 &&& 0.2      &    300 (0.61) & .9976(.0019) & .9990(.0011) & .9983(.0012) & .008(.002) & 0.011 & 1.00 \\ 
 &&  & 0.5    &    300 (0.76) & .9979(.0017) & .9983(.0019) & .9981(.0014) & .006(.002) & 0.009 & 1.00 \\ 
 &&  & 1.0    &    301 (1.1) & .9982(.0016) & .9960(.0033) & .9971(.0019) & .006(.002) & 0.008 & 1.00 \\ \hline
\multirow{10}{*}{0.2}  
 &\multirow{15}{*}{BRLVOF$_{ind}$}& \multirow{3}{*}{1} 
    & 0.05  &   257 (4.29) & .8240(.0112) & .9620(.0105) & .8875(.0082) & .045(.011) & 0.049 & 1.00 \\ 
 &&  & 0.1  &     257 (4.24) & .8250(.0114) & .9635(.0102) & .8887(.0083) & .040(.011) & 0.042 & 1.00 \\  
 &&& 0.2    &     256 (4.2) & .8267(.0113) & .9676(.0096) & .8915(.0080) & .031(.011) & 0.033 & 1.00 \\  
 &&  & 0.5  &     256 (4.05) & .8324(.0112) & .9760(.0082) & .8983(.0075) & .015(.011) & 0.016 & 1.00 \\ 
 &&  & 1.0  &     259 (4.21) & .8428(.0112) & .9759(.0084) & .9043(.0072) & .006(.011) & 0.007 & 1.00 \\  \cline{3-11}
 
 && \multirow{3}{*}{2} 
     & 0.05  &  255 (4.15) & .8198(.0113) & .9661(.0096) & .8868(.0081) & .043(.011) & 0.047 & 1.00 \\ 
 &&  & 0.1   &    255 (4.17) & .8208(.0114) & .9675(.0094) & .8880(.0081) & .039(.011) & 0.043 & 1.00 \\    
 &&& 0.2     &    254 (4.15) & .8236(.0114) & .9711(.0089) & .8911(.0078) & .032(.011) & 0.035 & 1.00 \\    
 &&  & 0.5   &    256 (4.12) & .8314(.0114) & .9760(.0082) & .8977(.0075) & .022(.011) & 0.024 & 1.00 \\    
 &&  & 1.0   &    262 (4.41) & .8466(.0111) & .9706(.0094) & .9042(.0074) & .018(.011) & 0.02 & 1.00 \\   \cline{3-11} 
 && \multirow{3}{*}{4} 
     & 0.05   & 257 (4.87) & .8141(.0117) & .9592(.0106) & .8792(.0090) & .038(.012) & 0.042 & 1.00 \\ 
 &&  & 0.1    &   258 (4.14) & .8160(.0118) & .9596(.0090) & .8805(.0082) & .034(.012) & 0.038 & 1.00 \\   
 &&& 0.2      &   256 (4.15) & .8193(.0116) & .9664(.0086) & .8859(.0078) & .030(.012) & 0.034 & 1.00 \\   
 &&  & 0.5    &   256 (4.19) & .8304(.0113) & .9724(.0087) & .8957(.0075) & .024(.011) & 0.026 & 1.00 \\   
 &&  & 1.0    &   262 (4.49) & .8463(.0112) & .9682(.0098) & .9030(.0074) & .021(.011) & 0.023 & 1.00 \\   \hline
\multirow{10}{*}{0.4} 
 &\multirow{15}{*}{BRLVOF$_{ind}$}& \multirow{3}{*}{1} 
     & 0.05   &  225 (6.85) & .6386(.0142) & .8455(.0214) & .7422(.0136) & .066(.014) & 0.072 & 1.00 \\ 
 &&  & 0.1    &    226 (6.66) & .6542(.0147) & .8700(.0210) & .7466(.0135) & .050(.015) & 0.055 & 1.00 \\   
 &&& 0.2      &    225 (6.65) & .6566(.0148) & .8802(.0203) & .7510(.0129) & .042(.015) & 0.045 & 0.99 \\  
 &&  & 0.5    &    218 (6.24) & .6650(.0145) & .9173(.0171) & .7708(.0116) & .019(.015) & 0.02 & 1.00 \\ 
 &&  & 1.0    &    221 (6.19) & .6821(.0145) & .9264(.0163) & .7854(.0111) & .009(.014) & 0.009 & 1.00 \\  \cline{3-11} 
 && \multirow{3}{*}{2} 
     & 0.05  & 220 (6.79) & .6383(.0149) & .8760(.0212) & .7373(.0132) & .063(.015) & 0.069 & 1.00 \\ 
 &&  & 0.1   &   217 (6.69) & .6409(.0146) & .8872(.0203) & .7439(.0127) & .053(.015) & 0.059 & 1.00 \\  
 &&& 0.2     &   215 (6.62) & .6451(.0148) & .8998(.0191) & .7511(.0123) & .044(.015) & 0.049 & 1.00 \\  
 &&  & 0.5   &   215 (6.34) & .6580(.0148) & .9201(.0170) & .7669(.0116) & .031(.015) & 0.034 & 1.00 \\  
 &&  & 1.0   &   224 (6.62) & .6827(.0146) & .9158(.0177) & .7819(.0113) & .027(.015) & 0.03 & 1.00 \\  \cline{3-11} 
 && \multirow{3}{*}{4} 
     & 0.05    & 212 (6.51) & .6327(.0146) & .8976(.0193) & .7419(.0123) & .051(.015) & 0.057 & 1.00 \\ 
 &&  & 0.1     &   213 (7.49) & .6353(.0146) & .8983(.0207) & .7433(.0132) & .049(.015) & 0.054 & 1.00 \\ 
 &&& 0.2       &   214 (6.23) & .6410(.0145) & .9047(.0181) & .7491(.0119) & .042(.015) & 0.047 & 1.00 \\ 
 &&  & 0.5     &   215 (6.36) & .6579(.0145) & .9191(.0175) & .7665(.0115) & .033(.015) & 0.038 & 1.00 \\ 
 &&  & 1.0     &   224 (6.69) & .6841(.0146) & .9157(.0178) & .7827(.0112) & .028(.015) & 0.032 & 1.00 \\  \hline
\end{tabular}
}
}
\end{table}
\FloatBarrier

\section{Convergence diagnostics for real data analysis}
\begin{figure} [H]
\centering
\begin{subfigure}{1.06\textwidth}
  \centering
  \includegraphics[width=.45\textwidth]{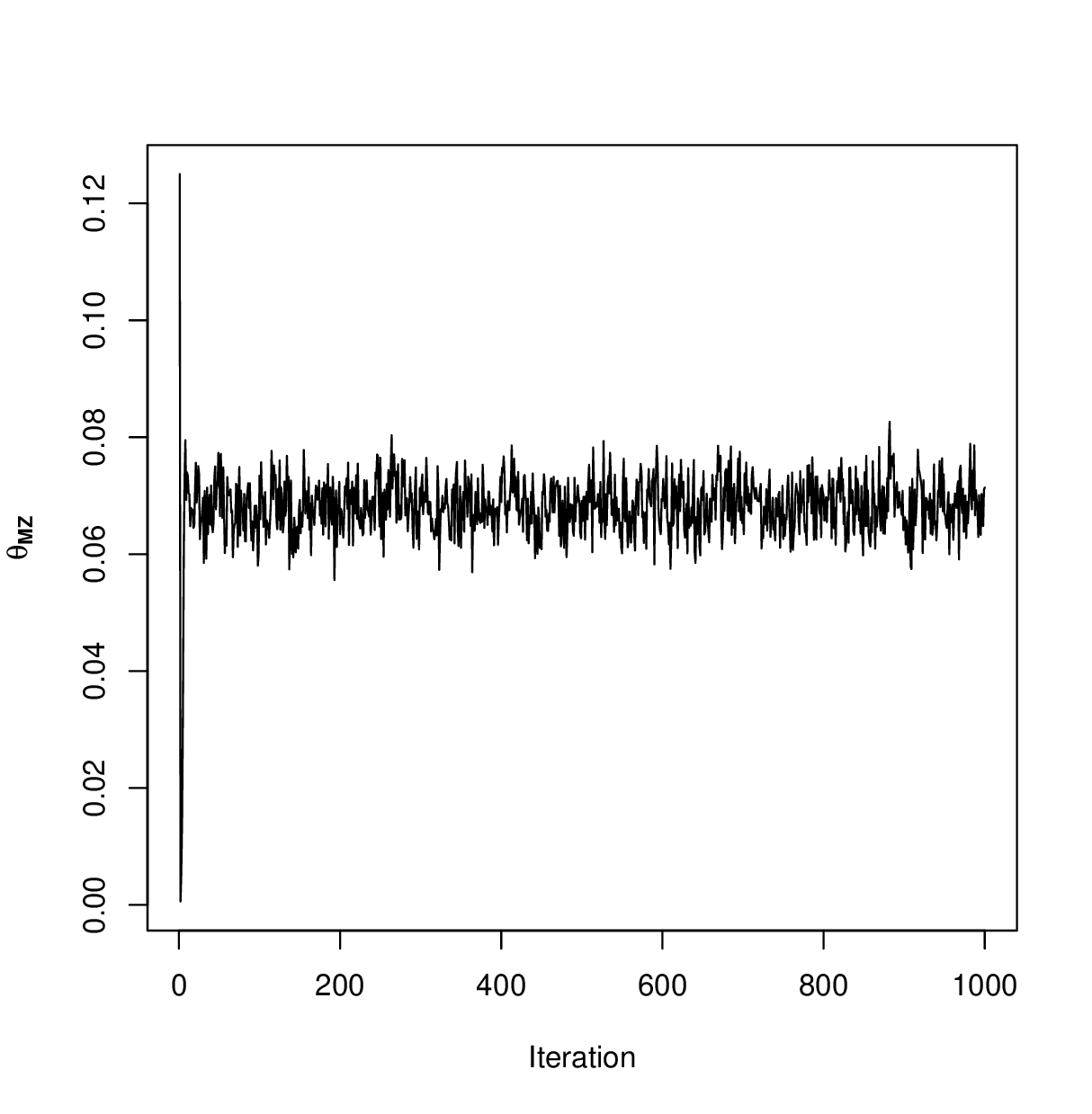}
  \includegraphics[width=.45\textwidth]{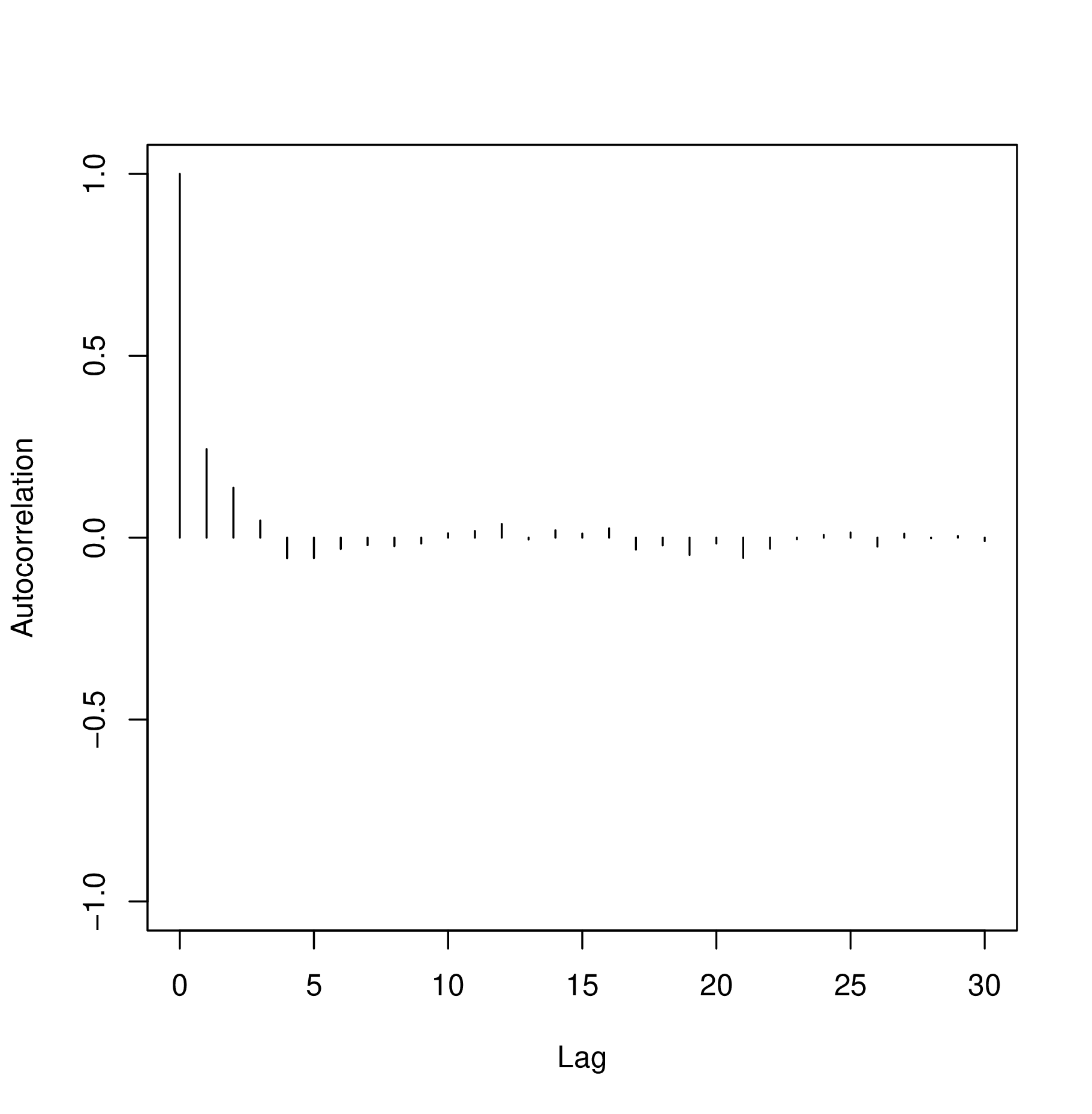} \\
   \includegraphics[width=.45\textwidth]{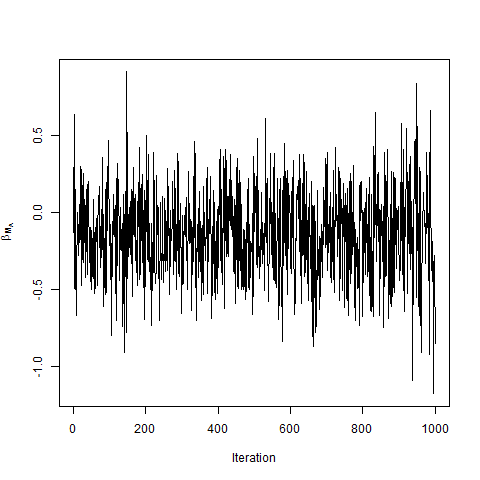}
     \includegraphics[width=.45\textwidth]{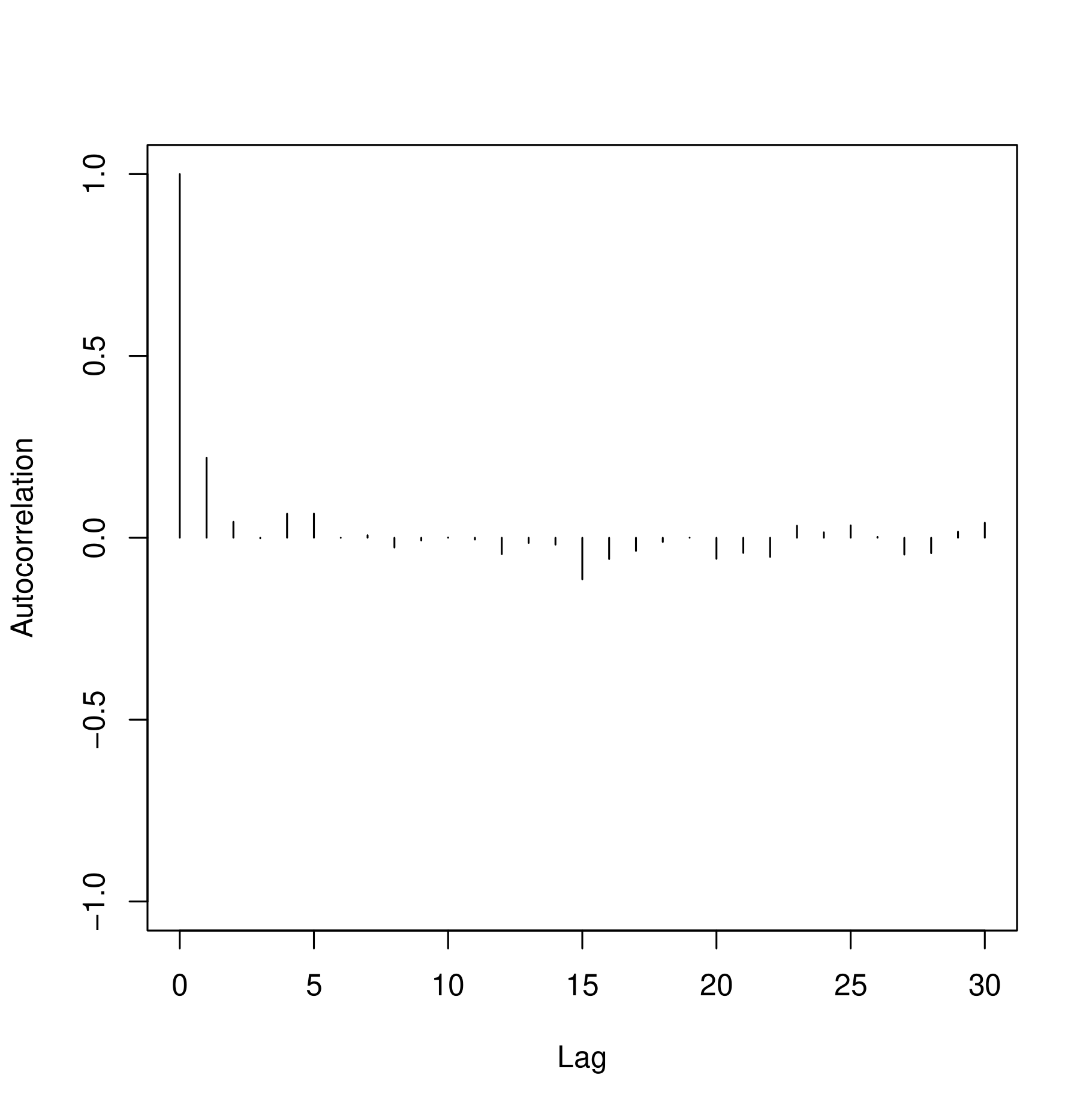}
  \label{fig:sub1}
\end{subfigure}
\caption{The upper panel depicts the trace and auto-correlation plots under BRLVOF$^{NS}$, for the component of $\theta_{MZ}$ that describes agreement on the first five digits of the zip code. The lower panel depicts the trace and auto-correlation plots under BRLVOF$^{NS}$, for the component of $\beta_M$ that describes the correlation between the ADL score and anaemia.}
\label{fig:test}
\end{figure}
\FloatBarrier

\begin{figure} [H]
\centering
\begin{subfigure}{1.06\textwidth}
  \centering
  \includegraphics[width=.45\textwidth]{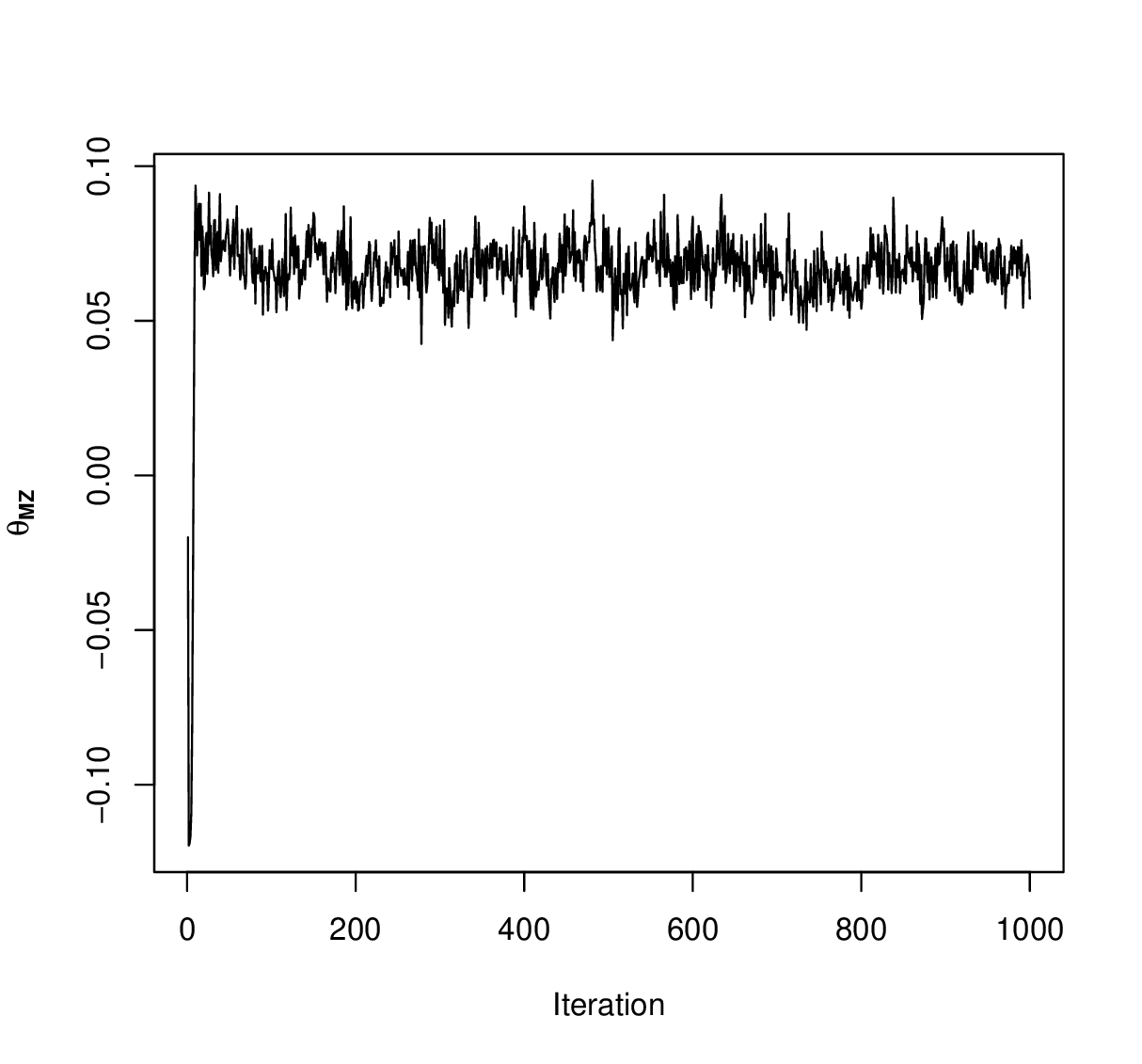}
  \includegraphics[width=.45\textwidth]{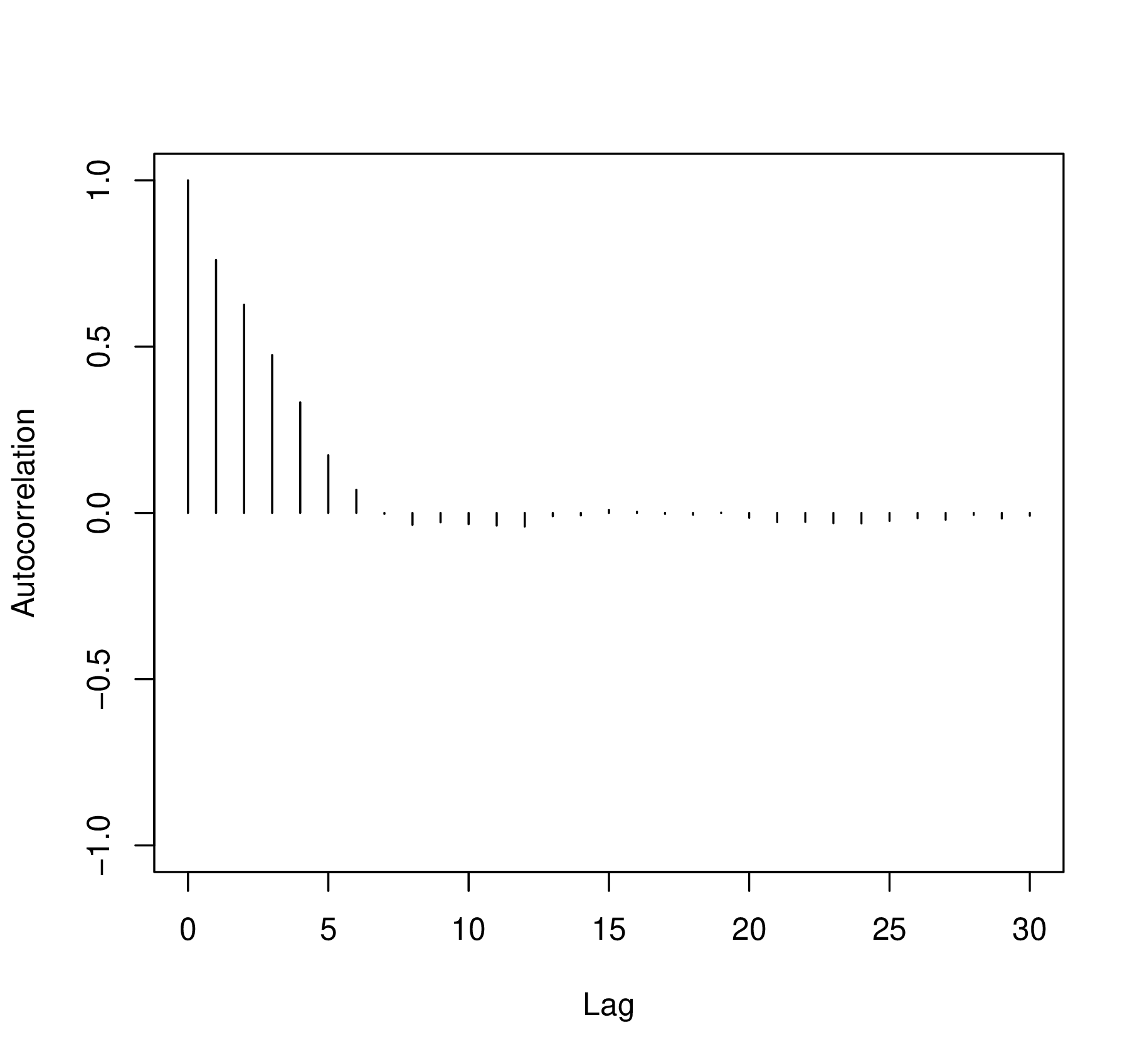} \\
   \includegraphics[width=.45\textwidth]{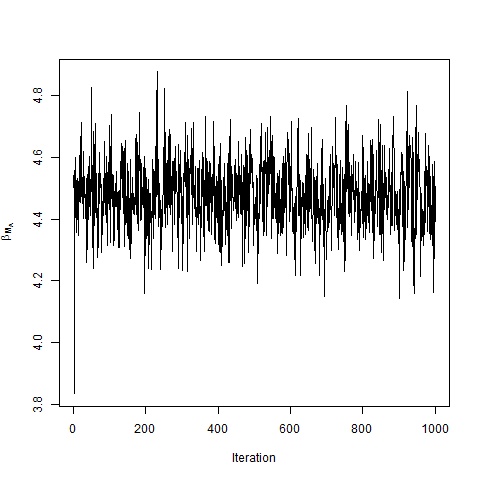}
     \includegraphics[width=.45\textwidth]{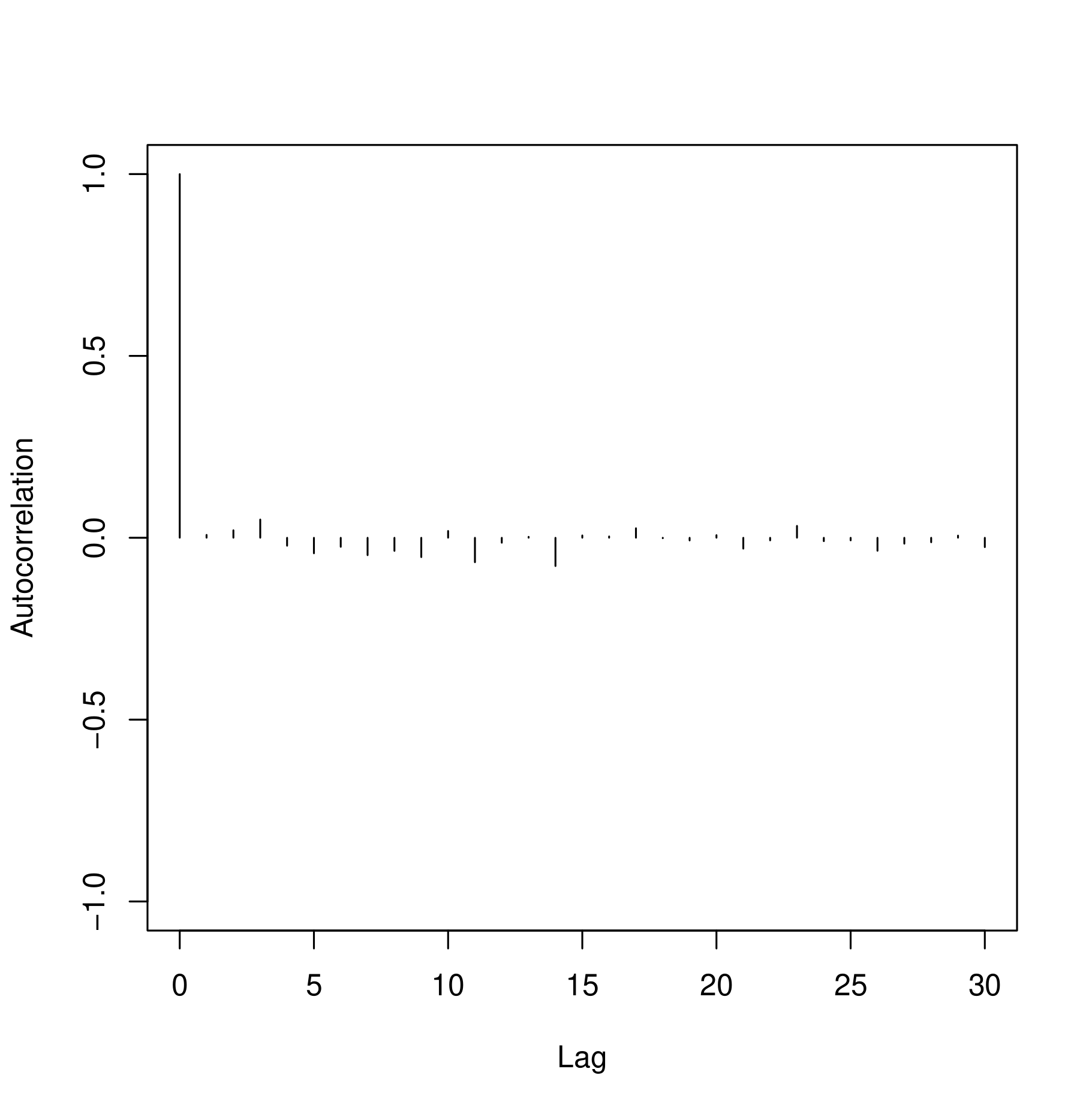}
  \label{fig:sub1}
\end{subfigure}
\caption{The upper panel depicts the trace and auto-correlation plots under BRLVOF$^{S}$, for the component of $\theta_{MZ}$ that describes agreement on the first five digits of the zip code. The lower panel depicts the trace and auto-correlation plots under BRLVOF$^{S}$, for the component of $\beta_M$ that describes the correlation between the ADL score and anaemia.}
\label{fig:test}
\end{figure}
\FloatBarrier


\section{Simulation with blocking}
We present a simulation scenario that incorporates blocking, and show the possible advantages of assuming that the variables exclusive to one files are also correlated among non-links. We consider files $\mathbf{A}$ and $\mathbf{B}$ of sizes 500 and 1000, respectively.
\FloatBarrier
\begin{table}[t] 
\caption {Average TPR, PPV, F1, bias, and RMSE across varying error rates under $BRLVOF$ and $BRLVOF_{Ind}$.} \label{tabbl}
\centering
\begin{tabular}{cc|ccccc}
$\epsilon$ & Method  & $\overline{TPR}$ & $\overline{PPV}$ & $\overline{F1}$ & $\overline{Bias}$ & RMSE \\ \hline
  \multirow{2}{*}{0.0} &$BRLVOF$ &  { .9991(.0014)} & .9923(.0049) & .9987(.0034) & .042(.007) & 0.043 \\
 & $BRLVOF_{Ind}$ & { .9879(.0016)} & .9814(.0046) & .9800(.0025) & .050(.009) & 0.051 \\ \hline
  \multirow{2}{*}{0.2}& $BRLVOF$ &  { .9791(.0016)} & .9760(.0067) & .9767(.0025) & .071(.007) & 0.072 \\
 & $BRLVOF_{Ind}$ & { .9701(.0013)} & .9718(.0023) & .9789(.0023) & .080(.007) & 0.080 \\ \hline
 \multirow{2}{*}{0.4}  & $BRLVOF$ &  { .9103(.0013)} & .9107(.0023) & .9009(.0024) & .100(.009) & 0.098 \\
 & $BRLVOF_{Ind}$ & {.9101(.0012)} & .9100(.0023) & .9009(.0015) & .120(.008) & 0.099 \\ \hline
\hline
\end{tabular}
\end{table}
\FloatBarrier
Both datasets contain $S = 250$ blocks. Consequently, there are 2 records within each block in file $\mathbf{A}$, and 4 records within each block in file $\mathbf{B}$. The total number of true links is 250, so that there is one true link per block. For record $j$ in the $q^{th}$ block in file $\mathbf{B}$, we generate $\mathbf{X_{Bj}}$ as
\begin{align}
 \mathbf{X_{B_j}} = -1+\mathbb{I}(S_j = q) + \alpha_j,
\end{align}
where $\alpha_j \overset{iid}\sim N(0,1)$, and $\mathbb{I}(.)$ is an indicator function that takes the value 1 if its argument holds, and 0 otherwise. We generate $X_{Aj} \sim N(10 + \beta_M\mathbf{X}_{Bj},0.1)$ if $j \in \mathbf{M}$, and $X_{Aj} \sim N(5 + \delta_U\mathbf{X}_{Bj},0.1)$ if $j \in \mathbf{U}$. We set $\beta_M = 6$ and $\delta_U=0.5$. We consider the same linking variables as in Section 4.1 of the main text, namely an individual's gender, ZIP code, and date of birth (DOB). Each linking variable is randomly perturbed with probability $\epsilon = (0.0,0.2,0.4)$. 

Table \ref{tabbl} displays results undfer BRLVOF when modeling associations between $\mathbf{X_{A}}$ and $\mathbf{X_{B}}$ among the unlinked record pairs ($BRLVOF$), and when assuming that they are independent ($BRLVOF_{Ind}$). For both procedures, we generate 1000 MCMC samples and discard the first 100. When $\epsilon=0$, $BRLVOF$ shows improvements over $BRLVOF_{Ind}$ in terms of $\overline{TPR}$, $\overline{PPV}$, $\overline{F1}$, as well as $\overline{Bias}$ and RMSE. These gains are more subtle, but noticeable, as the error level $\epsilon$ increases.


\printbibliography